\documentclass[5p,times]{elsarticle}

\usepackage{amsmath,amssymb,graphicx}
\usepackage{mathrsfs}

\usepackage[pdftex,dvipsnames]{xcolor}
\usepackage{xargs} 
\usepackage{hyperref}
\hypersetup{pdflang={English},pdfauthor={QuAIL},colorlinks=true}

\usepackage{ulem}

\newcommand{\newcite}[1]{\textcolor{Cerulean}{[\cite{#1}]}}

\usepackage{comment}
\usepackage[colorinlistoftodos,prependcaption,textsize=tiny]{todonotes}
\newcommandx{\unsure}[2][1=]{\todo[linecolor=red,backgroundcolor=red!25,bordercolor=red,#1]{#2}}
\newcommandx{\change}[2][1=]{\todo[linecolor=blue,backgroundcolor=blue!25,bordercolor=blue,#1]{#2}}
\newcommandx{\info}[2][1=]{\todo[linecolor=OliveGreen,backgroundcolor=OliveGreen!25,bordercolor=OliveGreen,#1]{#2}}
\newcommandx{\improvement}[2][1=]{\todo[linecolor=Plum,backgroundcolor=Plum!25,bordercolor=Plum,#1]{#2}}

\newcommandx{\thiswillnotshow}[2][1=]{\todo[disable,#1]{#2}}

\begin{document}

\hypersetup{
  linkcolor=ForestGreen,
  urlcolor=RoyalBlue,
  citecolor=Cerulean
}

\begin{frontmatter}

\title{Assessing and Advancing the Potential of Quantum Computing: A NASA Case Study}

\author[1]{Eleanor G. Rieffel\corref{cor1}}
\ead{eleanor.rieffel@nasa.gov}

\author[4,3]{Ata Akbari Asanjan}
\author[1,3]{M. Sohaib Alam}
\author[1,2]{Namit Anand}
\author[1,3]{David E. {Bernal Neira}}
\author[1,3]{Sophie Block}
\author[1,2]{Lucas T. Brady}
\author[1,2]{Steve Cotton}
\author[1,3]{Zoe Gonzalez Izquierdo}
\author[1]{Shon Grabbe}
\author[1,3]{Erik Gustafson}
\author[1,3]{Stuart Hadfield}
\author[1,3]{P. Aaron Lott}
\author[1,3]{Filip B. Maciejewski}
\author[1,2]{Salvatore Mandr\`a}
\author[1,3]{Jeffrey Marshall}
\author[1,2]{Gianni Mossi}
\author[1,2]{Humberto Munoz Bauza}
\author[1]{Jason Saied}
\author[1,3]{Nishchay Suri}
\author[1,3]{Davide Venturelli}
\author[1,3]{Zhihui Wang}

\author[1]{Rupak Biswas}

\cortext[cor1]{Corresponding author}

\affiliation[1]{organization={Quantum Artificial Intelligence Lab. (QuAIL), Exploration Technology Directorate}, 
    addressline={NASA Ames Research Center},
    city={Moffett Field}, 
    state={CA},
    postcode={94035}, 
    country={USA}}

\affiliation[2]{organization={KBR}, 
    addressline={601 Jefferson St.},
    city={Houston}, 
    state={TX},
    postcode={77002}, 
    country={USA}}

\affiliation[3]{organization={USRA Research Institute for Advanced Computer Science (RIACS)}, 
    city={Mountain View}, 
    state={CA},
    postcode={94043}, 
    country={USA}}

\affiliation[4]{organization={Data Science Group (DSG), Exploration Technology Directorate}, 
    addressline={NASA Ames Research Center},
    city={Moffett Field}, 
    state={CA},
    postcode={94035}, 
    country={USA}}

\date{\today}

\begin{abstract}

Quantum computing is one of the most enticing computational paradigms with the potential to revolutionize diverse areas of future-generation computational systems. While quantum computing hardware has advanced rapidly, from tiny laboratory experiments to quantum chips that can outperform even the largest supercomputers on specialized computational tasks, these noisy-intermediate scale quantum (NISQ) processors are still too small and non-robust to be directly useful for any real-world applications. In this paper, we describe NASA’s work in assessing and advancing the potential of quantum computing. We discuss advances in algorithms, both near- and longer-term, and the results of our explorations on current hardware as well as with simulations, including illustrating the benefits of algorithm-hardware co-design in the NISQ era. This work also includes physics-inspired classical algorithms that can be used at application scale today. We discuss innovative tools supporting the assessment and advancement of quantum computing and describe improved methods for simulating quantum systems of various types on high-performance computing systems that incorporate realistic error models. We provide an overview of recent methods for benchmarking, evaluating, and characterizing quantum hardware for error mitigation, as well as insights into fundamental quantum physics that can be harnessed for computational purposes.

\end{abstract}


\end{frontmatter}

\tableofcontents

\section{Introduction}
\label{sec:Introduction}

We describe work by 
NASA's Quantum Artificial Intelligence Laboratory (QuAIL) and collaborators on assessing algorithms and architectures across application areas targeted by NASA. 
This broad range of research activities includes developing fundamental theory, compilation and error mitigation, quantum machine learning, optimization, and simulation algorithms, as well as benchmarking and analysis of algorithms and hardware.
Much of this work is motivated and enabled via close collaborations with industry, government, and academic groups building quantum hardware, and illustrates the benefits of algorithm-hardware co-design in the near-term Noisy-Intermediate Scale Quantum (NISQ) era \cite{Preskill2018quantumcomputingin}.
This work builds on prior QuAIL team work \newcite{Rieffel_2019, Biswas_2017}  
and focuses on computational challenges in optimization, machine learning, simulations, tools supporting evaluation and algorithmic development, and theory and numerical work that deepens understanding of physical mechanisms that can be harnessed for quantum computation. 
For clarity, citations in double brackets, [[]], highlight works involving QuAIL team members, while citations in single brackets, [], refer to third-party works.

{Throughout our work, hardware-algorithm co-design is always present, with algorithmic needs feeding into the hardware design, hardware capabilities, and limitations feeding into algorithm design, and hardware noise characterization feeding into error mitigation techniques and the development of noise-aware algorithms.}
Key to our research has been open quantum system modeling, analysis, and simulation, leading to an improved understanding of physical implementation and co-design across multiple architectures, including gate-model, measurement-based,  and quantum annealing (QA)  architectures. Much of this work has been focused on available and proposed NISQ architectures, enabling empirical investigations for testing and developing theories and guiding new algorithm and hardware co-design approaches based on effective and efficient application-driven requirements to guide between where hardware stands and purely theoretically motivated architectures and algorithms.
This work enabled a deeper understanding of the mechanisms underpinning quantum annealing, such as tunneling, thermalization, and many-body delocalization, and their interaction with embedding parameters and annealing schedules. 
Early work on annealing architectures has been extended to and built upon to analyze multiple gate-based architectures and algorithms.

As part of the NASA community, the QuAIL team has had access to domain experts with a variety of challenging computational problems. These interactions, and those with partners, have led to work developing benchmark instance sets capturing aspects of computational challenges arising in space, aeronautics, and earth science missions, as well as in the exploration of quantum, quantum-classical hybrid, and physics-inspired approaches to tackling these problems. Some example application areas include planning and scheduling, anomaly detection, fault diagnosis, autonomous air traffic management, machine learning, and simulation of materials, chemistry, and high energy physics.    
The QuAIL team has led and supported multiple benchmarking and analysis activities focused on evaluating quantum architectures and approaches, comparing quantum and quantum-classical hybrid algorithms against state-of-the-art classical approaches, and providing resource estimation for running such algorithms at application scale on future quantum architectures. 

There are recurring themes and lessons learned from our recent work, which include:
\begin{itemize}
\item benefits of hardware-algorithm co-design in both the near-term NISQ era and the long-term fault tolerant era;
\item usefulness of experimentation on NISQ devices, from error characterization to insights into quantum and quantum-classical hybrid algorithms in the fault-tolerant and application-scale regimes;
\item  physics-inspired classical algorithms can be used at application scale today;
\item synergistic advances of classical and  quantum machine learning (QML) research, including quantum-ready classical algorithms that can be used today but are ready for replacement of some classical subroutines with quantum subroutines once quantum  hardware matures sufficiently;
\item the use of classical high-performance computing (HPC) for validation and benchmarking of quantum devices, quantum simulations, and algorithm exploration;
\item importance of and methods for generating hard problem instances for benchmarking quantum, quantum-inspired, and classical devices. 
\end{itemize}

This paper is organized as follows.
Sec.~\ref{sec:QAlg} surveys recent work on quantum algorithms for optimization and sampling.
Quantum-compatible machine learning work is covered in Sec.~\ref{sec:QMLAlg}.
We discuss applications of quantum computing to simulating quantum systems, including quantum chemistry, materials, condensed matter, high energy, and fundamental physics in Sec. \ref{sec:QCforSim}.
Tools for supporting quantum computing investigations are described in Sec. \ref{sec:Tools}.
Lastly, Sec.~\ref{sec:MechForQC} focuses on work associated with mechanisms for quantum computation.

\section{Quantum Optimization Algorithms and Sampling}
\label{sec:QAlg}
Quantum algorithm research can be broken into two broad categories: algorithms focused on near-term devices and applications and algorithms focused on long-term fault-tolerant computers. Near-term algorithm research enables demonstrations of quantum computing capabilities on current devices and the testing of error characterization, mitigation, and correction protocols within certain algorithms. Of particular interest is algorithmic work on near-term devices that can give insight into quantum algorithms that will be relevant in the fault-tolerant era. 
Hardware-algorithm co-design is of particular importance in the pre-fault-tolerant era and in the early fault-tolerant era when resources will be heavily constrained. To date, a number of quantum algorithms have been proposed for combinatorial optimization problems (e.g.,~\cite{Farhi_2001,farhi2014quantum,farhi2015quantum,Cerezo2021,Montanaro2020,Alexandru_2020}
). Not only does the size of current quantum hardware pose a challenge, but they are highly susceptible to various sources of error. Designing algorithms that can be run on the limited hardware of today and give insight into the design of quantum algorithms that will be effective in the fault-tolerant era is a challenge. 
Not only do the limitations and constraints of current quantum hardware influence algorithmic work, but algorithmic considerations also impact hardware designs and priorities. 
This cycle of co-design will first be illustrated in the setting of quantum optimization.

Quantum algorithms come in many different forms, from gate-based models to quantum annealing, and are diverse enough in scope to also include sampling methods and binary decision problems.
Quantum computers are always hybrid in nature; they rely on classical controllers. Many of the most effective algorithms, even in the future, are expected to be hybrid, integrating communication between the quantum and classical processors during algorithm execution. 
Optimization will provide our first and most extensive look at hybrid and well as fully quantum algorithms. 
A broad recent overview of quantum optimization~\newcite{abbas2023quantum} examines potential opportunities and challenges across many approaches and application areas.

A new paper, \newcite{bernal2024utilizing}, surveys HPC and quantum computing methods and perspectives for optimization.

\subsection{QAOA Analysis and Development}
The Quantum Approximate Optimization Algorithm (QAOA), a prominent near-term quantum optimization algorithm \cite{farhi2014quantum}, was extended to the Quantum Alternating Operator Ansatz by the QuAIL group 
\newcite{Hadfield_2019,Rieffel_2019}.
This algorithm uses a bang-bang quantum circuit that alternates between pulses related to the problem Hamiltonian and a mixer Hamiltonian. The pulse lengths, variational parameters optimized by an outer classical loop, aim to produce a state minimizing the problem Hamiltonian's energy. Analytical parameter determination remains an open question, with ongoing advancements including hardware-algorithm co-design strategies that can extract the full potential of currently available quantum devices.

In the cited work \newcite{Sud2022_parameter},
advancements in parameter setting strategies for QAOA were introduced, focusing on constraint satisfaction problems with polynomial growth in distinct cost values. The model, rooted in empirical QAOA observations, demonstrated effectiveness in the numerical evaluation of the heuristic for MaxCut on unweighted Erd\"os-R\'enyi random graphs, finding that for three levels of QAOA, the heuristic matches approximation ratios previously achieved with globally optimized methods. Additionally, for levels up to 20, the heuristic showed parameters with monotonically increasing approximation ratios, indicating its scalability and potential efficacy in deeper QAOA applications.

Despite much effort, obtaining rigorous performance bounds for quantum algorithms such as QAOA beyond relatively basic cases remains challenging. 
In \newcite{wang2018quantum}, analytical results were obtained for MaxCut problems on graphs utilizing the lowest depth realization of QAOA. 
It was shown that the expected value of each term in the cost function, corresponding to the graph edges, depends only on the local neighborhood of each edge. This property extends to deeper QAOA circuits but over larger neighborhoods. Similar notions of local algorithms exist classically.

In the recent work of \newcite{Marwaha2022}, the power of both quantum and classical local algorithms for approximately solving Max kXOR, a natural 
generalization of MaxCut from graphs to hypergraphs was considered. 
In Max kXOR, each constraint is the
XOR of exactly k variables and a parity bit. 
On instances with either random signs (parities) or no overlapping clauses and D+1 clauses per variable. This is analogous to triangle-free and regular graph instances of MaxCut, respectively. The expected satisfying fraction of the depth-1 QAOA is exactly calculated and compared with a generalization of
the classical local threshold algorithm from \cite{hirvonen2014large}. Notably, the quantum algorithm outperforms the threshold algorithm for $k > 4$.
On the other hand, potential difficulties for QAOA to achieve computational quantum advantage on this problem are highlighted. We first compute a tight upper bound on the maximum satisfying fraction of nearly all large random regular Max kXOR instances by numerically calculating the ground state energy density P(k) of a mean-field k-spin glass \cite{sen2018optimization}. The upper bound grows with k much faster than the performance of both local algorithms. A new obstruction result for low-depth quantum circuits (including QAOA), when $k = 3$, generalizing a result of \cite{Bravyi2020} for the $k = 2$ case, is also identified, and it is conjectured that a similar obstruction exists for all k. Such obstruction results are important steps towards a better understanding of the power of quantum circuits for optimization problems.

Several different QAOA approaches to solving the combinatorial circuit fault diagnostic (CCFD) problem are introduced and analyzed in \newcite{Leipold2022}. Comparing 
these approaches on a family of dense and highly connected circuits supports the intuition that approaches that are more closely tailored to exploiting the structure of the underlying optimization problems can have better performance than more general approaches.

In \newcite{Hadfield2023a}, a framework for analyzing layered quantum algorithms such as quantum alternating operator ansätze is developed. The framework relates quantum cost gradient operators, derived from the cost and mixing Hamiltonians, to classical cost difference functions that reflect the cost function neighborhood structure of the problem at hand. Exact general expressions for expectation values are derived as series expansions in the algorithm parameters, cost gradient operators, and cost difference functions. This enables novel interpretability and insight into QAOA behavior in various parameter regimes, including but not limited to small-parameter regimes. For single-level QAOA1, it is shown that the leading-order changes in the output probabilities and cost expectation value explicitly for arbitrary cost functions, demonstrating that, for sufficiently small positive parameters, probability flows from lower to higher cost states on average. This result is leveraged as it is shown that QAOA always beats random guessing. Several example applications of the framework, including Quadratic Unconstrained Binary Optimization (QUBO) problems and variants of MaxSat, are also considered, as well as the extension to mixing unitaries beyond the transverse-field mixer for constrained optimization problems~\newcite{Hadfield_2019}.

To better understand the behavior of QAOA,  an approach to analyzing deep circuits with gradually varying unitaries is introduced in \newcite{Kremenetski2023} and note several surprising phenomena. First, the ground state of the mixer Hamiltonian directly connecting to a highly excited state of the cost Hamiltonian will result in poor QAOA performance even in the adiabatic limit. Secondly, shallower circuits are shown to outperform deeper ones when the parameters are larger due to inherently non-adiabatic effects. Lastly, these phenomena, along with small parameter approximations~\newcite{Hadfield2023a} are shown to explain a general qualitative feature in performance \newcite{kremenetski2021quantum} for deep QAOA circuits with slowly varying parameters.

The real-world limitations of NISQ processors,
(e.g.,~the number of qubits, native gate sets, limited connectivity, effects of noise) have led the quantum computing community to design hardware-efficient ans\"atze, an example of hardware-algorithm co-design.
In Ref.~\newcite{LaRose2022}, a new ansatz that combines the mixing and phase-separation operators into a more general two-parameter family of operators is introduced, which is referred to as ``Quantum Alternate Mixer-Phaser Ansatz" (QAMPA). The primary motivation for this new ansatz is to reduce the depth of circuits, which should improve the performance on NISQ processors. For fully connected binary quadratic optimization problems, the circuits compile to roughly half the depth of standard QAOA on quantum processing units (QPUs) with nearest-neighbor connectivity. Also, our numerical noiseless simulations demonstrate that QAMPA performs almost on par with standard QAOA in parameter regimes that are achievable in current hardware.
In Ref.~\newcite{maciejewski2023design}, a Time-Block ansatz is proposed that requires only linear connectivity between qubits. Over-parametrization allows for more refined control over the circuits' sublayers. 
As an additional contribution, the ordering of gates is used as a variational parameter, which is shown to greatly improve the performance in practice.
Reported experimental implementation by the QuAIL team in collaboration with Rigetti Computings on Rigetti's QPU pushes the state-of-the-art to a new scale, demonstrating variational quantum-classical optimization on 50-qubit systems using up to $\sim 5,000$ two-qubit gates to approximately solve the fully-connected Sherrington-Kirpatrick model.
Despite such complexities, the algorithmic performance gains are maintained over a random guessing solver given the same number of function calls. Furthermore, an increase in performance with circuit depth is observed.

In Ref.~\newcite{maciejewski2024ndar}, the QuAIL team developed a novel meta-algorithm, Noise-Directed Adaptive Remapping (NDAR) for improved quantum optimization in the presence of certain types of hardware noise.
The NDAR involves a feedback loop, where each step is a stochastic optimization, such as QAOA implemented on QPU. 
The algorithm assumes a known "attractor`` state to which the noisy dynamics of a quantum device converges with time and noise accumulation.
This is fulfilled experimentally in the case of, for example, amplitude damping, a type of noise prevalent in superconducting architectures.
The aim of NDAR is to iteratively improve alignment between the low-energy states of the problem Hamiltonian and the attractor state of the noise.
This is achieved by using the best solution from the previous step to re-map the problem Hamiltonian solved in the next step using so-called bitflip (or spin-reversal) gauge transformation.
The obtained problem is equivalent to the original one, but the attractor state of the noise now becomes a better approximation to the ground state of the cost Hamiltonian.
In cited Ref. \newcite{maciejewski2024ndar}, the team benchmarked NDAR experimentally on the newest generation of Rigetti's superconducting chips, Ankaa-2.
The reported approximation ratios are within the range $0.9$-$0.96$ for random, fully connected graphs on $n=82$ qubits, using only depth $p=1$ QAOA applied in conjunction with NDAR.
This compares to $0.34$-$0.51$ for vanilla $p=1$ QAOA with the same number of function calls. 
As such, the obtained results are among the most complex and best-performant quantum optimization experiments performed to date (see, for example, \newcite{abbas2023quantum} for an overview of recent experimental progress).

Applying QAOA algorithms to problems with constraints presents an implementation challenge for near-term quantum resources. Ref.~\newcite{Hadfield_2019} presented a general framework for designing constraint-specific mixers.  In Ref.~\newcite{Wang2020}, strategies for enforcing hard constraints by using XY-Hamiltonians as mixing operators (mixers) are explored.  Under the constraint that an integer variable admitting $\kappa$ discrete values, encoded into qubits through the one-hot-encoding, XY Hamiltonian that preserves the total $Z$ is a natural and efficient mixer choice.  Despite the complexity of simulating the XY model, it is demonstrated that certain classes of the mixer Hamiltonian can be implemented without Trotter error in depth $O(\kappa)$. General strategies for implementing QAOA circuits on all-to-all connected hardware graphs and linearly connected hardware graphs inspired by fermionic simulation techniques are also specified. The algorithmic performance is validated on graph coloring problems that are known to be challenging for a given classical algorithm. The general strategy of using XY-mixers is borne out numerically, demonstrating a significant improvement over the general X-mixer.  The generalized W-state, an eigenstate of the XY mixer, which also corresponds to the symmetric (to particle permutation) state in the feasible subspace, yields better performance than easier-to-generate classical initial states when XY mixers are used.  The XY mixer is further examined for graph coloring problem in the presence of noise in Ref.~\newcite{Streif2021}.  The evolution, which in the noiseless case would preserve the subspace defined by the symmetry, could break the symmetry and introduce non-valid states.  The probability of staying in such symmetry-preserved subspaces under noise was analyzed, providing an exact formula for local depolarizing noise. These findings are applied to benchmark, under depolarizing noise, the symmetry robustness of XY-QAOA, which has local particle number conserving symmetries. The influence of the choice of the problem encoding on the symmetry robustness was also analyzed to provided guidance on realistic considerations in implementing symmetry-specific mixers for QAOA applications.

\subsection{Entanglement verification in QAOA}
\label{sec:entVerf}

A key ingredient of quantum algorithms is entanglement, although much about entanglement in quantum algorithms remains unknown. Verifying entanglement can be exponentially expensive. 
When information about the states to be verified is known, it may be possible to identify entanglement witnesses, operators that have bounds on their expectation values for all non-entangled
states, and have expectation values that violate those bounds in the state whose entanglement is to be verified. A famous example is the observables involved in  Bell inequalities. In the case of states arising in  QAOA applied to MaxCut, Ref.~\newcite{Sohaib2022a} constructed entanglement witnesses that have only twice, or at most thrice, the number of terms as the underlying problem Hamiltonian. Moreover, in order to verify entanglement, one needs only measure QAOA-MaxCut states in 3 bases, measuring all qubits in the $X$, $Y$, or $Z$ basis. The work included experiments on Rigetti quantum processors that demonstrate this approach for verifying entanglement. The same work also developed signatures of coherence, another key property of quantum behavior, and demonstrated that the states obtained by running QAOA for MaxCut on Rigetti QPUs   
maintain coherence for sizable problems of interest.

\subsection{Iterative Quantum Algorithms}

Iterative quantum algorithms were originally proposed by \citet{Bravyi2020} in the form of recursive QAOA (RQAOA), an extension based on QAOA.  This algorithm solves MaxCut problems by using a quantum algorithm to identify the most probable edge to fix before reducing the problem size by fixing that edge.  Our work extended this into a full framework of iterative quantum algorithms where the algorithm can be broken down into three parts \newcite{Brady2023, dupont2023quantum}:
\begin{enumerate}
    \item Preparation Rule: prepare some information about the system.  This can be preparing a quantum state using an optimization algorithm, sampling using a classical Monte Carlo, calculating graph properties, or anything else to gather information.
    \item Selection Rule: using the information from the previous step, rank features of the problem, selecting a feature that you are most confident about altering or fixing.
    \item Reduction Rule: Take the given feature and eliminate it from the problem. Usually, this involves fixing some information, such as collapsing two graph nodes into one.  This reduction should be tracked and can be back-tracked through at the end of the procedure.
\end{enumerate}
After each round of this procedure, the problem will be smaller and smaller until the optimal solution for this smaller problem is tractable to find.  Then, this solution can be taken to back-track through the reduction rules to recover a candidate solution for the full problem.

In Ref.~\newcite{Brady2023}, these techniques are applied to a Maximum Independent Set problem, developing several variants of the Iterative Quantum Algorithm for this problem.  Most notably, path-sum analysis techniques are applied to prove that an Iterative Quantum Algorithm using depth $p=1$ QAOA as its preparation rule only considers the same graph properties (i.e. degrees of all the nodes) as a classical greedy algorithm.  Numerical simulations further verify that this quantum algorithm always makes the same selection choices that a classical greedy algorithm would, giving it the same performance and provable bounds as the classical algorithm.  This is significant because it puts lower bounds and classical guarantees on the behavior of the quantum algorithm.
At higher depths, $p>1$, the classical side no longer offers such guarantees, but our numerical results show that the quantum algorithms outperform the classical greedy algorithm.

In~\newcite{dupont2023quantum}, the QuAIL team worked with our hardware collaborator Rigetti Computing to introduce an iterative quantum heuristic optimization algorithm to solve combinatorial optimization problems. 
The quantum algorithm is implemented on Rigetti's programmable superconducting quantum system using up to 72 qubits for solving paradigmatic Sherrington-Kirkpatrick Ising spin glass problems and observing an absolute performance better than a random sampling baseline.

Together with the Rigetti Computing team, we collaborated in \newcite{dupont2024quantum} to propose a hybrid quantum-classical algorithm to solve the MaxCut problem on 3-regular graphs up to several thousand variables. Inspired by semidefinite programming and considering the structure of the problem, the proposed method solves problems beyond the number of qubits in Rigetti's quantum computer. This approach was compared to state-of-the-art classical methods and achieved an average performance of 99\% on a large set of instances, making it competitive with respect to high-performance classical methods.

\subsection{Quantum Annealing Advances}
\label{sec:updatesqa}

Quantum annealing has been a major research focus area of the QuAIL team since its earliest days \newcite{Rieffel_2019, Biswas_2017}.
Quantum annealing \cite{Kadowaki1998,Farhi2000} is a general and universal \cite{Aharonov2008} setting for quantum computation that relies on continuous time evolution of a quantum system under mixing and problem Hamiltonians.  Quantum annealing is similar to QAOA discussed before but relies on slow evolutions and quantum adiabaticity. 
This evolution can often be realized on physical devices that require embedding of the problem into the native structure of the device, and often fine tuning of the annealing schedule is needed to realize improvements.
Quantum annealing has so far demonstrated quantum speedups over classical optimization algorithms only in very limited circumstances. One reason is the need for \emph{minor embedding} which maps optimization problems of interest (e.g. SAT, vertex-cover) to the type of problem natively optimized by a quantum annealer, an Ising Hamiltonian with connectivity determined by the architecture, 
which usually is an Ising spin Hamiltonian of the form
\begin{equation}\label{eq:spin_hamiltonian}
H = -\sum_{\langle i,j \rangle \in E} J_{ij} \sigma_i^z \sigma_j^z - \sum_{i \in V} h_i \sigma_i^z.
\end{equation}
The qubit operators $\sigma^z_i$ label the vertices of a graph $G=(V,E)$ and the two-qubit interaction terms $J_{ij} \sigma_i^z \sigma_j^z$ are associated to the edges $\langle i,j \rangle \in E$ of $G$.
While the coupling constants, $J_{ij}$, and the local fields, $h_i$, can be tuned to a degree, the graph $G$ itself is fixed once and for all by the architecture of the machine and is limited by engineering considerations.
If the target cost function of the problem of interest is not of the form in Eq. \eqref{eq:spin_hamiltonian}, one needs to devise a way of mapping it to a cost function that the machine can accept.
This usually requires using auxiliary qubits and representing the logical values of the original problem as long chains of physical qubits coupled together.
This mapping to a more complicated system was observed to significantly reduce the performance of quantum annealers compared to their performance on their native problem due to the freezing of such chain in the ``wrong'' configurations, in a way that the quantum annealer is no longer able to change their states dynamically.

Many practical optimization problems have high connectivity.
Minor embedding addresses this mismatch by effectively increasing the connectivity by grouping qubits together to act as one logical qubit, and adding terms to the cost function that penalizes configurations in which the qubits making up a logical qubit do not align. Parameters are chosen so that the ground state is preserved by the embedding. 

In \newcite{Izquierdo2021}, certain practical aspects of embedding for optimization problems through experimental demonstrations on a commercial quantum annealer were investigated.
The effects of embedding and its interplay with other annealing parameters (particularly those related to the annealing schedule) on the performance of the device for solving optimization problems were studied, aiming to provide both deeper insights into the physics of quantum annealing devices as well as pragmatic recommendations for their use. Two performance-enhancing methods previously applied to native optimization problems were adapted and shown to be advantageous also for embedded ones. A novel approach to enable gauge transformations for problems with the qubit coupling strength $J$ in an asymmetric range is introduced, making it amenable to embedded problems. An annealing schedule with an appropriately located pause was also confirmed to still improves performance in the embedded case and explored how the optimal location shifts with the magnitude of the ferromagnetic coupling $|J_F|$, thus extending the theoretical picture for pausing and thermalization in quantum annealing to the embedded case.

Our follow-up work \newcite{Izquierdo2022} delved deeper into this theory and practical recommendations through demonstrations of an updated annealing architecture and several problem classes requiring embedding. Various aspects of the physics-based picture previously explored were confirmed, and taken further by identifying certain characteristics of an optimization problem that are predictive of its hardness for currently available quantum annealing devices. Based on these results,  a set of qualitative guidelines for parameter setting in quantum annealers was presented that is expected to be particularly useful for users coming into quantum annealing from other areas of optimization, and that should help maximize the performance of these devices without requiring a prohibitively large amount of resources.

Also focusing on practical aspects and existing annealing devices is~\newcite{Pokharel2023}, in which the performance of quantum annealers has improved since their inception was explored, with a comparative study carried out throughout the years using each new available processor. This work highlights how the improvements in connectivity, leading to smaller embeddings, boost the optimization capabilities of these machines.

While the embedding process for optimization is sound by construction, the same is not true for sampling. Ref.~\newcite{Marshall2020} showed that even with an annealer that returns a perfectly classical Boltzmann (thermal) distribution of the final Hamiltonian $e^{- H/T}$, the process of embedding drastically alters the statistics. Not only is the distribution `hotter' (larger $T$), it is also strictly not Boltzmann anymore. A follow-up work \newcite{Marshall2022a} applied a similar analysis of the quantum version of the problem (sampling a quantum Boltzmann distribution, where the Hamiltonian has non-zero off-diagonals), which showed similar effects; quantum observables were distorted by the embedding, including shifting the location of a phase transition. The message in these works is clear: annealers with naive embedding can not be used for classical or quantum thermal sampling.

While these sampling capabilities are still limited, they can be useful for certain scenarios. For native problems (i.e. not needing embedding), it was demonstrated in~\newcite{Gonzalez2020} that using a pause and quench schedule, can be used to probe the thermal properties of the strictly quantum Hamiltonian that describes the system at the time of the pause, allowing us to distinguish between graphs whose classical Ising spectra are the same, which could not be done with a standard annealing schedule.
The theoretical work \newcite{Knysh2020} demonstrates that embedding chains typically crop up as harmful Griffiths phases but can also be used as a resource to balance out singularities in the logical problem, changing its universality class.

Ref \newcite{bernal2020integer} used constrained programming to find provably optimal embeddings for quantum annealers.
A mixed-integer programming formulation was also developed, and a comparison of these approaches was performed to find embeddings of various problems in D-Wave's quantum annealers. Although these approaches are generally not capable of finding embeddings as efficiently as the heuristics, these methods were well suited for those cases that were pathologically difficult for the heuristic methods, 

In quantum annealing, finding a good schedule is often essential to obtaining quantum advantage. Like finding good parameters in QAOA,  optimizing the schedule can be itself a difficult task.  Therefore, it is essential to create ans\"atze that can be effectively optimized with as few variational parameters as possible.  QAOA serves as an effective ansatz in a gate-based computer, but a quantum annealer has additional degrees of freedom in its schedule.  Ref.~\newcite{unsal2022b} introduced a new parameterization of the quantum annealing schedule based on clipped polynomials.
 
In addition to introducing this schedule parameterization, a benchmarking against against QAOA was performed, introducing notions of emulation time. 
Additionally, the ability of this polynomial schedule to numerically emulate QAOA was shown, producing an equivalent or better final state using the same number of variational parameters.
Further, the polynomial schedule can produce states that are strictly better than those produced by QAOA.  
This shows increased power for annealing schedules, but it requires much more experimental and physical control. 

Ref.~\newcite{Pintos2023} derived a version of quantum speed limits specific to quantum annealing. 
The work explored this bound in several toy problems and showed that this bound is saturable for some quantum annealing procedures, including the Hamming spike, the $p$-spin model, and the adiabatic unstructured search.  These problems all represent extreme toy problems, so further refinements are needed to make this bound saturable for more realistic and less symmetric problems.

Further advances related to quantum annealing are discussed in Secs.~\ref{sec:relaxation} and \ref{sec:accelquanttunel}.

\subsection{Quantum Optimization Applications}
\label{sec:optimization_applications}

The QuAIL group has a long history of working on planning-related problems \newcite{Rieffel_2019, Biswas_2017}. Extending our previous work for planning problems, \newcite{Stollenwerk2020} presents novel efficient QAOA constructions for optimization problems over proper colorings of chordal graphs. The effectiveness of these constructions was demonstrated using the flight-gate assignment problem, in which flights are assigned to airport gates to minimize the total transit time of all passengers. Feasible assignments correspond to proper graph coloring of a conflict graph derived instance-wise from the input data. Further, \newcite{Stollenwerk2020b} investigated the feasibility of applying quantum annealing to solve a simplified air traffic management problem (strategic conflict resolution) for wind-optimal trajectories. Our mapping is performed through an original presentation of the conflict-resolution problem in terms of a conflict graph, where the nodes of the graph represent flights and the edges represent a potential conflict between flights.

\label{sec:constrainedprogramming}
In a pair of papers, \newcite{booth2020quantum, Booth2021}, we were able to speed up certain highly utilized subroutines for solving constraint satisfaction problems, given access to a fault-tolerant quantum computer. One key result leveraged quantum search to accelerate filtering for the `alldifferent' constraint (all variables take on unique values). This work showed how to incorporate quantum filtering algorithms into a hybrid classical-quantum backtracking search protocol. Resource estimates suggest constraint programming is a promising candidate application for early fault-tolerant quantum computers.

In \newcite{Shaydulin2021}, the relationship between QAOA and the underlying symmetries of the objective function is explored, particularly in hard problem classes where a nontrivial symmetry subgroup can be obtained efficiently.
The work shows how symmetries of the objective function imply invariant measurement outcome probabilities across states connected by such symmetries, independent of the choice of algorithm parameters or number of layers. 
Using this result, machine learning techniques can predict QAOA performance based on symmetry considerations. Numerical evidence suggests that a small set of graph symmetry properties suffices to predict the minimum QAOA depth required to achieve a target approximation ratio on the MaxCut problem in a practically relevant setting where QAOA parameter schedules are constrained to be linear.

\subsection{Distributed Quantum Computing}

We have also looked at distributed graph problems, in which a network of 
computers, each with some information about the graph, collaboratively work to compute a global graph property.
Ref. \newcite{Kerger2023a} presents two algorithms in the Quantum CONGEST-CLIQUE model of distributed computation that succeed with high probability: one for producing an approximately optimal Steiner tree and one for producing an exact spanning arborescence of minimum weight, the analog of a minimum spacing tree in a directed graph.

The CONGEST distributed computational model allows messages of limited size to be transmitted within a network described by a communication graph of size $n$ in a series of rounds to address a computational problem. The size limitation for such messages is \(O(\log(n))\) bits at each edge of the communication graph per round. The communication graph in the CONGEST-CLIQUE model is fully connected. In the quantum CONGEST-CLIQUE model, at most \(O(\log n)\), classical and quantum bits (qubits) can be communicated across each edge of the communication graph per round.

Each of these algorithms uses \(O(n^{1/4}\log^k n)=\tilde{O}(n^{1/4})\) rounds of communication and \(\tilde{O}(n^{9/4})\) messages, achieving a lower round and message complexity than any known algorithm in the classical CONGEST-CLIQUE model.
At a high level, these results are achieved by combining classical algorithms with fast quantum subroutines.
These asymptotic speedups further contribute to understanding what problems can be solved more efficiently when quantum communication is allowed in this CONGEST-CLIQUE model of distributed computation.

As a final result, the pre-factors accompanying the polylogarithmic terms derived in the complexity analysis were estimated.
By performing such an analysis, it was discovered that, although providing an asymptotic speedup, these distributed quantum methods would only surpass the performance of trivial classical algorithms when the graphs to be processed (i.e., the network of quantum computers) would have over \(10^{21}\) nodes.
Such results only motivated us to work on improving these distributed algorithms, for which our current work addresses generalizations of these methods.

\subsection{Quantum Algorithm Design and Problem Encoding}
QAOA and other quantum approaches to optimization can also be applied to problems on d-level variables. Building off of previous work that considered the qubit case in detail \newcite{Hadfield_2019,hadfield2021representation}, Ref~\newcite{sawaya2023encoding} presents a toolkit for design and analysis of discrete (i.e., integer-based) optimization problems, wherein the problem and corresponding algorithmic primitives are expressed using a quantum intermediate representation that is encoding-independent. This compact representation often facilitates efficient problem compilation and comparison between different encoding choices, for example between variants of binary and unary encodings for fixed d. Numerical studies comparing several qubit encodings exhibit a number of preliminary trends toward guiding the choice of encoding for a particular set of hardware, problem, and algorithm. 
For moderate sized up to 16-level quantum variables, low-depth mixing operators are constructed, demonstrating that binary encodings remain amenable for QAOA.
 
Diagrammatic representations of quantum circuits offer novel approaches to their design and analysis. Extensions of the ZX-calculus~\cite{van2020zx} especially suitable for parameterized quantum circuits are proposed in \newcite{stollenwerk2022diagrammatic}, in particular for computing observable expectation values as functions of the quantum circuit parameters, which are important algorithmic quantities in a variety of applications ranging from combinatorial optimization to quantum chemistry. 
In particular, formal rules for dealing with linear combinations of ZX-diagrams are given, where the relative complex-valued scale factors of each diagram must be kept track of, in contrast to most previously applications. 
The diagrammatic approach is shown to offer useful insights into algorithm structure and performance through direct application to several example applications drawn from the literature including realizations of hardware-efficient ansatze and QAOA, where calculations can become more intuitive and potentially easier to approach systematically than by alternative means. Recent related work has further considered applying similar diagrammatic approaches to the study of barren plateaus~\cite{zhao2021analyzing} and to quantum machine learning~\cite{toumi2021diagrammatic}.

\subsection{Physics Inspired Heuristics}

One of the first and most significant boons of quantum computing has been its inspiration and advancement of classical computing methods.
Physics-inspired heuristics provide completely classical methods of solving hard problems in new ways.
As part of our work in this area, a set of tools and methods that not only implement some of these physics- and quantum-inspired heuristics, but also integrate them with other solution techniques to address challenging computational tasks, using quantum computing only as a source of inspiration have been developed.

As mentioned above, there is a significant effort from the QuAIL team to understand quantum algorithms for optimization, namely quantum annealing and QAOA.
These methods are designed to find ground states of transverse field Ising problems, which, from an optimization standpoint, correspond to QUBO problems.
A classical algorithm to address these problems is parallel tempering, a modification of simulated annealing that has proved to be one of the strongest contenders when comparing the performance of quantum methods for optimization.
An efficient Python implementation of the parallel tempering algorithm has been provided as part of the open-source project PySA \newcite{Mandra_PySA_Fast_Simulated_2023}.

Another contribution is in the tools required to address constrained optimization problems using physics- and quantum-inspired heuristic methods.
Although there are known mappings of constrained optimization problems to the QUBO formalism, which is amenable for these heuristics and other quantum algorithms, their efficient implementation is challenging.
The group has also developed tools for the community trying to use and understand these methods when facing applications, usually represented through constrained optimization.
These tools are provided as software written in the Julia programming language, under the umbrella of the package QUBO.jl \newcite{Xavier_Qubo_jl_2023}.
Within it, there are tools for reformulating mixed-integer nonlinear programming problems into QUBO, connecting these instances with classical and quantum solvers, and analyzing both the instances and results provided by the solvers.

Although mapping constraints as penalizations might be sufficient for their enforcement, this method requires finding these penalization factors, which might lead to challenging numerical issues for both classical and quantum methods.
A way of leveraging these Ising solvers, potentially quantum, while enforcing the constraints exactly for mixed-binary quadratic programs was proposed.
This method relies on a copositive reformulation of the problem and the application of a cutting-plane algorithm, which generates increasingly accurate linearizations of the problem, for which center solutions are computed classically, derived from the possibly suboptimal solutions of QUBO problems found by the Ising solvers.
This method, described in \newcite{brown2023copositive}, also provides convergence guarantees to provable optimal solutions despite the heuristic nature of the Ising solvers, and for a test case of finding the Max-clique of random graphs, it is able to match and even surpass state-of-the-art solvers for these constrained problems.

On the other hand, the group has also worked on the simulation of coherent Ising machines (CIM), a non-conventional architecture for solving Ising problems heuristically.
The CIM dynamics are described by a set of ordinary differential equations, which can also address continuous non-convex quadratic optimization.
The dynamics of such continuous variable CIMs (CV-CIM) use optical pulses to perform the optimization.
Updates to those pulses are computed via stochastic gradient descent.
Classical optimization techniques have been used to improve the dynamic simulations of CV-CIM \newcite{Brown2024_accelerating}, namely momentum and Adam updates, to address the weaknesses of stochastic gradient descent found in classical optimization.
Through this modification, the CV-CIM's convergence, sample diversity, and stability can be significantly improved.

Continuing with the CIM work, the physics and the performance of this class of machines, which are faithfully described by a set of coupled ordinary differential equations ODEs, was investigated as part of the NTT Phi Lab collaboration and the NTT NSF Expeditions program on Coherent Ising Machines led by Stanford University. The fast clock speed and relatively mature engineering in the field of photonics allows for the practical possibility of these systems, either simulated in CPUs, GPUs, FPGAs or realized via integrated photonic circuits to be deployed as components to solve contrived problems that have crucial time-threshold, such as the MIMO wireless Maximum-Likelihood decoding problem already subject of Ref.~\newcite{Kim2021} (more on this below). Indeed, elaborated fully-electronic or opto-electronic implementation of the CIM machines are at the moment among the most performant Ising solvers known for a variety of challenging benchmarks, according to a recent review~\newcite{mohseni2022ising}.
In \newcite{singh2021ising}, the team investigate the basic numerical implementation of the CIM equations for the purpose of wireless decoding, setting up problems using the Rayleigh fading channel model and calculating resource requirement for a high-speed Ising machine to increase the overall throughput of current state-of-art network implementations. It is noteworthy that the work spurred multiple derivative improvements from other groups, and led to current related lines of research active, such as a full-FPGA optimized implementation and a neural-network representation of the CIM using Deep Operator Networks~\newcite{taassob23}.

The sign problem is one of the biggest impediments to simulating quantum systems using Monte Carlo techniques.  When Monte Carlo techniques are used to simulate a Hamiltonian or system with a sign problem, the pseudo-probabilities for the Monte Carlo method oscillate rapidly between positive and negative values (or arbitrary complex phases when simulating real-time evolution), making evaluation of averages and integrals exponentially difficult.  Most Monte Carlo techniques for simulating quantum systems are, therefore, limited to sign-problem-free Hamiltonians, a vanishingly small fraction of all possible systems.

There are methods for mitigating but not solving the sign problem, such as one already used in high energy physics, involving integration along Lefschetz Thimbles \cite{witten2010new,witten2011analytic}.  These thimbles are a generalizations of the complex plane of stationary phase.  In broad terms, these methods seek to evaluate a highly oscillatory integral with a sign problem by deforming the surface of integration into the complex hyper-plane and deforming the surface iteratively to try to get it to flow to a Lefschetz Thimble.  The idea then is that on this thimble, the integral is easy to evaluate since its phase is no longer oscillatory.  These methods are themselves highly computationally taxing, and at best they can mitigate the slow-down of the sign problem to a lesser exponential factor.  Our work with collaborators \newcite{Mooney2022}, sought to extend these ideas from continuous variable high energy physics to the discrete variables of quantum spin systems.  This process is non-trivial since the complexification requires continuous variables, requiring us to look at over-parameterized continuous bases for quantum spin systems, such as the spin coherent states, which themselves require approximations that are sometimes ill-justified for low spin number.  Our work successfully implemented these Lefschetz Monte Carlo methods and demonstrated their effectiveness for a small three-spin system with a sign problem.

An example of physics-inspired heuristic applied to real-world application is the use
of the parallel tempering algorithm for soft MIMO decoding in 5G (\texttt{ParaMax}) \newcite{Kim2021}.
Here, we demonstrate that \texttt{ParaMax} can achieve near optimal maximum-likelihood
throughput performance in the Large MIMO regime, Massive MIMO systems where the base station
has additional RF chains, to approach the number of base station antennas, in order
to support even more parallel spatial streams. These results are achieved by 
using the ParaMax Ising Solver (PMIS). It is based on simulated annealing, 
featuring a parallel tempering algorithm highly-tailored to optimize the Ising model of MIMO detection.
While the front-end of our PMIS implementation is in Python, 
the core is completely written in $C^{++}$. To further maximize the performance
of PMIS to satisfy limited processing time in wireless standards, the following
innovations have been implemented: the use of static memory to improve compile-time optimization; the use of parameter pack expansion to unroll vector-vector and matrix-vector multiplications; and the use of SIMD instructions to further improve operations like vector-vector and matrix-vector multiplications.

One of the main reasons classical heuristics fail to identify solutions of optimization problems
is being stuck in local minima \newcite{BENDALL2006288}. Thermal cycling is an optimization algorithm
based on simulated annealing that uses cycles of heating and quenching,
while following a decreasing temperature schedule, to escape
local minima and improve the performance of local heuristics \newcite{Mobius1997}.

In \newcite{Barzegar2021}, a comprehensive parameter tuning of the algorithm was performed and it was 
demonstrated that it competes closely with other state-of-the-art algorithms such as parallel
tempering with isoenergetic cluster moves \newcite{zhu2015efficient,mandra2016strengths,mandra2018deceptive}, while overwhelmingly outperforming more simplistic 
heuristics such as simulated annealing. In details, the thermal cycling algorithm works by preparing
an ensemble $\Omega$ 
of $N_p$ lower energy states among $N_0$ quenched
random configurations. Starting from the initial inverse temperature $\beta_i = 0$, 
the above pool of states is annealed toward a final inverse temperature of $\beta_f$
in $N_T$ steps. At any given temperature step, a given number of states in $\Omega$
are update using Metropolis (heating) followed by an immediate quench via a local search
method (cooling). If any of the new states have an energy smaller than its original state,
such states are updated in $\Omega$. In practice, the above process
steers the ensemble toward the low-lying states while ensuring that meta-stable configurations do not hinder the dynamics. The temperature is then reduced, and the cycles start over.

The main advantage of algorithms like the thermal cycling is the possibility
of using a variety of variable-update classes in the quenching
phase. By using carefully designed updates, the exploration of the exponentially many meta-stable
can be avoided in favor of lower-energy states. However, considering the extra computational
cost associated with tailored updates, it is important to considering the trade-off between
a better exploration of the configuration landscape and the induced overhead.
By accurately choosing the optimal update strategy, it was demonstrated that the thermal cycling algorithm is a competitive heuristic 
when complex structures are present, avoiding being trapped in local minima.
While the thermal cycling algorithm performs the worst when narrow barriers are present,
where algorithms like simulated quantum annealing and isoenergetic cluster
updates perform the best, it was observed that the thermal cycling algorithm
works the best on dense graphs with wide barriers.

\section{Quantum-compatible Machine Learning Algorithms}

\label{sec:QMLAlg}
Over the past few years, significant strides in the field of machine learning (ML) have been made, particularly in integrating quantum computing with ML algorithms (QML). This fusion has been explored in both discriminative models like classification and generative models, focusing on understanding and leveraging quantum probability distributions. The central challenge has been effectively integrating samples from quantum processors into state-of-the-art ML models, such as Variational AutoEncoders (VAE) and Generative Adversarial Networks (GAN). The QuAIL team has worked with collaborators, from domain experts to hardware groups, to advance these algorithms. These efforts on specific applications in aeronautics and Earth science including image segmentation and modeling for wildfire detection, time-series analysis for flight anomaly detection, and parameter learning for material science modeling. In all cases, there is a tight connection between the algorithm and the hardware, illustrating the potential for specialized quantum hardware that have less stringent requirements than universal quantum processors and so could be built more easily and put to use earlier than universal processors, and also the potential of algorithms that can be tailored to emerging quantum hardware. This connection enables a virtuous cycle of hardware-algorithm co-design.

This section delves into these hybrid ML models, examining how quantum samples, particularly from Boltzmann Machines, are integrated within the latent spaces of these models. In several instances, these quantum-compatible methods, based on discrete latent space models, have provided significant value over standard approaches, demonstrating continued opportunity in QML research. However, the utilization of samples from noisy, sparse latent distributions, characteristic of modern quantum hardware, remains a challenge. Overcoming these hurdles is vital, especially considering the demands of high-performance computing environments.

There are four distinct paradigms for how to combine machine learning with quantum computing:
\begin{enumerate}
    \item \textbf{Classical-Classical (CC):} This approach leverages quantum-inspired methods within discrete latent space models, showcasing how quantum influence can enhance traditional algorithms.
    \item \textbf{Classical-Quantum (CQ):} Exemplified by models such as the Quantum-assisted Variational Autoencoder (QVAE), this paradigm involves the use of quantum computing elements to process and interpret classical data.
    \item \textbf{Quantum-Classical (QC):} Models optimizing quantum heuristics with classical machine learning techniques and reflects the application of classical ML techniques to optimize or interpret results obtained from quantum simulations or quantum processors. This paradigm reflects the integration of quantum computational power in solving complex quantum problems, which are then analyzed and optimized using classical ML techniques. 
    \item \textbf{Quantum-Quantum (QQ): } Although not explored in this section, this approach signifies the complete integration of quantum data with quantum computational methods.
\end{enumerate}
With the exception of subsection \ref{sec:meta}, this section mostly concentrates on advancements in CQ paradigm. 
 
\subsection{Quantum-Assisted Variational Autoencoder} \label{sec:QVAE}
Both quantum and classical Boltzmann Machine (BM) are able to model powerful and flexible probability distributions. However, training these models classically is challenging and time-consuming. Quantum annealers, regarded as analog quantum devices, can effectively simulate Quantum BMs (QBMs). Ref. \newcite{Gao2020} introduced a quantum-assisted VAE (QVAE), which incorporates samples from a quantum annealer into the prior of a discrete VAE. The QBM component of the model features a tunable transverse field, enhancing the prior's expressiveness and allowing for a more focused latent distribution without compromising the network's learning ability. This model excels at handling high-dimensional data spaces, a significant challenge for classical machine learning algorithms \newcite{Gao2020}. 

The QVAE's encoder comprised two fully-connected layers leading into a hierarchical posterior, facilitating complex posterior distributions and a tighter Evidence lower Bound (ELBO). To prevent overfitting, the decoder has a simpler structure with two fully-connected layers \newcite{Gao2020}.

An integral part of the study was comparing the performance of a Restricted BM (RBM), a classically simulated QBM, and a QBM trained with samples drawn from the D-Wave 2000Q quantum annealer. This device, featuring 2048 qubits each of degree 6, required embedding any RBM larger than 6 nodes per side with a coupling strength of -1. The logical value of a variable at the end of an anneal was determined by a majority vote, adding a layer of robustness to the modeling process \newcite{Gao2020}.

After training, the QVAE's encoder indexed the dataset efficiently without extra discretization due to its binary latent space. The encoding algorithm clustered similar objects under identical bit strings and an inverted index was constructed to map each bit string to its corresponding data points. Since each item is processed once and independently, the original data can be removed from the main memory, reducing memory usage. For queries, the QVAE encodes the item and sorts occupied bit strings by their Hamming distance to the query embedding, balancing memory efficiency and search resolution. The results were significant \newcite{Gao2020}:
\begin{itemize}
\item \textbf{Embedded Proximity and Hamming Distance:} Experiments validated that Hamming distance in the compressed space effectively approximates Euclidean distance in the original space, as shown through k-Approximate Nearest Neighbor Search (ANNS) on the Moderate Resolution Imaging Spectroradiometer (MODIS) dataset. This underscores the efficiency of quantum-assisted techniques in high-dimensional data retrieval.
\item \textbf{Impact of the Transverse Field:} Adjusting the transverse field parameter influenced the distribution's characteristics and search speed, with optimal speedup observed at certain ranges. This highlights the precise control quantum models offer in data representation and optimization.
\item \textbf{Memory Consumption:} The quantum model showed superior memory efficiency over methods like Hierarchical Navigable Small World (HNSW) and Localality Sensitive Hashing (LSH), particularly for large datasets like the complete MODIS dataset, underscoring the models practicality and scalability.
\end{itemize}
The effective utilization of quantum annealing in handling latent spaces by the QVAE indicates potential quantum enhancements for other ML models, like the supervised U-Net-VAE hybrid used in Ref. \newcite{Asanjan2023} for wildfire detection. This model, integrating a U-Net architecture with a VAE, comprises four submodels: (1) a prior network, (2) a posterior network, (3) a U-Net network for feature extraction, and (4) a combination network. It is effective in generating stochastic wildfire segmentations and simulating unknown wildfire scenarios. Incorporating quantum-sourced samples could further enhance its predictive accuracy and robustness, particularly in analyzing high-dimensional satellite imagery, offering a promising direction for future research in remote sensing and wildfire analysis \newcite{Asanjan2023}.

\subsection{Optimizing Adversarial Networks Through Quantum Annealing} \label{sec:QAAAN}
In exploring the integration of quantum annealing with deep learning architectures, Ref. \newcite{Wilson2021} introduced the Quantum-assisted Associative Adversarial Network (QAAAN). This model incorporates quantum annealing to train a BM that optimizes the feature distribution extracted by the discriminator network in the adversarial model. The unique capabilities of quantum computing are leveraged to find more effective representations in the latent space, thereby enhancing the generation of realistic data. This approach, particularly significant for its exploration of reparametrization of discrete variables, demonstrates a crucial step in integrating quantum models with traditional neural networks \newcite{Wilson2021}.

In the QAAAN study, the researchers investigated various topologies for the probabilistic graphical models in the latent space, including complete, symmetric bipartite, and Chimera topologies.  It is noteworthy that the choice of topology affects both the model's learning rate and quality, similar to how transverse fields influenced the QVAE’s performance in Ref. \newcite{Wilson2021b}. This insight is crucial for future developments in quantum-enhanced deep learning, as it highlights the need for careful consideration of model architecture to harness the full potential of quantum computing.

The QAAAN's exploration of topological variations extends to practical performance metrics. The researchers utilized the Inception Score and the Frechet Inception Distance to assess the model's ability to generate realistic data. This evaluation was pivotal in establishing the QAAAN's efficacy in real-world applications. The research also demonstrated the model's scalability by successfully applying it to complex datasets, such as the LSUN bedrooms dataset. This scalability indicates the potential for quantum-assisted models in handling large-scale, high-dimensional data challenges \newcite{Wilson2021}.

\subsection{Advancements in Quantum-Inspired Generative Models: RBMs in VAEs and Invertible Flows} \label{sec: RBM_Enhanced}
Researchers in \newcite{oconnor2021rbmflow} introduced models that mark a significant advancement in combining Energy Based Models (EBMs) with Invertible Flows (IFs), specifically RBM-Flow and D-Flow. These non-autoregressive IFs enable enable exact likelihood training and efficient sampling without requiring a discriminator network, which is a notable difference from VAEs, autoregressive models, and GANs, respectively \newcite{oconnor2021rbmflow}. 

RBM-Flow, a subclass of EBM-Flow, leverages RBM as its trainable, base distribution, thus enhancing the model's capacity to capture complex data distributions. D-Flow, an evolution of RBM-Flow, zeroes all couplings in the latent RBM, allowing the encoding of global features as discrete labels in the latent space, offering a structured approach for meaningfully handling high-dimensional date. The usage of RBMs in RBM-Flow primarily enhances their energy-based properties and discrete nature to boost the generative capabilities of the IF framework \newcite{oconnor2021rbmflow}. 

\subsection{Quantum-Compatible Discrete Deep Generative Models} \label{sec:anomaly}
In \newcite{templin2023anomaly}, deep generative learning models are developed and applied to the task of anomaly detection in a commercial flight-operations dataset consisting of multivariate time series. Specifically, the performance of three unsupervised deep generative models are explored which consist of variational autoencoders with Gaussian, Bernoulli, and Boltzmann priors. Two of the VAEs contained discrete latent variables (DVAEs), one with a factorized Bernoulli prior and one with a RBM prior. The work demonstrated the competitiveness of a discrete deep generative model with its Gaussian counterpart for anomaly detection tasks.Also of note, the DVAE model with RBM prior can be readily integrated with quantum sampling by replacing its generative process with measurements of quantum states obtained from quantum hardware devices, such as a quantum annealer or gate-model device.

\subsection{Quantum-Enhanced Optimization of Quantum Heuristics} \label{sec:meta}
VQAs, categorized under quantum heuristics, show great potential for practical quantum computing applications. The optimization of these algorithms for effective hardware performance is a critical area of focus. Ref. \newcite{Wilson2021b} assessed the efficacy of a Long Short Term Memory (LSTM) recurrent neural network model (the meta-learner) in optimizing two quantum heuristics, comparing its performance traditional optimizers (Bayesian optimization, evolutionary strategies, L-BFGS-B and Nelder-Mead). 

The meta-learner outperformed traditional optimizers, such as Bayesian optimization, evolutionary strategies, L-BFGS-B, and Nelder-Mead, demonstrating superior performance in noisy environments. For example, starting from the around the same point in Fermi-Hubbard model problems, the L-BFGS-B performance reduced by 0.35 whereas the meta-learner only reduced by 0.2 \newcite{Wilson2021b}, indicating that meta-learning will be especially useful in noisy near-term quantum heuristics implemented on hardware. In addition to robustness, the meta-learner showed a higher frequency of reaching near-optimal solutions than the next best optimizer (evolutionary strategies) in noisy simulations \newcite{Wilson2021b}. 

Looking ahead, the continued improvement of meta-learning methods is anticipated. Despite the current lack of investigation into their performance scaling to larger problem sizes, largely due to the challenges in simulating large quantum systems, the potential for these methods on hardware implementations is significant. The meta-learner introduced by \newcite{Wilson2021b} applies a single model across various parameters (a 'coordinatewise' approach). Envisioning a 'qubitwise' approach, where distinct models are trained for each qubit's parameters in a given hardware graph, could open up new optimization avenues. Such a method might account for the unique physical characteristics of each qubit, potentially leading to more finely-tuned optimizations tailored to specific hardware environments. This direction underscores the importance of developing meta-learning methods that not only adapt to the complexities of quantum problems but also leverage the peculiarities of quantum hardware.

Ref.~\newcite{Wilson2021b} also notes the different implementations of QAOA used for Graph Bisection and MAX-2-SAT. Understanding the impact of mixer and initial state variations on performance, as well as characterizing the relative power of different QAOA mixers, remains an open area of research.

\section{Quantum Computing for Simulation} 
\label{sec:QCforSim}
Our recent work related to quantum simulations for condensed matter physics, quantum computing for material discovery, quantum simulation for chemistry, and open quantum system simulation are highlighted in this section. Among the many advancements showcased in this section are recent works that use variational techniques for both weakly and strongly correlated systems that are able to reach chemical accuracy with a minimal number of time steps; extension of existing neural network techniques for solving electronic structure problems; and a further reminder of the importance of classical HPC computing methods for benchmarking and quantum chemistry. 

\subsection{Quantum Simulation for Condensed Matter Physics}
\subsubsection{Kitaev Spin Models}
Quantum spin systems, notably the Kitaev honeycomb model, are promising candidates for demonstrating the potential capabilities of quantum computers over classical ones in applications beyond simulation. Kitaev's influential work \cite{KITAEV20062} introduced this well-studied model, which is especially attractive for quantum algorithm research due to its simple lattice structure and the simplicity with which its Hamiltonian can be expressed in terms of Pauli operators. 

In the study \newcite{Li2023}, the team successfully prepared the ground state of the Kitaev honeycomb model using the Hamiltonian variational ansatz (HVA) \cite{PhysRevA.92.042303}. The model's square octagon lattice version was found to be particularly well-suited for implementation on Rigetti's Aspen-9 quantum processor. In practical experiments, their small-scale demonstrations of developed techniques, enhanced with error mitigation strategies, yielded results that closely aligned with theoretical predictions.

Further exploration into the Kitaev honeycomb spin model revealed an alternative approach \newcite{Jahin2022} , wherein the Hamiltonian is expressed purely in terms of fermionic degrees of freedom. This alternative fermionic description of the spin-system not only confirmed the model's exact solvability in the absence of interactions, but also led to the development of tailored variational quantum eigensolver (VQE) ansatze for different types of interactions. This approach allowed for accurate ground state preparations that aligned with theoretical results and demonstrated that for certain interactions, the number of required qubits could be cut in half \newcite{Jahin2022}.

\subsubsection{Fermi-Hubbard Model}
The Fermi-Hubbard Model (FHM) is crucial in quantum physics for exploring phenomena like superconductivity and insulating phases in materials. It describes the complex interactions of electrons moving within a lattice, offering critical insights into these states.

A significant study \cite{arute2020observation} simulated the dynamics of a one-dimensional FHM using a 16-qubit digital superconducting quantum processor. The simulation revealed distinct spreading velocities for charge and spin densities in highly excited regimes, surpassing the limits of the conventional quasiparticle model. To mitigate systematic errors and decoherence, the researchers developed an accurate gate calibration procedure and employed a series of error-mitigation techniques. These procedures allowed for the faithful simulation of the model's time evolution, even with over 600 two-qubit gates in the circuits. This work represents a crucial step towards practical quantum simulation of strongly correlated phenomena using current quantum devices, such as the classically hard 2D FHM  \cite{arute2020observation}.

Subsequently, \newcite{levy2022solving} explored a quantum annealing protocol for the FHM using a novel fermion-to-qubit encoding, which simplifies the FHM representation and enhances simulation. The proposed driver solutions targets high fidelity in various FHM interaction regimes, especially in intermediate and large U/t regimes. The protocol is designed for coherent evolution fermionic logical subspace, particularly suited for accurate simulation \newcite{levy2022solving}.

The study discussed potential scalability and hardware noise challenges, noting the lack of currently suitable hardware. However, the research suggested that advancements in quantum annealing hardware could enable practical implementation, positioning this method as a promising contender to gate-based algorithms for quantum advantage in non-fault tolerant regimes \newcite{levy2022solving}.

\subsection{Quantum Computing for Material Discovery}
\label{sec:QCforMaterDisc}
\subsubsection{Advancements in State Preparation Techniques}
Quantum computations offer significant advancements in simulating quantum many-body systems, with various approaches focusing on the efficient preparation of ground or excited states.

In the study \newcite{Kremenetski2021a}, researchers explored a novel method for simulating the time dynamics of Adiabatic State Preparation (ASP). ASP, which is an alternative approach to preparing ground states to study their wave-function and energy, starts with a known, easily prepared ground state Hamiltonian  and evolves it slowly to the target Hamiltonian, maintaining the system in its ground state, as per the quantum adiabatic theorem (ensuring the system remains in the ground state throughout the evolution).  The approach employs an adaptive sampling configuration interaction scheme, enhancing the efficiency of preparing ground and excited states, and is particularly effective in avoiding the complexities of optimization in noisy environments.

In \newcite{Woitzik_2020}, a systematic investigation into VQEs for determining ground-state energies and properties of two-dimensional model fermionic systems was presented. The research focuses on the efficiency of different entangler blocks in the VQE and how they impact the convergence to the system's ground state with minimal gate operations, crucial for NISQ devices. The study also monitors the entanglement during optimization using the concurrence measure and investigates the scaling of the VQE circuit depth as a function of the desired energy accuracy \newcite{Woitzik_2020}. In related work, \newcite{khan2023preoptimizing} presented and benchmarked an approach for finding good starting parameters for parameterized quantum circuits by classically simulating VQE by approximating the parameterized quantum circuit as a matrix-product-state (MPS) with a limited bond dimension.

Lastly, in \newcite{klymko2021real} a detailed analysis of variational quantum phase estimation (VQPE), a method rooted in real-time evolution for estimating ground and excited states on near-term quantum hardware. The study establishes the theoretical foundation of VQPE and highlights its efficiency in creating compact variational expansions for solving strongly correlated Hamiltonians. Central to VQPE is a set of equations with a geometrical interpretation, crucial for time evolution grid setup to decouple eigenstates from time-evolved expansion states. This connects VQPE to classical filter diagonalization algorithms. The paper also introduces a unitary formulation of VQPE, reducing the required quantum measurements, and analyzes the impact of noise, significantly improving upon previous considerations. Through both numerical simulations and a hardware implementation for the transverse field Ising model, the paper demonstrates VQPE's efficiency for both weakly and strongly correlated systems, reaching chemical accuracy with a minimal number of time steps \newcite{klymko2021real}.

\subsubsection{Many-body Scar Preparation}
Many body scars are an interesting phenomenon that seemingly violate the eigenstate thermalization hypothesis, similar to the phenomenon of many-body localization. Such scarred states give rise to long-lived coherent dynamics in an otherwise thermalizing system and may provide a path towards physical realizations of quantum memory. 

In Ref.~\newcite{Gustafson2023c}, state preparation protocols for scarred eigenstates and their superpositions that enable their dynamical simulation on quantum computers were proposed. In addition to deterministic methods that made use of the structure of such states and their superpositions as well as their tensor network representations that scale linearly with the system size, it was also found that there were stochastic circuits which can prepare certain target states in constant depth, though with exponentially vanishing post-selection probability. A mix of variational and MPS inspired circuits that were found numerically to construct the target states with great accuracy were employed. The team also provided proof-of-principle state-preparation demonstrations on superconducting quantum hardware. 

Quantum many-body scar states are highly excited eigenstates of many-body systems that exhibit atypical entanglement and correlation properties relative to typical eigenstates at the same energy density. Scar states also give rise to infinitely long-lived coherent dynamics when the system is prepared in a special initial state having finite overlap with them. When perturbed, the dynamics of such systems are generally hard to simulate classically, but efficient with quantum computers.

\subsubsection{Advancements in Hamiltonian Simulation Techniques}
The work presented in the paper \newcite{chamaki2022selfconsistent} is a significant stride in the realm of Hamiltonian simulations for chemistry applications, addressing the critical issue of gate complexity. Unlike traditional methods that primarily rely on the Hamiltonian itself, leading to approximations that may not fully capture the essential aspects of a problem, SQuISH innovatively uses both the Hamiltonian and an approximate eigenstate. This dual approach enables the creation of a truncated Hamiltonian with reduced complexity, enhancing the efficiency and accuracy of simulations. Integrating this with the ongoing efforts to minimize gate complexity, the Self-consistent Quantum Iteratively Sparsified Hamiltonian (SQuISH) algorithm stands out for its ability to balance computational efficiency with the depth of simulation, making it a pivotal development in quantum computing for chemistry and material science applications \newcite{chamaki2022selfconsistent}.

\subsection{Quantum Simulation for Chemistry}
\label{sec:QSforChem}
{In recent years, neural networks have provided an alternate framework for solving electronic structure problems, including applications in directly solving the time-independent Schr\"odinger equation to find approximate ground states of atoms and molecules (see for example, \cite{vonLilienfeld2020}). A notable example of such a neural network is Fermi Net \cite{Pfau_2020}. In \newcite{wilson2021simulations}, Fermi Net was improved with Diffusion Monte Carlo (DMC). Additionally, this work introduced several modifications to the network and optimization approaches that reduce the number of required resources while maintaining or improving the modeling performance.

In related work, \newcite{Wilson_2023} developed a periodic neural network \textit{Ansatz} for variationally finding the ground-state wavefunction of the homogeneous electron gas (HEG), which is a fundamental model of condensed matter theory. This work extended previous applications of neural networks to molecular systems with methods for handling periodic boundary conditions and includes two notable changes to improve performance: splitting the pairwise streams by alignment and generating backflow coordinates for the orbitals from the network.

A broader perspective on quantum computing for chemical engineering ~\newcite{Bernal2022,bernal2023impact,bernal2024utilizing} introduces concepts for an audience of chemical engineers and discusses three potential first applications for this field: computational chemistry, optimization, and machine learning.

\subsubsection{Advances in classical solvers for chemistry Hamiltonians} 

The electronic structure problem---i.e., the determination of the many-body wavefunction of all electrons in the field of fixed nuclear charges---is the central problem of quantum chemistry and it is most rigorously solved by finding the exact low-lying eigenstates of the electronic many-body Hamiltonian. Of course, because the number of basis states grows exponentially as the number of ways of arranging the electrons in the chosen one-particle orbital basis set, exact solutions are intractable on classical computing hardware for all but the simplest chemical systems. This, and the central importance of this problem, make it an ideal simulation target for quantum computers.  
Nevertheless, despite the promise/hope of quantum hardware-based solution methods, classical HPC computing methods remain crucial: (i) for benchmarking eventual fault-tolerant quantum hardware on realistic problem instances, and (ii) for preparing initial/starting states on quantum hardware which cannot be too far off from the true solution for the usual quantum algorithm to run efficiently (or at all). 

Thus, an ongoing focus of work in the group is to improve state-of-the-art classical approaches to the electronic structure problem, in particular, to develop new HPC-friendly algorithms which leverage the notion that, even for chemistry applications, one does not need to solve the exponentially-scaling eigenproblem to machine precision and, moreover, for initial state preparation, relaxed accuracy requirements may be more acceptable still.  
In this spirit, a truncated version of the traditional Davidson method for diagonalizing chemistry Hamiltonians has been developed \newcite{Cotton2022} which generates ``chemically-accurate'' solutions (answers within a chemically-relevant 1.6 milli-Hartrees of the exact solution) for the ground and first few low-lying excited electronic eigenstates at vastly reduced classical computational cost versus a full diagonalization, and that is competitive with other state-of-the-art approximate eigensolvers. Work-in-progress in the group involves developing efficiently parallel distributed-memory versions of the methodology which may be quite effective due to the structure of the newly-developed classical algorithms and their implementation.

\subsection{Open Quantum System Simulation}
Realistic quantum systems inevitably interact with external degrees of freedom (the environment), which leads to a non-unitary open system evolution.
This presents a challenge for the simulation of many relevant physical phenomena on conventional quantum computers that natively implement unitary gates. For this reason, there has been much less research into the emulation of open systems on unitary based quantum computers.
In a recent work 
\newcite{Suri2023twounitary}, presents an analysis of the resources required for prominent techniques to implement non-unitary operators, some of which incurred a prohibitive classical overhead. Improvements to these algorithms were made, by proposing a fully quantum method without such limitations. Additionally, a quantum algorithm that can decompose any $d$ dimensional operator into two separate unitaries, which can then be implemented deterministically, was proposed.

\subsection{Quantum simulation of High Energy Physics}
High energy physics provides an interesting avenue for quantum simulation and possible demonstration of quantum advantage. In this arena, several fundamental aspects of nature are described by quantum field theories, and in particular gauge theories. As a starting point however, it is worthwhile to consider simple quantum field theories, and in \newcite{kurkcuoglu2022quantum}, members of the team investigated how to simulate a simple scalar field on qudit hardware.

While many proposed methods exist for simulating gauge theories describing high energy physics, a computationally promising direction is to approximate continuous gauge groups with discrete subgroups, thereby translating the problem to simulating discrete gauge theories, which are more readily mapped onto digital quantum hardware.
Discrete groups are represented by mapping elements of the group to bit strings stored on registers of qubits.
It was shown in \cite{Lamm:2019bik} that 4 primitive group operations are necessary for Trotterized time evolution of pure gauge theories: group inversion, multiplication, phasing, and Fourier transformation.
Members of the team derived quantum circuits of the unitary operators for the dihedral groups ($\mathbb{D}_N$) \newcite{Alam2022b}, $\mathbb{Z}_2$ lattice gauge theory\newcite{charles2023simulating}, as well as a subgroup of SU(2), the binary octahedral group \newcite{Gustafson:2023kvd}.

In \newcite{Alam2022b}, members of the team laid out the framework for constructing quantum gates for discrete groups by enumerating the gates for $\mathbb{D}_N$, a simple non-Abelian group. While $\mathbb{D}_N$ theories map to the group $U(1) \times \mathbb{Z}_2$ in the large $N$ limit, which may by itself be of limited interest, the tools and methods developed extend to future work. In addition to the quantum circuits derived in \newcite{Gustafson:2023kvd}, the team found that many of the operators can be decomposed similarly to the classical fast Fourier transform; the operations can be broken down into operations on a series of subgroups. The study also investigated the resource costs for time evolution and compared it to the best reported standard; the team found this representation of an SU(2) gauge theory reduced the required T-gates for simulation by 8 orders of magnitude.

\section{Tools for Quantum Computing Investigations}
\label{sec:Tools}

Validation and benchmarking of quantum devices are critical for the development of such emerging technologies. Modeling quantum effects in software allows one to verify not only that the behavior is as expected but also to aid in the design of new devices. A prime example of this is a joint work between Google and NASA from 2019 and 2020, respectively, \newcite{Arute2019, Villalonga_2020}, which established for the first time a quantum over classical advantage, which used a combination of classical simulation (led by NASA) to verify experimental results. The team continues to develop technologies for quantum simulation in various settings and architectures and also to design methods for benchmarking and characterization.

\subsection{Classical Methods for Simulating Quantum Circuits}

Simulating deep quantum circuits with many qubits is a daunting task. Indeed, while 
quantum resources only scale polynomially with the number of qubits and the depth of the quantum
circuits, classical resources scale \emph{exponentially}. To overcome this limitation and further
raise the bar for a quantum advantage, many classical methods have been developed to tackle
different class of quantum circuits. For instance, it is well known that quantum circuits
composed of Clifford gates only \cite{gottesman1998fault,aaronson2004improved} can be simulated in polynomial time. Indeed, since
Clifford gates are nothing else than the stabilizers of the Pauli group, only a polynomial amount
of classical resources are required to store the full quantum state. However, since Clifford gates are neither
classical universal (that is, not all classical circuits can be described using Clifford gates) nor
quantum universal (that is, not all quantum circuits can be described using Clifford gates),
their application is limited to error-correction algorithms. Nevertheless, if the quantum circuit
is \emph{almost} composed of Clifford gates, the classical computational cost is exponential
in the number of non-Clifford gates only \cite{aaronson2004improved,bravyi2016improved}. 

In \newcite{Mi2021}, a technique called Clifford ``expansion'' was used to classically simulate large 
out-of-time-order correlators with a few non-Clifford gates, and compare the numerical results
with the experimental ones. While the Clifford expansion technique allow the simulation of deep circuits
with many qubits with a few non-Clifford gates, the method becomes quickly impractical.
While alternative methods that use approximate magic states \cite{bravyi2016improved} or noisy gates \cite{aharonov2023polynomial} may help
to mitigate the computational cost, Clifford expansion is limited to those quantum circuits with a limited
number of non-Cliffords.

On the other hand, methods based on tensor network contraction \cite{markov2008simulating} have been proven to be
of practical use to benchmark NISQ devices \cite{morvan2023phase,arute2020observation,kechedzhi2024effective,PhysRevX.9.021033,gray2021hyper,pan2022solving}. The basic idea
behind the tensor network contraction is to represent an arbitrary quantum circuits as a network
where each gate is a different tensor. Therefore, the calculation of amplitudes reduces to the
contraction of the tensor network \cite{markov2008simulating}. While the Clifford expansion method is limited by the 
number of non-Cliffords, the tensor network contraction is limited by a quantity called
``tree-width'', which is closely related to how close the network is to a tree. That is,
if the tensor network representing the quantum circuit is tree-like, 
its tree-width will be close to zero. On the contrary,
if the quantum circuit has many entangling gates and its tensor network representation is close
to a fully-connected graph, its tree-width will be of the order of the number of qubits. 
As demonstrated in \cite{markov2008simulating}, it is possible to deterministically simulate a quantum circuit
in a time that grows exponentially with the tree-width. However, finding the tree-decomposition
of a graph that minimize the tree-width is an NP-Complete problem \cite{arnborg1987complexity}. Moreover, 
the tree-width does not take into account the fact that, during the contraction, some tensors may
exceed the available memory. To mitigate this problem, techniques like slicing
have been introduced \cite{gray2021hyper,morvan2023phase,PhysRevX.9.021033,pan2022solving}, which add an exponential overhead to the contraction but keep the tensor contraction within the maximum amount of memory available.

Another widely used method to simulate quantum circuits is based on the MPS 
approximation \cite{verstraete2008matrix,vidal2008class}. An MPS is a quantum state of the form
\begin{equation}
|\psi\rangle = \sum_{\{s\}} \operatorname{Tr}\left[A_1^{(s_1)} A_2^{(s_2)} \cdots A_N^{(s_N)}\right] |s_1 s_2 \ldots s_N\rangle,
\end{equation}
where $A_i^{(s_i)}$ are square matrices of order $\chi$ (called local dimension), and the indices $s_i$ go over states in the computational basis. The order $\chi$ represents how much the qubits are \emph{entangled}:
more precisely, a system of non-entangled qubits can be represented by an MPS with $\chi = 1$. On the
other hand, a system of highly-entangled qubits will require $\chi$ being exponentially large
with the number of entangled qubits. Since the simulation cost is proportional to $\chi$, approximate MPS simulations can be performed by capping the largest allowed $\chi$ \newcite{morvan2023phase}.

Large-scale quantum computation requires encoding `logical qubits' using an error correction code. Due to a favorable topology, the surface code is a promising candidate for superconducting qubits. One challenge for superconducting qubits, however, is properly isolating the lowest two energy levels of an anharmonic oscillator, with leakage referring to the presence of a population outside of this subspace (which occurs via several physical mechanisms). The simulation of small error correction codes is also critical to aid in processor design, however, each additional level one wishes to simulate incurs an exponential overhead (e.g., for a single leaked state, a fully quantum simulation of $n$ qubits scales as $O(3^n)$ instead of $O(2^n)$). In \newcite{Marshall2023b} it was shown that in error correction codes under realistic situations, an approximation can be employed to faithfully simulate a system with leakage (of any number of leaked states), but still retaining the $O(2^n)$ scaling. 
Given access to HPC resources, this framework enabled a quantum simulation of a distance 5 surface code with leakage.
Moreover, this approximation allows one to justify a stronger approximation (Pauli twirling), which can then allow leakage to be simulated \textit{efficiently} (polynomially in $n$) via the stabilizer formalism.

\subsection{Photonics}
A heralded source of highly pure and indistinguishable photons is a prerequisite for many quantum information processing tasks in a photonic (e.g.,~linear optical) setting. Indeed, the degree of distinguishability is related to the classical hardness of Boson sampling \cite{Renema2018}, and the ability to generate entangled photonic states. One promising approach to this end is to use photon distillation, first observed in \cite{Sparrow2017} (by Sparrow and Birchall), and improved upon by \newcite{Marshall2022b}, where linear optics and post-selection is used to filter out erroneous photons in a heralded manner (in contrast to using spectral filters), in principle, to arbitrarily high purity/indistinguishability. 
This was recently improved in Refs. \newcite{saied2024general}, \cite{somhorst2024photon}, in which more general families of distillation protocols were given, leading to greater efficiency (i.e., linearly scaling resource requirements). 

Another key task for the development of quantum computing primitives is simulation, which is notoriously challenging due to the largeness of the Hilbert space; for $n$ photons in $m$ modes, it scales as $d_{n,m}:={n+m-1 \choose n}$. Indeed, the `naive' approach to simulation a la the second quantization scales linearly in the dimension. However in many situations much more efficient representations can be found (e.g.,~in certain cases a polynomially scaling simulation can be performed, similar in spirit to the stabilizer representation \cite{Rahimi-Keshari2016}). Ref.~\newcite{Marshall2023a} showed that by realizing Fock states as an approximate superposition of coherent states, a scaling equivalent to qubit systems, $O(2^n)$, can be achieved in the linear optical setting, for computing transition amplitudes. Moreover, such an approach is in many cases trivially parallelizable, since each coherent state in the representation can be updated independently of the others.

Unitary designs provide a systematic way to approximate Haar-random unitaries \cite{mele2023introduction} and have a multitude of applications in quantum computing such as randomized benchmarking \cite{dankert2009exact}, classical shadow tomography \cite{huang_predicting_2020}, etc. In a recent work by the NASA team, the structure of unitary designs in photonic architectures was investigated \newcite{Saied2024_spie}. Linear optical unitaries can \textit{only} form a \(1\)-design and are limited in their capabilities for various applications that requite \(t \geq 2\)-designs. It was proven that adding a single `SNAP' gate to the group of linear optical unitaries transforms them into a universal gate set for quantum computing and hence allows for the realization of \textit{arbitrary} \(t\)-designs in these systems. However, the structure of linear optics is such that the SNAP gate is a \textit{nonlinear} interaction, which is necessarily probabilistic. Understanding the cost of these nonlinear operations is an interesting research direction for the future.

\subsection{Quantum Computing Benchmarking}

As part of the effort to measure the performance of quantum computing algorithms, it was necessary to provide a set of benchmarking instances representative of various features of application problems and analysis tools able to synthesize the results in a way that makes them interpretable for users and developers of these computational tools.
As highlighted throughout the manuscript, the Hamiltonian formalism provides a well-suited way of incorporating practical applications for quantum computers.
Addressing this challenge is also an active area of research that is currently being pursued, and an initial contribution has been the release of the library of Hamiltonians, HamLib \newcite{sawaya2023hamlib}.
HamLib provides several hundreds of Hamiltonians representing instances of computational problems such as binary variable optimization, including MaxCut, Max-k-Cut, and QMaxCut; discrete variable optimization including the Max-k-SAT and the traveling salesperson problem; condensed matter physics models, including the transverse field Ising model, the Heisenberg model, Fermi-Hubbard model, Bose-Hubbard model; and problems in chemistry such as molecular electronic structure and molecular vibrational structure problems.
This work aims to provide researchers with prepared and mapped problem instances, which allow for reproducibility and standardization in research studies and further tests and developments of quantum algorithms.

Among these tests, Ref.~\newcite{lubinski2023optimization} showcases a collaborative effort focused on using optimization applications as quantum performance benchmarks.
In this work, the canonical instances of MaxCut were considered and addressed via QAOA, implemented in different hardware realizations of qubits, including superconducting and ion qubits and classical simulations, and QA implemented in D-Wave quantum annealers.
This work highlights the differences in performance between these algorithms and hardware implementations and provides an example of performance benchmarking of quantum computers by addressing an application.

To facilitate the benchmarking analysis, our team has also been working on implementing analysis tools, drawing inspiration from classical optimization and optimization research.
By considering parametrized stochastic optimization solvers as a mathematical framework able to describe QA and QAOA solution methods, among other classical and quantum approaches to optimization, a set of analytical techniques was developed to visually inform about the solver performances and aid in the parameter tuning procedure.
The benchmarking method is described in Ref.~\newcite{bernal2024benchmarking_short} and its implementation is available in Ref.~\newcite{E_Bernal_Neira_Stochastic_Benchmark_toolkit_2023}.

\subsection{Noise Modeling and Characterization}
The modeling and characterization of noise in near-term quantum computational devices are critical to aid in the design of such devices, with the aim of achieving fault tolerance. One key aspect to this is using the right metric; whilst full-scale state or process tomography would provide, in principle, all information about a system, it is intractable at modest sizes. It is, therefore, necessary to design metrics that distill the relevant information from the state/process but are still efficiently accessible experimentally. A common metric is the average fidelity; the overlap of the evolved state with the ideal target state, uniformly averaged over all initial states. However, this fidelity is only a single element in the $4^n\times 4^n$ process matrix describing the noisy evolution. In \newcite{Wudarski2020}, it was shown that averaging initial states non-uniformly allows for a fidelity metric that is determined by many elements of the transfer matrix. Thus, it can capture additional properties that the standard average fidelity is blind to.

Related to this discussion, in \newcite{Sud2022}, it was shown that approximations of global noise channels to local ones allow for an efficient protocol to extract the noise parameters of the process matrix. This work was targeted at repetitive circuits, such as QAOA, where the noise after each round can be estimated, with an experimental demonstration using Rigetti's quantum processor. 

Additionally, noise specifically in QAOA was studied in \newcite{Marshall2020b}, which gave analytical predictions of performance in the presence of local gate noise. This work additionally studied the trade-off between deeper circuits, which ideally do better at optimizing the cost function, but at the expense of more noise acting in the system, which tends to do the opposite.

Another aspect restricting quantum control are frequency fluctuations. In particular, low frequency (such as $1/f$ or other temporally correlated) noise has been analyzed \newcite{Wudarski2023a} in terms of noise statistics in a repetitive sequence of Ramsey pulses. Moreover, the same Ramsey protocol can be exploited in the nonergodic limit of the noise (shorter than the correlation length) \newcite{Wudarski2023b} to provide supplemental properties of noise, that are otherwise washed out at in the ergodic regime. 
Work led by the Google Quantum AI team \newcite{McCourt2023} showed experimentally that non-Gaussianity in low-frequency noise can be observed in superconducting qubits, by performing dynamical decoupling (DD) sequences. Additionally, in \newcite{evert2024syncopated} the QuAIL team worked with Rigetti to use DD to characterize undesired two-qubit couplings as well as the underlying single-qubit decoherence and in suppressing them. A syncopated DD technique is developed that protects against decoherence and selectively targets unwanted two-qubit interactions.

\subsection{Quantum Error Correction and Mitigation}
Quantum error correction (QEC) threshold theorems show that increasing the size (distance) of a QEC code can enable arbitrarily small logical error rates. However, a low enough (code dependent) physical error rate is required in order to guarantee one is in this regime; otherwise the additional qubits can actually cause the logical error rate to \textit{increase}. It was therefore a remarkable result by Google Quantum AI (with assistance in noisy simulations from NASA) in 2023 that showed experimental evidence of a logical error rate reduction by going from a distance 3 to distance 5 surface code \cite{Acharya2023}, a necessary step in the path towards fault tolerance. 

One of the remarkable advances in error correction in recent years was the discovery of Floquet codes with dynamical logical qubits. In these codes, the protected subspace changes dynamically, with the logical information updating so as to remain in the protected subspace. The first examples of such codes were honeycomb codes, with the techniques generalizing to trivalent lattices. Recently, members of the QuAIL team developed a Floquet code on a rectangular lattice \newcite{Alam2024Dynamical}.

In~\newcite{hu2022logical}, logical shadow tomography (LST), a error-mitigation technique that is inspired by the QEC theory and more suitable for implementation on near-term quantum computers was developed. 
LST is a technique to estimate error-mitigated expectation values on noisy quantum computers. 
The framework is a marriage and generalization of two powerful error mitigating methods: the subspace expansion and virtual distillation.
Our technique performs shadow tomography on a logical state to produce a memory-efficient classical reconstruction of the noisy density matrix. 
Using efficient classical post-processing, one can mitigate errors by projecting a general nonlinear function of the noisy density matrix into the code space. 
The method was shown to be favorable in the quantum and classical resources overhead. 
Our technique requires reasonable number of samples, $O(4^k)$, to estimate a logical Pauli observable with $[[N,k]]$ error correction code.
Relative to virtual distillation, this technique can compute powers of the density matrix without additional copies of quantum states or quantum memory. 

In~\newcite{akhtar2024dualunitary} we introduced a new classical shadow tomography protocol based on dual-unitary brick-wall circuits that we termed ``dual-unitary shadow tomography'' (DUST). We first showed that even at the \(2\)-qubit level, dual-unitaries outperform random Clifford unitaries in that they have a smaller shadow norm (and hence require fewer samples to estimate observables). This is intimately related to the fact that dual-unitary circuits have maximal operator entanglement while random Clifford circuits do not. We generalize this observation to the multiqubit case and by using a transfer matrix method, show that the fast-thermalizing properties of dual-unitary circuits make them better at predicting large operators compared to shallow brick-wall Clifford circuits. In fact, DUST outperforms standard Clifford shadows at any depth \(t\) for predicting extensive observables, and this effect is even more pronounced for small system sizes.

In recent work~\newcite{Suri2024}, an error mitigation method for certain realistic noise models that can be applied given partial knowledge of noise to achieve lower biases than typically possible with other methods was presented. 
Subspaces that decay uniformly were obtained and used to perform error mitigated quantum computation. The expectation values of dynamics encoded in such subspaces are unbiased estimators of noise-free expectation values.  
This theory was applied to a system of qubits and qudits undergoing relaxation with varying decay rates and show that such subspaces can be used to obtain unbiased estimation of the true expectation value up to second order variations in the decay rates. Given varying decay rates, this method was implemented in analog and circuit-model settings, improving on dual-rail qubits, and given partial knowledge of noise, show that it can outperform probabilistic error cancellation.

\subsection{Pulse Control}

Together with colleagues at the University of Michigan and Argonne National Lab, the application of various classical optimization techniques to pulsed or bang-bang controls for quantum problems was studied.  This work originally focused on various classical methods applied to the quantum control setting \newcite{Fei2023binarycontrolpulse} but has been expanded to considering new heuristics for reducing the number of control switches \newcite{fei2023switching} and to considering unitary control errors \newcite{fei2024binary}.  This work all relies on a continuous relaxation of the control problem, followed by various techniques for rounding and fixing the controllers to be binary pulses, akin to what would appear in QAOA.  Constraints on the length of the pulses and the switching time are also considered and incorporated into this framework.  Combinations of various classical techniques are used in attempts to improve on performance while satisfying physical and noise constraints on the system.  These techniques are applied to toy problems, such as minimizing the energy of a transverse field Ising model, compiling quantum circuits, and constructing quantum gates from physical controllers.

In \newcite{ozguler2022numerical}, the problem of co-designing algorithms on novel Circuit Quantum Electrodynamics (cQED) architectures enabled by 3D superconducting radio-frequency cavities such as those developed at SQMS~\newcite{alam2022quantum} is tackled numerically. This is accomplished by setting up a pulse-engineering approach to compiling the primitive gates of QAOA in qudit systems. The customized use of the {\tt{juqbox}} quantum optimal control package allowed the synthesis of a single round of the Max-k-Cut QAOA algorithm for arbitrary parameters at high-fidelity, on a realistic noiseless model describing the Hamiltonian of one or two cavity modes.
In subsequent work, other SQMS researchers generalized the model including the effect of additional modes~\cite{xu2022neural}.

\subsection{Generating Hard Problem Instances}

With the development of more competitive classical 
\cite{zhu2015efficient}, quantum \cite{Arute2019,morvan2023phase,kim2023evidence},
or ``quantum-inspired'' \cite{kim2021physics, aramon2019physics} heuristics,
it is becoming of utmost importance to identify classes of instances which 
are well suited to transversally benchmark such devices. While it is well known
that random instances have a hardness threshold \cite{mertens2006threshold,mezard2002analytic,mezard2002random,kirkpatrick1994critical}, it is a major challenges 
to designing planted problems with known solutions which are hard to optimize: 
indeed, planting a desired solution may affect the energy landscape of the instance,
making it more convex and therefore easier to explore. In the past few years,
multiple planting methods have been developed, including ``quiet'' planting
techniques \cite{qu1999hiding, barthel2002hiding, krzakala2009hiding,
zdeborova2011quiet, krzakala2012reweighted, sicuro2021planted}, ``patch'' planting \cite{Hen2015, Marshall2016}, and the use of cryptographic
protocols \newcite{mandrà2023generating}. Hard instances can also be generated
from more practical applications such planning problems~\cite{hen2014phase}.

Ref. \newcite{mandrà2023generating} proposed a novel way of randomly generating hard instances of disordered Ising Hamiltonians with unique planted ground states by casting the public key of the McEliece cryptographic protocol \cite{McEliece} as an interacting system of Ising spins, so that finding the ground state of the resulting Hamiltonian is equivalent to breaking the associated McEliece cryptosystem by only knowing its public key, while the private key allows a party who knows it to recover the planted ground state. Thanks to this equivalence, the security of the McEliece cryptosystem in the typical case is leveraged in order to produce Ising Hamiltonians whose ground state is computationally hard to find. Secondly, by using mathematical methods originating in the statistical physics of disordered systems (see \cite{MezardMontanariBook} for an accessible introduction) the authors have studied in simplified settings the effect of random energy scrambling on easy-to-optimize energy landscapes, following the notion that in the McEliece cryptosystem the public and the private keys are connected through random permutations. It was shown that the energy-scrambling can introduce a phenomenon known as “clustering” into the energy landscape of the Hamiltonian: close enough to the ground-state, the states with a given energy density organize themselves in well-separated groups (\emph{i.e.} ``clusters''), indicating a complicated geometry for the energy landscape of the kind that common local-search heuristics find it difficult to explore.

In an effort to provide a unified cross-platform framework for generating random instances
with planted solutions, Ref.~\newcite{perera2021chook} devised \texttt{chook}, an open-source and user-friendly
library to generate random instances with known solutions and tunable hardness. \texttt{chook}
is designed to be extensible and updated to include new developed planting methods. At
the time of writing, \texttt{chook} includes generators such as the Wishart planting \cite{hamze2020wishart},
equation planting \cite{hen2019equation}, deceptive cluster loop planting \newcite{mandra2018deceptive}, as well as planted solutions
for higher-order (beyond quadratic) binary optimization problems.

\subsection{Randomized Benchmarking}
Randomized benchmarking (RB) is a powerful method for determining the error rate of experimental quantum gates \cite{dankert2009exact}. 
As a resource-efficient alternative method to quantum process tomography, RB also provides an estimate for the average fidelity that is independent of state preparation and measurement (SPAM) errors.
RB characterizes gates that form a group $G$ that form a unitary 2-design in a resource-scalable fashion. Character RB can benchmark more general gates using techniques from representation theory but has only been explored for ``multiplicity-free'' groups. In the work of \newcite{Claes2021}, the character RB theory to explicitly treat non-multiplicity-free groups was developed. Our rigorous derivations enabled us to provide conditions under which instantiations of the framework yield practical RB protocols. This generalized approach with applications to three distinct situations of practical interest: benchmarking of gates with subspace-preserving properties, characterization of leakage, and benchmarking of the matchgate group was demonstrated. In all three cases, compared to existing theories, our method requires similar resources but either provides a more accurate estimate of gate fidelity or applies to a more general group of gates.

While the structure of unitary \(2\)-designs for multiqubit systems is well understood by now, little is known about exact \(2\)-designs for \textit{multiqudit} systems, especially for non-prime-power dimensions \cite{zhu_multiqubit_2017, webb_clifford_2016, kueng_qubit_2015}. In an upcoming work, a class of weighted state \(2\)-designs for qudits in arbitrary dimensions \newcite{Anand2024_qudit_designs} that allow us to devise several new protocols for RB, shadow tomography, etc., in various qudit architectures.

\subsection{Master Equation Emulation}
Prior work by Marshall et al. in \newcite{Marshall2017} led to an experimental implementation from the Levenson-Falk group at USC, showing a high degree of control of a non-Markovian master equation using superconducting qubits \newcite{Vlachos2022}. It was further demonstrated, as predicted in \newcite{Marshall2017}, that introducing non-Markovian noise to a Markovian bath can actually, somewhat counter intuitively, \textit{increase} the coherence time of a qubit. 

\section{Mechanisms for Quantum Computation}
\label{sec:MechForQC}

This section describes a broad set of topics of a more advanced nature, often of a more fundamental nature all generally related to quantum information and its ties to quantum computation. Most of this work is more theoretical in nature, such as work on the spread of information through a quantum system or on how basis dependent properties interact during quantum evolution.  There are also more practical focuses such as how to design logic gates and to translate quantum effects into equivalent classical enhancements and architectures for large scale quantum computing. We begin with a discussion of work related to fundamentals of quantum theory and the use of a quantum computer in testing the boundaries of quantum theory.

\subsection{Wigner Friend Scenario}

Bell's theorem enable tests of quantum theories that later underpinned quantum information processing. As mentioned in Sec.~\ref{sec:entVerf}, it gives rise to entanglement witnesses which are one way of investigating quantum entanglement in quantum states, including those arising at each stage of an algorithm or protocol. Recently, researchers realized that the Wigner friend scenario, an equally intriguing but less known thought experiment than Schr\"odinger's cat, can be extended to give rise to new inequalities, local friendlienss inequalities, that have weaker assumptions than those in Bell's theorem. These inequalities provide new experimental horizons, extending from an already realized proof-of-principle experiment in which a single photon plays the role of Wigner's friend to a full experiment involving a quantum computer running a human-level artificial intelligence with a requirement of space-like separation between the parties of the distance between the moon and earth or more. In \newcite{Wiseman2023thoughtfullocal}, Rieffel together with two Australian colleagues estimated the resources for this full experiment and helped refine the statement of a new ``local friendliness" theorem. While both a sufficiently large and robust quantum computer and a human-level AI have not been realized today, they appear technically feasible within the next decades. The full experiment provides a challenging target, with related nearer-term feasible experiment targets  that test the boundaries of quantum theory being explored.

\subsection{Information Scrambling}

Information scrambling is a fundamental mechanism by which closed, interacting quantum many-body systems spread information throughout their nonlocal degrees of freedom \cite{swingle_unscrambling_2018,xu2023scrambling}. 
Scrambling phenomena often underlies methods to demonstrate and benchmark quantum computational capabilities.
While typically characterized via the decay of ``out-of-time-ordered correlators'' (OTOCs) \cite{kitaev_simple_2015}, this delocalization manifests itself via two distinct processes: (i) ``operator spreading,'' which corresponds to the increasing support of local operators under Heisenberg evolution \cite{nahum2018operator} and (ii) ``operator entanglement generation'' which represents the increasing complexity of Heisenberg-evolved operators, quantified via the inability to express them as a simple tensor products of local operators \cite{zanardi_entanglement_2001}.

The dynamics of operator spreading and operator entanglement growth have turned out to be crucial in understanding minimal models of quantum chaos such as ``dual-unitary circuits'' \cite{PhysRevX.9.021033} and ``Sachdev-Ye-Kitaev'' (SYK) type models \cite{kitaev_simple_2015}, as well as simple models for the black hole information paradox such as the Hayden-Preskill protocol \cite{Hayden_2007}. Beyond these somewhat exotic applications, these two mechanisms are also intimately related to classical simulability of quantum dynamics \cite{Dubail_2017}. For example, in local Hamiltonian systems on a lattice, operator spreading is controlled via Lieb-Robinson (LR) bounds \cite{bravyi-lieb-robinson-2006}, which provides an estimate for the maximal locality one needs to consider in order to approximate Heisenberg-evolved operators well. In Ref. \newcite{Mi2021}, LR bounds were used to introduce a ``lightcone filter'' which allowed them to improve experimental accuracy in measuring OTOCs. In contrast, operator entanglement is a nonlocal observable that does not directly follow such a (exponential) bound. However, in the same spirit as above, the growth of operator entanglement is directly related to the simulability of Heisenberg operators via a matrix product operator (MPO) representation \cite{Dubail_2017}. This simulation cost shows up in tensor network contraction approaches to classically simulating quantum dynamics \newcite{Mandra2021}.

In a work led by the Google Quantum AI team, both operator spreading and operator entanglement growth were observed in a \(53\)-qubit superconducting quantum processor \newcite{Mi2021}. This demonstrated a high degree of control, verified by classical simulations from the NASA QuAIL team. 
They considered two different classes of random unitaries (by either using ${\rm SWAP}$ gates or $\sqrt{\rm SWAP}$ gates) to experimentally realize different rates of operator spreading and operator entanglement growth. 
Furthermore, they introduced models to simulate the growth of OTOCs in large-scale systems. A key observation here was that the classical resources required to simulate operator growth are much tamer compared to those required for simulating operator entanglement, consistent with previous numerical and analytical studies.

These experiments have motivated further study on the connection between OTOCs and operator entanglement, which was rigorously established in Ref. \newcite{styliaris_information_2021}. A series of works generalized this connection to open quantum systems \newcite{PhysRevA.103.062214} and systems at finite temperature \newcite{Anand2022brotocsquantum}, and placed it in an operator algebraic framework \newcite{PhysRevA.107.042217}. Building on these theoretical and experimental advances, the QuAIL team along with researchers at USC, further generalized these results to include non-Hermitian Hamiltonians, which typically violate LR bounds, and represent effective models for measurement-induced phase transitions \newcite{Barch2023}.

\subsection{Stoquasticity}

Stoquasticity and the sign problem are related to the difficulty of classically simulating quantum phenomena. A problem or Hamiltonian with no sign problem is simulatable using classical Monte Carlo techniques, but when a sign problem occurs, the pseudo-probabilities in the quantum Monte Carlo technique start oscillating rapidly between positive and negative, leading to an exponential slow down in the integration/averaging.  The sign problem occurs in most quantum Hamiltonians, but stoquastic Hamiltonians are a notable class where no sign problem occurs.  Stoquasticity is a basis-dependent property of the Hamiltonian and is characterized by the Hamiltonian being represented by a matrix whose off-diagonal entries are all non-positive.

When a single Hamiltonian is being simulated, a stoquastic basis always exists (e.g., the eigenbasis of the Hamiltonian), but finding a stoquastic (or more generally a sign problem free) basis, also known as ``curing'' the sign problem is an NP-hard task \cite{Marvian_2019}.  However, some quantum algorithms, such as quantum annealing and QAOA are characterized by the interplay and evolution under two distinct and non-commuting Hamiltonians.  In Ref.~\newcite{Bringewatt2022}, the question of whether a single basis exists under which multiple Hamiltonians are stoquastic was explored.  If such a basis, coined the ``simultaneous stoquastic'' basis exists, then it would be possible to simulate a quantum annealing or QAOA task using Monte Carlo techniques with no sign problem.  Our results, based off the theory of unitary similarity, show that for any given set of Hamiltonians of size $d\geq 2$ and dimension $N>2$, the probability that a simultaneously stoquastic basis exists is vanishingly small.  Furthermore, a simple no go result for testing this based off the commutator of the Hamiltonians was provided.  This could potentially have further applications in the computational complexity of simulating such Hamiltonian processes where no such basis exists.

\subsection{Relaxation Mechanisms in Quantum and Classical Transverse-Field Annealing}
\label{sec:relaxation}
Prior work in Ref.~\newcite{Marshall2019} showed that strategically introducing a pause in a quantum annealing schedule can increase the performance by several orders of magnitude. In that work, it was posited that thermal relaxations were occurring during the pause after excitations from driving through the minimum gap. A follow-up study in \newcite{Albash2021} showed that the same behavior could be found in a purely classical analog of quantum annealing (SVMC), thus providing more insights into the origin of this effect (i.e., it is not a uniquely quantum phenomenon). This work also gave a general condition on pausing for it to yield an improvement agnostic to the particular system.

\subsection{Acceleration of Quantum Tunneling in Spin Systems}
\label{sec:accelquanttunel}

Expanding on the original work on Random-Frequency Quantum Annealing (RFQA) \cite{Kapit_2021}, the QuAIL team and their collaborators have explored ways of mitigating the decay of performance in currently-available quantum annealers caused by minor embedding. The freezing of the physical qubit chains that represent logical qubits is suppressed by adding independently oscillating local fields with a distribution of frequencies to the usual transverse-field annealing setup. This off-equilibrium process allows a chain to escape its frozen state via a quantum tunneling event induced by the proliferation of weak resonances introduced by the oscillatory fields, thus reducing the effect of frozen chains. This was studied in \newcite{2306.10632} using the 1D transverse-field Ising chain as a model for a chain of spins used in the embedding of logical problems.
A follow-up work \newcite{2311.17814} designed to shed light on the physics behind this process showed how the same results can be engineered in a driven Floquet system, where additional fields oscillate with the same frequency. Moreover, it uncovered the existence of some novel off-equilibrium phenomenology of these physical systems. Using the Transverse-Field Ising models (TFIM) in 1D and 2D as testing grounds, the authors showed three off-equilibrium regimes as a function of the drive strength: for weak drives, the system exhibits exponentially decaying tunneling rates but robust magnetic order, typical of the (undriven) dynamics in the ordered phase of models exhibiting spontaneous symmetry breaking such as the TFIM. In the crossover regime at intermediate drive strength, one observes polynomial decay of tunneling alongside vanishing magnetic order, and at very strong drive strengths both the Rabi frequency and time-averaged magnetic order are approximately constant with increasing system size. 

\subsection{Engineering Topology}
Exotic topological states of matter such as the non-Abelian anyons exhibit braiding statistics foundational to topological fault-tolerant quantum computation and can emerge in systems displaying the fractional quantum Hall effect (FQHE). Topologically non-trivial bands when made flat (degenerate) can lead to FQHE states. In recent work by  Suri et al.~\newcite{PhysRevB.108.155409}, a method to create bands with higher order topology which is not possible to obtain otherwise for example by a simple magnetic field was proposed. Such higher order topological bands when made flat can lead to potentially more exotic non-Abelian anyons useful for quantum computation. 

\subsection{Compilation approaches}

Compiling quantum circuits to realistic hardware architectures generally involves two steps, circuit routing and gate synthesis. 
In \newcite{Do2020}, earlier work, \newcite{venturelli2017_ijcai2017p620}, was extended, which recognized that routing could be viewed as temporal planning task, and thus state-of-the-art temporal planners applied,  to route circuits that implement QAOA for Graph Coloring problems. 
On the gate synthesis side, Ref.~\newcite{Alam2023}, 
studied the case of single qubit quantum compilation and state preparation by modeling the task as a Markov Decision Process, and solving it using dynamical programming methods.
This work demonstrates that classical reinforcement learning techniques can produce shorter gate sequences than have been identified before, and can also generate gate sequences that work well in the presence of noise. These results represent an important advance in methods for quantum compilation that work in the presence of the physical constraints of NISQ-era hardware.

\subsection{Quantum Computing Architectures}

Quantum algorithms are usually framed in terms of qubits in the gate based circuit model. Physical realizations of qubits take the form of several different implementations, such as superconducting hardware, trapped ions, neutral atoms etc. Recently, advances on the experimental side have opened the avenue to control the quantum state of superconducting cavity modes coupled to a qubit-like transmon. In the Fock basis, this in turn enables the control and manipulation of qudits, larger dimensional analogues of qubits. In \newcite{alam2022quantum}, it was describe how such a platform may be realized through 3D superconducting radio frequency (SRF) cavities, which typically have large Q factors. Several algorithms of interest in high energy physics (HEP) involve bosonic degrees of freedom, which are often more easily mapped onto qudits rather than qubits.

A potentially promising approach for assembling future large scale-quantum computers is discussed in \newcite{gold2021entanglement}. Here a modular solid state architecture with deterministic inter-module coupling between four physically separate integrated chips is described, and the quality of the inter-module entanglement is confirmed by a demonstration of Bell-inequality violation for disjoint pairs of entangled qubits across the four separate silicon dies.

\section{Future Outlook}
\label{sec:FutureOutlook}

Many industry, government, and academic groups have ambitious programs to increase the size and improve the robustness of quantum processors. For the next several years, however, a gap will remain between the capabilities of the hardware and the capabilities needed to run quantum and quantum-classical algorithms at application-scale. While performing spectacular one-off experiments are tempting, the field will advance by prioritizing experiments and theory that inform efforts to reach application scale in a robust, repeatable way. The challenges that remain are significant enough to require broad collaborations and open sharing of information.  

As the hardware matures, we will have increasing opportunities for exploration of algorithms, significantly beyond what has been possible up to now. For many of the most computationally intensive state-of-the-art practical algorithms, performance must be evaluated empirically – theoretical analysis is just too hard. For this reason, there are machine learning, SAT, and planning and scheduling competitions with sets of challenge problems on which algorithms can compete empirically, and within industry and government, heuristic algorithms, without proofs of their performance, are frequently used to attack practical problems. Currently, evaluation of the performance of quantum algorithms is restricted to small cases, given the size and non-robustness of current quantum processors and that generically there is exponential overhead in simulating quantum algorithms on classical computers, quickly exceeding the capabilities of even the world’s largest supercomputers. While the size and robustness of quantum processors are expected to improve steadily in the coming years, they will remain resource limited in the next decades. Algorithm-hardware co-design will be critical to reap the most benefit from hardware advances.

Co-design is also needed between hardware and error mitigation, error correction, native gate choices, and compilation of quantum algorithms to native gates. Next generation quantum systems will provide information that will enable richer error models. These models will inform error correction which can in turn inform hardware design priorities and architecture choices. As quantum computers scale up, they will necessarily involve multiple QPUs, first resembling clusters and then supercomputers, but with quantum as well as classical processors, where the classical processors will not only be used for control of quantum processors, but also for decoding as part of error correction, and often as co-processors in hybrid quantum-classical algorithms. We expect research into future HPC systems and quantum systems to mutually inform each other. Further research directions include understanding the space of architectural designs and co-designing distributed algorithms, error correction layers, and classical and quantum communication and synchronization with architectural choices. It is an open question how heterogeneous these quantum computers will be. It is possible that they will include multiple types of quantum processors, superconducting, ion-trap, photonic, neutral atom, etc., and will likely include a variety of specialized quantum components such as quantum memory or magic state factories or special-purpose quantum processors for, say, quantum Fourier transforms. Specialized quantum processors that have less stringent requirements than universal quantum computers could potentially be built more easily and put to use earlier than universal processors. Resource estimation for specific problems of application interest, including resource estimates for newly considered problems to improved estimates for already identified problems, will help guide hardware, error correction, and architectural priorities. 

Quantum computing is one of the most enticing computational paradigms, with the potential to revolutionize diverse areas of future-generation computational systems. The NASA QuAIL team looks forward to continuing to advance the understanding of the potential of these systems in collaborations with others in this diverse field. We hope this glimpse of the variety of the current research and the many open areas will further entice both young people at the beginning of their careers and professionals in adjacent fields to join this fascinating effort.

\section{Acknowledgements}
\label{sec:Acknowledgements}

The authors are grateful for our many collaborators who have made much of this work possible. We are grateful for support from NASA Ames Research Center, NASA Advanced Exploration systems (AES) program, NASA Earth Science Technology Office (ESTO), NASA Transformative Aeronautic Concepts Program (TACP), and NASA Space Communications and Navigation (SCaN). 
We are grateful for the support of the U.S. Department of Energy, Office of Science, National Quantum Information Science Research Centers, Superconducting Quantum Materials and Systems Center (SQMS) under the contract No. DE-AC02-07CH11359 and Co-Design Center for Quantum Advantage (C2QA) under Contract No. DE-SC0012704, from DARPA under IAA 8839, Annex 114, 128, 129, 130, SAA2-403688, and SAA2-403746, 
from Google LLC under SSA-403706, and from NTT Research Inc. under SAA2-403506.
A.A.A., M.S.A, D.E.B.N., S.B., E.G., Z.G.I, S.H., P.A.L, F.M., J.M., N.S., D.V., and Z.W. are supported by NASA Academic Mission Services (NAMS), contract number NNA16BD14C. 
N.A., H.M.B, L.T.B., S.C., S.M., and G.M. are KBR employees working under the Prime Contract No. 80ARC020D0010 with NASA.

\section{About the NASA QuAIL Team}
\label{sec:QuAILTeam}
The mandate of the NASA’s Quantum Artificial Intelligence Laboratory (QuAIL), located at NASA Ames in the heart of Silicon Valley, is to assess and advance the potential impact of quantum computers on computational problems that will be faced by NASA missions in decades to come. 
QuAIL team members, physicists, computer scientists, mathematicians, and engineers, come from a wide variety of backgrounds with complementary expertise that supports the interdisciplinary nature of quantum computing research. The team has strong collaborations with application domain experts in and outside NASA. QuAIL is a theory and numerical group that has forged close collaborations with groups implementing quantum hardware. 

\bibliographystyle{elsarticle-num-names}

\bibliography{references_with_doi}

\begin{thebibliography}{210}
\expandafter\ifx\csname natexlab\endcsname\relax\def\natexlab#1{#1}\fi
\providecommand{\url}[1]{\texttt{#1}}
\providecommand{\href}[2]{#2}
\providecommand{\path}[1]{#1}
\providecommand{\DOIprefix}{doi:}
\providecommand{\ArXivprefix}{arXiv:}
\providecommand{\URLprefix}{URL: }
\providecommand{\Pubmedprefix}{pmid:}
\providecommand{\doi}[1]{\href{http://dx.doi.org/#1}{\path{#1}}}
\providecommand{\Pubmed}[1]{\href{pmid:#1}{\path{#1}}}
\providecommand{\bibinfo}[2]{#2}
\ifx\xfnm\relax \def\xfnm[#1]{\unskip,\space#1}\fi
\bibitem[{Preskill(2018)}]{Preskill2018quantumcomputingin}
\bibinfo{author}{J.~Preskill},
\newblock \bibinfo{title}{Quantum {C}omputing in the {NISQ} era and beyond},
\newblock \bibinfo{journal}{{Quantum}} \bibinfo{volume}{2} (\bibinfo{year}{2018}) \bibinfo{pages}{79}. \URLprefix \url{https://doi.org/10.22331/q-2018-08-06-79}. \DOIprefix\doi{10.22331/q-2018-08-06-79}.
\bibitem[{Rieffel et~al.(2019)Rieffel, Hadfield, Hogg, Mandrà, Marshall, Mossi, O’Gorman, Plamadeala, Tubman, Venturelli, Vinci, Wang, Wilson, Wudarski, and Biswas}]{Rieffel_2019}
\bibinfo{author}{E.~G. Rieffel}, \bibinfo{author}{S.~Hadfield}, \bibinfo{author}{T.~Hogg}, \bibinfo{author}{S.~Mandrà}, \bibinfo{author}{J.~Marshall}, \bibinfo{author}{G.~Mossi}, \bibinfo{author}{B.~O’Gorman}, \bibinfo{author}{E.~Plamadeala}, \bibinfo{author}{N.~M. Tubman}, \bibinfo{author}{D.~Venturelli}, \bibinfo{author}{W.~Vinci}, \bibinfo{author}{Z.~Wang}, \bibinfo{author}{M.~Wilson}, \bibinfo{author}{F.~Wudarski}, \bibinfo{author}{R.~Biswas}, \bibinfo{title}{{From Ansätze to Z-Gates: A NASA View of Quantum Computing}}, \bibinfo{publisher}{IOS Press}, \bibinfo{year}{2019}. \URLprefix \url{http://dx.doi.org/10.3233/APC190010}. \DOIprefix\doi{10.3233/apc190010}.
\bibitem[{Biswas et~al.(2017)Biswas, Jiang, Kechezhi, Knysh, Mandrà, O’Gorman, Perdomo-Ortiz, Petukhov, Realpe-Gómez, Rieffel, Venturelli, Vasko, and Wang}]{Biswas_2017}
\bibinfo{author}{R.~Biswas}, \bibinfo{author}{Z.~Jiang}, \bibinfo{author}{K.~Kechezhi}, \bibinfo{author}{S.~Knysh}, \bibinfo{author}{S.~Mandrà}, \bibinfo{author}{B.~O’Gorman}, \bibinfo{author}{A.~Perdomo-Ortiz}, \bibinfo{author}{A.~Petukhov}, \bibinfo{author}{J.~Realpe-Gómez}, \bibinfo{author}{E.~Rieffel}, \bibinfo{author}{D.~Venturelli}, \bibinfo{author}{F.~Vasko}, \bibinfo{author}{Z.~Wang},
\newblock \bibinfo{title}{{A NASA perspective on quantum computing: Opportunities and challenges}},
\newblock \bibinfo{journal}{Parallel Computing} \bibinfo{volume}{64} (\bibinfo{year}{2017}) \bibinfo{pages}{81–98}. \URLprefix \url{http://dx.doi.org/10.1016/j.parco.2016.11.002}. \DOIprefix\doi{10.1016/j.parco.2016.11.002}.
\bibitem[{Farhi et~al.(2001)Farhi, Goldstone, Gutmann, Lapan, Lundgren, and Preda}]{Farhi_2001}
\bibinfo{author}{E.~Farhi}, \bibinfo{author}{J.~Goldstone}, \bibinfo{author}{S.~Gutmann}, \bibinfo{author}{J.~Lapan}, \bibinfo{author}{A.~Lundgren}, \bibinfo{author}{D.~Preda},
\newblock \bibinfo{title}{{A Quantum Adiabatic Evolution Algorithm Applied to Random Instances of an NP-Complete Problem}},
\newblock \bibinfo{journal}{Science} \bibinfo{volume}{292} (\bibinfo{year}{2001}) \bibinfo{pages}{472--475}. \URLprefix \url{https://www.science.org/doi/abs/10.1126/science.1057726}. \DOIprefix\doi{10.1126/science.1057726}.
\bibitem[{Farhi et~al.(2014)Farhi, Goldstone, and Gutmann}]{farhi2014quantum}
\bibinfo{author}{E.~Farhi}, \bibinfo{author}{J.~Goldstone}, \bibinfo{author}{S.~Gutmann}, \bibinfo{title}{{A Quantum Approximate Optimization Algorithm}}, \bibinfo{year}{2014}. \href{http://arxiv.org/abs/1411.4028}{{\tt arXiv:1411.4028}}.
\bibitem[{Farhi et~al.(2015)Farhi, Goldstone, and Gutmann}]{farhi2015quantum}
\bibinfo{author}{E.~Farhi}, \bibinfo{author}{J.~Goldstone}, \bibinfo{author}{S.~Gutmann}, \bibinfo{title}{{A Quantum Approximate Optimization Algorithm Applied to a Bounded Occurrence Constraint Problem}}, \bibinfo{year}{2015}. \href{http://arxiv.org/abs/1412.6062}{{\tt arXiv:1412.6062}}.
\bibitem[{Cerezo et~al.(2021)Cerezo, Arrasmith, Babbush, Benjamin, Endo, Fujii, McClean, Mitarai, Yuan, Cincio, and Coles}]{Cerezo2021}
\bibinfo{author}{M.~Cerezo}, \bibinfo{author}{A.~Arrasmith}, \bibinfo{author}{R.~Babbush}, \bibinfo{author}{S.~C. Benjamin}, \bibinfo{author}{S.~Endo}, \bibinfo{author}{K.~Fujii}, \bibinfo{author}{J.~R. McClean}, \bibinfo{author}{K.~Mitarai}, \bibinfo{author}{X.~Yuan}, \bibinfo{author}{L.~Cincio}, \bibinfo{author}{P.~J. Coles},
\newblock \bibinfo{title}{{Variational quantum algorithms}},
\newblock \bibinfo{journal}{Nature Reviews Physics} \bibinfo{volume}{3} (\bibinfo{year}{2021}) \bibinfo{pages}{625--644}. \URLprefix \url{https://doi.org/10.1038/s42254-021-00348-9}. \DOIprefix\doi{10.1038/s42254-021-00348-9}.
\bibitem[{Montanaro(2020)}]{Montanaro2020}
\bibinfo{author}{A.~Montanaro},
\newblock \bibinfo{title}{Quantum speedup of branch-and-bound algorithms},
\newblock \bibinfo{journal}{Phys. Rev. Res.} \bibinfo{volume}{2} (\bibinfo{year}{2020}) \bibinfo{pages}{013056}. \URLprefix \url{https://link.aps.org/doi/10.1103/PhysRevResearch.2.013056}. \DOIprefix\doi{10.1103/PhysRevResearch.2.013056}.
\bibitem[{Alexandru et~al.(2020)Alexandru, Bridgett-Tomkinson, Linden, MacManus, Montanaro, and Morris}]{Alexandru_2020}
\bibinfo{author}{C.-M. Alexandru}, \bibinfo{author}{E.~Bridgett-Tomkinson}, \bibinfo{author}{N.~Linden}, \bibinfo{author}{J.~MacManus}, \bibinfo{author}{A.~Montanaro}, \bibinfo{author}{H.~Morris},
\newblock \bibinfo{title}{Quantum speedups of some general-purpose numerical optimisation algorithms},
\newblock \bibinfo{journal}{Quantum Science and Technology} \bibinfo{volume}{5} (\bibinfo{year}{2020}) \bibinfo{pages}{045014}. \URLprefix \url{https://dx.doi.org/10.1088/2058-9565/abb003}. \DOIprefix\doi{10.1088/2058-9565/abb003}.
\bibitem[{Abbas et~al.(2023)Abbas, Ambainis, Augustino, B{\"a}rtschi, Buhrman, Coffrin, Cortiana, Dunjko, Egger, Elmegreen et~al.}]{abbas2023quantum}
\bibinfo{author}{A.~Abbas}, \bibinfo{author}{A.~Ambainis}, \bibinfo{author}{B.~Augustino}, \bibinfo{author}{A.~B{\"a}rtschi}, \bibinfo{author}{H.~Buhrman}, \bibinfo{author}{C.~Coffrin}, \bibinfo{author}{G.~Cortiana}, \bibinfo{author}{V.~Dunjko}, \bibinfo{author}{D.~J. Egger}, \bibinfo{author}{B.~G. Elmegreen}, et~al., \bibinfo{title}{{Quantum Optimization: Potential, Challenges, and the Path Forward}}, \bibinfo{year}{2023}. \href{http://arxiv.org/abs/2312.02279}{{\tt arXiv:2312.02279}}.
\bibitem[{{Bernal Neira} et~al.(2024){Bernal Neira}, Laird, Lueg, Harwood, Trenev, and Venturelli}]{bernal2024utilizing}
\bibinfo{author}{D.~E. {Bernal Neira}}, \bibinfo{author}{C.~D. Laird}, \bibinfo{author}{L.~R. Lueg}, \bibinfo{author}{S.~M. Harwood}, \bibinfo{author}{D.~Trenev}, \bibinfo{author}{D.~Venturelli},
\newblock \bibinfo{title}{{Utilizing modern computer architectures to solve mathematical optimization problems: A survey}},
\newblock \bibinfo{journal}{Computers \& Chemical Engineering}  (\bibinfo{year}{2024}) \bibinfo{pages}{108627}. \DOIprefix\doi{https://doi.org/10.1016/j.compchemeng.2024.108627}.
\bibitem[{Hadfield et~al.(2019)Hadfield, Wang, O’Gorman, Rieffel, Venturelli, and Biswas}]{Hadfield_2019}
\bibinfo{author}{S.~Hadfield}, \bibinfo{author}{Z.~Wang}, \bibinfo{author}{B.~O’Gorman}, \bibinfo{author}{E.~Rieffel}, \bibinfo{author}{D.~Venturelli}, \bibinfo{author}{R.~Biswas},
\newblock \bibinfo{title}{{From the Quantum Approximate Optimization Algorithm to a Quantum Alternating Operator Ansatz}},
\newblock \bibinfo{journal}{Algorithms} \bibinfo{volume}{12} (\bibinfo{year}{2019}) \bibinfo{pages}{34}. \URLprefix \url{http://dx.doi.org/10.3390/a12020034}. \DOIprefix\doi{10.3390/a12020034}.
\bibitem[{Sud et~al.(2024)Sud, Hadfield, Rieffel, Tubman, and Hogg}]{Sud2022_parameter}
\bibinfo{author}{J.~Sud}, \bibinfo{author}{S.~Hadfield}, \bibinfo{author}{E.~Rieffel}, \bibinfo{author}{N.~Tubman}, \bibinfo{author}{T.~Hogg},
\newblock \bibinfo{title}{{Parameter-setting heuristic for the quantum alternating operator ansatz}},
\newblock \bibinfo{journal}{Physical Review Research} \bibinfo{volume}{6} (\bibinfo{year}{2024}) \bibinfo{pages}{023171}.
\bibitem[{Wang et~al.(2018)Wang, Hadfield, Jiang, and Rieffel}]{wang2018quantum}
\bibinfo{author}{Z.~Wang}, \bibinfo{author}{S.~Hadfield}, \bibinfo{author}{Z.~Jiang}, \bibinfo{author}{E.~G. Rieffel},
\newblock \bibinfo{title}{{Quantum approximate optimization algorithm for MaxCut: A fermionic view}},
\newblock \bibinfo{journal}{Physical Review A} \bibinfo{volume}{97} (\bibinfo{year}{2018}) \bibinfo{pages}{022304}. \URLprefix \url{https://link.aps.org/doi/10.1103/PhysRevA.97.022304}. \DOIprefix\doi{10.1103/PhysRevA.97.022304}.
\bibitem[{Marwaha and Hadfield(2022)}]{Marwaha2022}
\bibinfo{author}{K.~Marwaha}, \bibinfo{author}{S.~Hadfield},
\newblock \bibinfo{title}{Bounds on approximating {M}ax {$k$}{XOR} with quantum and classical local algorithms},
\newblock \bibinfo{journal}{{Quantum}} \bibinfo{volume}{6} (\bibinfo{year}{2022}) \bibinfo{pages}{757}. \URLprefix \url{https://doi.org/10.22331/q-2022-07-07-757}. \DOIprefix\doi{10.22331/q-2022-07-07-757}.
\bibitem[{Hirvonen et~al.(2014)Hirvonen, Rybicki, Schmid, and Suomela}]{hirvonen2014large}
\bibinfo{author}{J.~Hirvonen}, \bibinfo{author}{J.~Rybicki}, \bibinfo{author}{S.~Schmid}, \bibinfo{author}{J.~Suomela}, \bibinfo{title}{{Large cuts with local algorithms on triangle-free graphs}}, \bibinfo{year}{2014}. \href{http://arxiv.org/abs/1402.2543}{{\tt arXiv:1402.2543}}.
\bibitem[{Sen(2018)}]{sen2018optimization}
\bibinfo{author}{S.~Sen},
\newblock \bibinfo{title}{{Optimization on sparse random hypergraphs and spin glasses}},
\newblock \bibinfo{journal}{Random Structures \& Algorithms} \bibinfo{volume}{53} (\bibinfo{year}{2018}) \bibinfo{pages}{504--536}. \DOIprefix\doi{https://doi.org/10.1002/rsa.20774}.
\bibitem[{Bravyi et~al.(2020)Bravyi, Kliesch, Koenig, and Tang}]{Bravyi2020}
\bibinfo{author}{S.~Bravyi}, \bibinfo{author}{A.~Kliesch}, \bibinfo{author}{R.~Koenig}, \bibinfo{author}{E.~Tang},
\newblock \bibinfo{title}{{Obstacles to Variational Quantum Optimization from Symmetry Protection}},
\newblock \bibinfo{journal}{Phys. Rev. Lett.} \bibinfo{volume}{125} (\bibinfo{year}{2020}) \bibinfo{pages}{260505}. \URLprefix \url{https://link.aps.org/doi/10.1103/PhysRevLett.125.260505}. \DOIprefix\doi{10.1103/PhysRevLett.125.260505}.
\bibitem[{Leipold et~al.(2022)Leipold, Spedalieri, and Rieffel}]{Leipold2022}
\bibinfo{author}{H.~Leipold}, \bibinfo{author}{F.~M. Spedalieri}, \bibinfo{author}{E.~Rieffel},
\newblock \bibinfo{title}{{Tailored Quantum Alternating Operator Ans\"atzes for Circuit Fault Diagnostics}},
\newblock \bibinfo{journal}{Algorithms} \bibinfo{volume}{15} (\bibinfo{year}{2022}). \URLprefix \url{https://www.mdpi.com/1999-4893/15/10/356}. \DOIprefix\doi{10.3390/a15100356}.
\bibitem[{Hadfield et~al.(2022)Hadfield, Hogg, and Rieffel}]{Hadfield2023a}
\bibinfo{author}{S.~Hadfield}, \bibinfo{author}{T.~Hogg}, \bibinfo{author}{E.~G. Rieffel},
\newblock \bibinfo{title}{Analytical framework for quantum alternating operator ansätze},
\newblock \bibinfo{journal}{Quantum Science and Technology} \bibinfo{volume}{8} (\bibinfo{year}{2022}) \bibinfo{pages}{015017}. \URLprefix \url{https://dx.doi.org/10.1088/2058-9565/aca3ce}. \DOIprefix\doi{10.1088/2058-9565/aca3ce}.
\bibitem[{Kremenetski et~al.(2023)Kremenetski, Apte, Hogg, Hadfield, and Tubman}]{Kremenetski2023}
\bibinfo{author}{V.~Kremenetski}, \bibinfo{author}{A.~Apte}, \bibinfo{author}{T.~Hogg}, \bibinfo{author}{S.~Hadfield}, \bibinfo{author}{N.~M. Tubman}, \bibinfo{title}{{Quantum Alternating Operator Ansatz ({QAOA}) beyond low depth with gradually changing unitaries}}, \bibinfo{year}{2023}. \href{http://arxiv.org/abs/2305.04455}{{\tt arXiv:2305.04455}}.
\bibitem[{Kremenetski et~al.(2021)Kremenetski, Hogg, Hadfield, Cotton, and Tubman}]{kremenetski2021quantum}
\bibinfo{author}{V.~Kremenetski}, \bibinfo{author}{T.~Hogg}, \bibinfo{author}{S.~Hadfield}, \bibinfo{author}{S.~J. Cotton}, \bibinfo{author}{N.~M. Tubman}, \bibinfo{title}{{Quantum Alternating Operator Ansatz (QAOA) Phase Diagrams and Applications for Quantum Chemistry}}, \bibinfo{year}{2021}. \href{http://arxiv.org/abs/2108.13056}{{\tt arXiv:2108.13056}}.
\bibitem[{LaRose et~al.(2022)LaRose, Rieffel, and Venturelli}]{LaRose2022}
\bibinfo{author}{R.~LaRose}, \bibinfo{author}{E.~Rieffel}, \bibinfo{author}{D.~Venturelli},
\newblock \bibinfo{title}{{Mixer-phaser Ans{\"a}tze for quantum optimization with hard constraints}},
\newblock \bibinfo{journal}{Quantum Machine Intelligence} \bibinfo{volume}{4} (\bibinfo{year}{2022}) \bibinfo{pages}{17}. \URLprefix \url{https://doi.org/10.1007/s42484-022-00069-x}. \DOIprefix\doi{10.1007/s42484-022-00069-x}.
\bibitem[{Maciejewski et~al.(2023)Maciejewski, Hadfield, Hall, Hodson, Dupont, Evert, Sud, Alam, Wang, Jeffrey, Sundar, Lott, Grabbe, Rieffel, Reagor, and Venturelli}]{maciejewski2023design}
\bibinfo{author}{F.~B. Maciejewski}, \bibinfo{author}{S.~Hadfield}, \bibinfo{author}{B.~Hall}, \bibinfo{author}{M.~Hodson}, \bibinfo{author}{M.~Dupont}, \bibinfo{author}{B.~Evert}, \bibinfo{author}{J.~Sud}, \bibinfo{author}{M.~S. Alam}, \bibinfo{author}{Z.~Wang}, \bibinfo{author}{S.~Jeffrey}, \bibinfo{author}{B.~Sundar}, \bibinfo{author}{P.~A. Lott}, \bibinfo{author}{S.~Grabbe}, \bibinfo{author}{E.~G. Rieffel}, \bibinfo{author}{M.~J. Reagor}, \bibinfo{author}{D.~Venturelli}, \bibinfo{title}{{Design and execution of quantum circuits using tens of superconducting qubits and thousands of gates for dense Ising optimization problems}}, \bibinfo{year}{2023}. \href{http://arxiv.org/abs/2308.12423}{{\tt arXiv:2308.12423}}.
\bibitem[{Maciejewski et~al.(2024)Maciejewski, Biamonte, Hadfield, and Venturelli}]{maciejewski2024ndar}
\bibinfo{author}{F.~B. Maciejewski}, \bibinfo{author}{J.~Biamonte}, \bibinfo{author}{S.~Hadfield}, \bibinfo{author}{D.~Venturelli},
\newblock \bibinfo{title}{{Improving Quantum Approximate Optimization by Noise-Directed Adaptive Remapping}},
\newblock \bibinfo{journal}{arXiv preprint arXiv:2404.01412}  (\bibinfo{year}{2024}). \DOIprefix\doi{10.48550/arXiv.2404.01412}.
\bibitem[{Wang et~al.(2020)Wang, Rubin, Dominy, and Rieffel}]{Wang2020}
\bibinfo{author}{Z.~Wang}, \bibinfo{author}{N.~C. Rubin}, \bibinfo{author}{J.~M. Dominy}, \bibinfo{author}{E.~G. Rieffel},
\newblock \bibinfo{title}{{$XY$ mixers: Analytical and numerical results for the quantum alternating operator ansatz}},
\newblock \bibinfo{journal}{Phys. Rev. A} \bibinfo{volume}{101} (\bibinfo{year}{2020}) \bibinfo{pages}{012320}. \URLprefix \url{https://link.aps.org/doi/10.1103/PhysRevA.101.012320}. \DOIprefix\doi{10.1103/PhysRevA.101.012320}.
\bibitem[{Streif et~al.(2021)Streif, Leib, Wudarski, Rieffel, and Wang}]{Streif2021}
\bibinfo{author}{M.~Streif}, \bibinfo{author}{M.~Leib}, \bibinfo{author}{F.~Wudarski}, \bibinfo{author}{E.~Rieffel}, \bibinfo{author}{Z.~Wang},
\newblock \bibinfo{title}{{Quantum algorithms with local particle-number conservation: Noise effects and error correction}},
\newblock \bibinfo{journal}{Phys. Rev. A} \bibinfo{volume}{103} (\bibinfo{year}{2021}) \bibinfo{pages}{042412}. \URLprefix \url{https://link.aps.org/doi/10.1103/PhysRevA.103.042412}. \DOIprefix\doi{10.1103/PhysRevA.103.042412}.
\bibitem[{Alam et~al.(2022)Alam, Wudarski, Reagor, Sud, Grabbe, Wang, Hodson, Lott, Rieffel, and Venturelli}]{Sohaib2022a}
\bibinfo{author}{M.~S. Alam}, \bibinfo{author}{F.~A. Wudarski}, \bibinfo{author}{M.~J. Reagor}, \bibinfo{author}{J.~Sud}, \bibinfo{author}{S.~Grabbe}, \bibinfo{author}{Z.~Wang}, \bibinfo{author}{M.~Hodson}, \bibinfo{author}{P.~A. Lott}, \bibinfo{author}{E.~G. Rieffel}, \bibinfo{author}{D.~Venturelli},
\newblock \bibinfo{title}{{Practical Verification of Quantum Properties in Quantum-Approximate-Optimization Runs}},
\newblock \bibinfo{journal}{Phys. Rev. Appl.} \bibinfo{volume}{17} (\bibinfo{year}{2022}) \bibinfo{pages}{024026}. \URLprefix \url{https://link.aps.org/doi/10.1103/PhysRevApplied.17.024026}. \DOIprefix\doi{10.1103/PhysRevApplied.17.024026}.
\bibitem[{Brady and Hadfield(2023)}]{Brady2023}
\bibinfo{author}{L.~T. Brady}, \bibinfo{author}{S.~Hadfield}, \bibinfo{title}{{Iterative Quantum Algorithms for Maximum Independent Set: A Tale of Low-Depth Quantum Algorithms}}, \bibinfo{year}{2023}. \href{http://arxiv.org/abs/2309.13110}{{\tt arXiv:2309.13110}}.
\bibitem[{Dupont et~al.(2023)Dupont, Evert, Hodson, Sundar, Jeffrey, Yamaguchi, Feng, Maciejewski, Hadfield, Alam et~al.}]{dupont2023quantum}
\bibinfo{author}{M.~Dupont}, \bibinfo{author}{B.~Evert}, \bibinfo{author}{M.~J. Hodson}, \bibinfo{author}{B.~Sundar}, \bibinfo{author}{S.~Jeffrey}, \bibinfo{author}{Y.~Yamaguchi}, \bibinfo{author}{D.~Feng}, \bibinfo{author}{F.~B. Maciejewski}, \bibinfo{author}{S.~Hadfield}, \bibinfo{author}{M.~S. Alam}, et~al.,
\newblock \bibinfo{title}{{Quantum-enhanced greedy combinatorial optimization solver}},
\newblock \bibinfo{journal}{Science Advances} \bibinfo{volume}{9} (\bibinfo{year}{2023}) \bibinfo{pages}{eadi0487}. \DOIprefix\doi{10.1126/sciadv.adi0487}.
\bibitem[{Dupont et~al.(2024)Dupont, Sundar, Evert, Neira, Peng, Jeffrey, and Hodson}]{dupont2024quantum}
\bibinfo{author}{M.~Dupont}, \bibinfo{author}{B.~Sundar}, \bibinfo{author}{B.~Evert}, \bibinfo{author}{D.~E.~B. Neira}, \bibinfo{author}{Z.~Peng}, \bibinfo{author}{S.~Jeffrey}, \bibinfo{author}{M.~J. Hodson}, \bibinfo{title}{{Quantum Optimization for the Maximum Cut Problem on a Superconducting Quantum Computer}}, \bibinfo{year}{2024}. \href{http://arxiv.org/abs/2404.17579}{{\tt arXiv:2404.17579}}.
\bibitem[{Kadowaki and Nishimori(1998)}]{Kadowaki1998}
\bibinfo{author}{T.~Kadowaki}, \bibinfo{author}{H.~Nishimori},
\newblock \bibinfo{title}{{Quantum annealing in the transverse Ising model}},
\newblock \bibinfo{journal}{Physical Review E} \bibinfo{volume}{58} (\bibinfo{year}{1998}) \bibinfo{pages}{5355--5363}. \URLprefix \url{https://doi.org/10.1103%2Fphysreve.58.5355}. \DOIprefix\doi{10.1103/physreve.58.5355}.
\bibitem[{Farhi et~al.(2000)Farhi, Goldstone, Gutmann, and Sipser}]{Farhi2000}
\bibinfo{author}{E.~Farhi}, \bibinfo{author}{J.~Goldstone}, \bibinfo{author}{S.~Gutmann}, \bibinfo{author}{M.~Sipser},
\newblock \bibinfo{title}{{Quantum Computation by Adiabatic Evolution}},
\newblock \bibinfo{journal}{arxiv}  (\bibinfo{year}{2000}). \URLprefix \url{https://arxiv.org/abs/quant-ph/0001106}. \href{http://arxiv.org/abs/quant-ph/0001106}{{\tt arXiv:quant-ph/0001106}}.
\bibitem[{Aharonov et~al.(2008)Aharonov, van Dam, Kempe, Landau, Lloyd, and Regev}]{Aharonov2008}
\bibinfo{author}{D.~Aharonov}, \bibinfo{author}{W.~van Dam}, \bibinfo{author}{J.~Kempe}, \bibinfo{author}{Z.~Landau}, \bibinfo{author}{S.~Lloyd}, \bibinfo{author}{O.~Regev},
\newblock \bibinfo{title}{{Adiabatic Quantum Computation Is Equivalent to Standard Quantum Computation}},
\newblock \bibinfo{journal}{SIAM Review} \bibinfo{volume}{50} (\bibinfo{year}{2008}) \bibinfo{pages}{755--787}. \URLprefix \url{https://doi.org/10.1137/080734479}. \DOIprefix\doi{10.1137/080734479}.
\bibitem[{Gonzalez~Izquierdo et~al.(2021)Gonzalez~Izquierdo, Grabbe, Hadfield, Marshall, Wang, and Rieffel}]{Izquierdo2021}
\bibinfo{author}{Z.~Gonzalez~Izquierdo}, \bibinfo{author}{S.~Grabbe}, \bibinfo{author}{S.~Hadfield}, \bibinfo{author}{J.~Marshall}, \bibinfo{author}{Z.~Wang}, \bibinfo{author}{E.~Rieffel},
\newblock \bibinfo{title}{{Ferromagnetically Shifting the Power of Pausing}},
\newblock \bibinfo{journal}{Phys. Rev. Appl.} \bibinfo{volume}{15} (\bibinfo{year}{2021}) \bibinfo{pages}{044013}. \URLprefix \url{https://link.aps.org/doi/10.1103/PhysRevApplied.15.044013}. \DOIprefix\doi{10.1103/PhysRevApplied.15.044013}.
\bibitem[{Gonzalez~Izquierdo et~al.(2022)Gonzalez~Izquierdo, Grabbe, Idris, Wang, Marshall, and Rieffel}]{Izquierdo2022}
\bibinfo{author}{Z.~Gonzalez~Izquierdo}, \bibinfo{author}{S.~Grabbe}, \bibinfo{author}{H.~Idris}, \bibinfo{author}{Z.~Wang}, \bibinfo{author}{J.~Marshall}, \bibinfo{author}{E.~Rieffel},
\newblock \bibinfo{title}{{Advantage of Pausing: Parameter Setting for Quantum Annealers}},
\newblock \bibinfo{journal}{Phys. Rev. Appl.} \bibinfo{volume}{18} (\bibinfo{year}{2022}) \bibinfo{pages}{054056}. \URLprefix \url{https://link.aps.org/doi/10.1103/PhysRevApplied.18.054056}. \DOIprefix\doi{10.1103/PhysRevApplied.18.054056}.
\bibitem[{Pokharel et~al.(2023)Pokharel, Izquierdo, Lott, Strbac, Osiewalski, Papathanasiou, Kondratyev, Venturelli, and Rieffel}]{Pokharel2023}
\bibinfo{author}{B.~Pokharel}, \bibinfo{author}{Z.~G. Izquierdo}, \bibinfo{author}{P.~A. Lott}, \bibinfo{author}{E.~Strbac}, \bibinfo{author}{K.~Osiewalski}, \bibinfo{author}{E.~Papathanasiou}, \bibinfo{author}{A.~Kondratyev}, \bibinfo{author}{D.~Venturelli}, \bibinfo{author}{E.~Rieffel},
\newblock \bibinfo{title}{{Inter-generational comparison of quantum annealers in solving hard scheduling problems}},
\newblock \bibinfo{journal}{Quantum Information Processing} \bibinfo{volume}{22} (\bibinfo{year}{2023}) \bibinfo{pages}{364}. \DOIprefix\doi{https://doi.org/10.1007/s11128-023-04077-z}.
\bibitem[{Marshall et~al.(2020)Marshall, Di~Gioacchino, and Rieffel}]{Marshall2020}
\bibinfo{author}{J.~Marshall}, \bibinfo{author}{A.~Di~Gioacchino}, \bibinfo{author}{E.~G. Rieffel},
\newblock \bibinfo{title}{Perils of embedding for sampling problems},
\newblock \bibinfo{journal}{Phys. Rev. Res.} \bibinfo{volume}{2} (\bibinfo{year}{2020}) \bibinfo{pages}{023020}. \URLprefix \url{https://link.aps.org/doi/10.1103/PhysRevResearch.2.023020}. \DOIprefix\doi{10.1103/PhysRevResearch.2.023020}.
\bibitem[{Marshall et~al.(2022)Marshall, Mossi, and Rieffel}]{Marshall2022a}
\bibinfo{author}{J.~Marshall}, \bibinfo{author}{G.~Mossi}, \bibinfo{author}{E.~G. Rieffel},
\newblock \bibinfo{title}{Perils of embedding for quantum sampling},
\newblock \bibinfo{journal}{Phys. Rev. A} \bibinfo{volume}{105} (\bibinfo{year}{2022}) \bibinfo{pages}{022615}. \URLprefix \url{https://link.aps.org/doi/10.1103/PhysRevA.105.022615}. \DOIprefix\doi{10.1103/PhysRevA.105.022615}.
\bibitem[{Gonzalez~Izquierdo et~al.(2020)Gonzalez~Izquierdo, Zhou, Markstr\"om, and Hen}]{Gonzalez2020}
\bibinfo{author}{Z.~Gonzalez~Izquierdo}, \bibinfo{author}{R.~Zhou}, \bibinfo{author}{K.~Markstr\"om}, \bibinfo{author}{I.~Hen},
\newblock \bibinfo{title}{Discriminating nonisomorphic graphs with an experimental quantum annealer},
\newblock \bibinfo{journal}{Phys. Rev. A} \bibinfo{volume}{102} (\bibinfo{year}{2020}) \bibinfo{pages}{032622}. \URLprefix \url{https://link.aps.org/doi/10.1103/PhysRevA.102.032622}. \DOIprefix\doi{10.1103/PhysRevA.102.032622}.
\bibitem[{Knysh et~al.(2020)Knysh, Plamadeala, and Venturelli}]{Knysh2020}
\bibinfo{author}{S.~Knysh}, \bibinfo{author}{E.~Plamadeala}, \bibinfo{author}{D.~Venturelli},
\newblock \bibinfo{title}{{Quantum annealing speedup of embedded problems via suppression of Griffiths singularities}},
\newblock \bibinfo{journal}{Phys. Rev. B} \bibinfo{volume}{102} (\bibinfo{year}{2020}) \bibinfo{pages}{220407}. \URLprefix \url{https://link.aps.org/doi/10.1103/PhysRevB.102.220407}. \DOIprefix\doi{10.1103/PhysRevB.102.220407}.
\bibitem[{Bernal et~al.(2020)Bernal, Booth, Dridi, Alghassi, Tayur, and Venturelli}]{bernal2020integer}
\bibinfo{author}{D.~E. Bernal}, \bibinfo{author}{K.~E. Booth}, \bibinfo{author}{R.~Dridi}, \bibinfo{author}{H.~Alghassi}, \bibinfo{author}{S.~Tayur}, \bibinfo{author}{D.~Venturelli},
\newblock \bibinfo{title}{{Integer programming techniques for minor-embedding in quantum annealers}},
\newblock in: \bibinfo{booktitle}{{Integration of Constraint Programming, Artificial Intelligence, and Operations Research: 17th International Conference, CPAIOR 2020, Vienna, Austria, September 21--24, 2020, Proceedings 17}}, \bibinfo{organization}{Springer}, \bibinfo{year}{2020}, pp. \bibinfo{pages}{112--129}.
\bibitem[{Unsal and Brady(2022)}]{unsal2022b}
\bibinfo{author}{C.~M. Unsal}, \bibinfo{author}{L.~T. Brady}, \bibinfo{title}{{Quantum Adversarial Learning in Emulation of Monte-Carlo Methods for Max-cut Approximation: QAOA is not optimal}}, \bibinfo{year}{2022}. \href{http://arxiv.org/abs/2211.13767}{{\tt arXiv:2211.13767}}.
\bibitem[{Garc\'{\i}a-Pintos et~al.(2023)Garc\'{\i}a-Pintos, Brady, Bringewatt, and Liu}]{Pintos2023}
\bibinfo{author}{L.~P. Garc\'{\i}a-Pintos}, \bibinfo{author}{L.~T. Brady}, \bibinfo{author}{J.~Bringewatt}, \bibinfo{author}{Y.-K. Liu},
\newblock \bibinfo{title}{{Lower Bounds on Quantum Annealing Times}},
\newblock \bibinfo{journal}{Phys. Rev. Lett.} \bibinfo{volume}{130} (\bibinfo{year}{2023}) \bibinfo{pages}{140601}. \URLprefix \url{https://link.aps.org/doi/10.1103/PhysRevLett.130.140601}. \DOIprefix\doi{10.1103/PhysRevLett.130.140601}.
\bibitem[{Stollenwerk et~al.(2020{\natexlab{a}})Stollenwerk, Hadfield, and Wang}]{Stollenwerk2020}
\bibinfo{author}{T.~Stollenwerk}, \bibinfo{author}{S.~Hadfield}, \bibinfo{author}{Z.~Wang},
\newblock \bibinfo{title}{{Toward Quantum Gate-Model Heuristics for Real-World Planning Problems}},
\newblock \bibinfo{journal}{IEEE Transactions on Quantum Engineering} \bibinfo{volume}{1} (\bibinfo{year}{2020}{\natexlab{a}}) \bibinfo{pages}{1--16}. \DOIprefix\doi{10.1109/TQE.2020.3030609}.
\bibitem[{Stollenwerk et~al.(2020{\natexlab{b}})Stollenwerk, O’Gorman, Venturelli, Mandrà, Rodionova, Ng, Sridhar, Rieffel, and Biswas}]{Stollenwerk2020b}
\bibinfo{author}{T.~Stollenwerk}, \bibinfo{author}{B.~O’Gorman}, \bibinfo{author}{D.~Venturelli}, \bibinfo{author}{S.~Mandrà}, \bibinfo{author}{O.~Rodionova}, \bibinfo{author}{H.~Ng}, \bibinfo{author}{B.~Sridhar}, \bibinfo{author}{E.~G. Rieffel}, \bibinfo{author}{R.~Biswas},
\newblock \bibinfo{title}{{Quantum Annealing Applied to De-Conflicting Optimal Trajectories for Air Traffic Management}},
\newblock \bibinfo{journal}{IEEE Transactions on Intelligent Transportation Systems} \bibinfo{volume}{21} (\bibinfo{year}{2020}{\natexlab{b}}) \bibinfo{pages}{285--297}. \DOIprefix\doi{10.1109/TITS.2019.2891235}.
\bibitem[{Booth et~al.(2020)Booth, O’Gorman, Marshall, Hadfield, and Rieffel}]{booth2020quantum}
\bibinfo{author}{K.~E. Booth}, \bibinfo{author}{B.~O’Gorman}, \bibinfo{author}{J.~Marshall}, \bibinfo{author}{S.~Hadfield}, \bibinfo{author}{E.~Rieffel},
\newblock \bibinfo{title}{Quantum-accelerated global constraint filtering},
\newblock in: \bibinfo{booktitle}{{Principles and Practice of Constraint Programming: 26th International Conference, CP 2020, Louvain-la-Neuve, Belgium, September 7--11, 2020, Proceedings 26}}, \bibinfo{organization}{Springer}, \bibinfo{year}{2020}, pp. \bibinfo{pages}{72--89}. \DOIprefix\doi{https://doi.org/10.1007/978-3-030-58475-7_5}.
\bibitem[{Booth et~al.(2021)Booth, O'Gorman, Marshall, Hadfield, and Rieffel}]{Booth2021}
\bibinfo{author}{K.~E.~C. Booth}, \bibinfo{author}{B.~O'Gorman}, \bibinfo{author}{J.~Marshall}, \bibinfo{author}{S.~Hadfield}, \bibinfo{author}{E.~Rieffel},
\newblock \bibinfo{title}{Quantum-accelerated constraint programming},
\newblock \bibinfo{journal}{{Quantum}} \bibinfo{volume}{5} (\bibinfo{year}{2021}) \bibinfo{pages}{550}. \URLprefix \url{https://doi.org/10.22331/q-2021-09-28-550}. \DOIprefix\doi{10.22331/q-2021-09-28-550}.
\bibitem[{Shaydulin et~al.(2021)Shaydulin, Hadfield, Hogg, and Safro}]{Shaydulin2021}
\bibinfo{author}{R.~Shaydulin}, \bibinfo{author}{S.~Hadfield}, \bibinfo{author}{T.~Hogg}, \bibinfo{author}{I.~Safro},
\newblock \bibinfo{title}{{Classical symmetries and the Quantum Approximate Optimization Algorithm}},
\newblock \bibinfo{journal}{Quantum Information Processing} \bibinfo{volume}{20} (\bibinfo{year}{2021}). \URLprefix \url{http://dx.doi.org/10.1007/s11128-021-03298-4}. \DOIprefix\doi{10.1007/s11128-021-03298-4}.
\bibitem[{Kerger et~al.(2023)Kerger, {Bernal Neira}, Gonzalez~Izquierdo, and Rieffel}]{Kerger2023a}
\bibinfo{author}{P.~Kerger}, \bibinfo{author}{D.~E. {Bernal Neira}}, \bibinfo{author}{Z.~Gonzalez~Izquierdo}, \bibinfo{author}{E.~G. Rieffel},
\newblock \bibinfo{title}{{Mind the $\tilde{{O}}$: Asymptotically Better, but Still Impractical, Quantum Distributed Algorithms}},
\newblock \bibinfo{journal}{Algorithms} \bibinfo{volume}{16} (\bibinfo{year}{2023}). \URLprefix \url{https://www.mdpi.com/1999-4893/16/7/332}. \DOIprefix\doi{10.3390/a16070332}.
\bibitem[{Hadfield(2021)}]{hadfield2021representation}
\bibinfo{author}{S.~Hadfield},
\newblock \bibinfo{title}{On the representation of {B}oolean and real functions as {H}amiltonians for quantum computing},
\newblock \bibinfo{journal}{ACM Transactions on Quantum Computing} \bibinfo{volume}{2} (\bibinfo{year}{2021}) \bibinfo{pages}{1--21}. \DOIprefix\doi{https://doi.org/10.1145/3478519}.
\bibitem[{Sawaya et~al.(2023)Sawaya, Schmitz, and Hadfield}]{sawaya2023encoding}
\bibinfo{author}{N.~P. Sawaya}, \bibinfo{author}{A.~T. Schmitz}, \bibinfo{author}{S.~Hadfield},
\newblock \bibinfo{title}{{Encoding trade-offs and design toolkits in quantum algorithms for discrete optimization: coloring, routing, scheduling, and other problems}},
\newblock \bibinfo{journal}{Quantum} \bibinfo{volume}{7} (\bibinfo{year}{2023}) \bibinfo{pages}{1111}. \DOIprefix\doi{https://doi.org/10.22331/q-2023-09-14-1111}.
\bibitem[{van~de Wetering(2020)}]{van2020zx}
\bibinfo{author}{J.~van~de Wetering},
\newblock \bibinfo{title}{{ZX}-calculus for the working quantum computer scientist},
\newblock \bibinfo{journal}{arXiv preprint arXiv:2012.13966}  (\bibinfo{year}{2020}). \href{http://arxiv.org/abs/2012.13966}{{\tt arXiv:2012.13966}}.
\bibitem[{Stollenwerk and Hadfield(2023)}]{stollenwerk2022diagrammatic}
\bibinfo{author}{T.~Stollenwerk}, \bibinfo{author}{S.~Hadfield},
\newblock \bibinfo{title}{{Diagrammatic Analysis for Parameterized Quantum Circuits}},
\newblock in: \bibinfo{booktitle}{{Proceedings 19th International Conference on Quantum Physics and Logic (QPL 2022), Wolfson College, Oxford, UK, 27 June - 1 July 2022}}, volume \bibinfo{volume}{394}, \bibinfo{organization}{Electronic Proceedings in Theoretical Computer Science}, \bibinfo{year}{2023}, pp. \bibinfo{pages}{262--301}. \URLprefix \url{http://cgi.cse.unsw.edu.au/~eptcs/paper.cgi?QPL2022.15}. \DOIprefix\doi{10.4204/EPTCS.394.15}.
\bibitem[{Zhao and Gao(2021)}]{zhao2021analyzing}
\bibinfo{author}{C.~Zhao}, \bibinfo{author}{X.-S. Gao},
\newblock \bibinfo{title}{{Analyzing the barren plateau phenomenon in training quantum neural networks with the ZX-calculus}},
\newblock \bibinfo{journal}{Quantum} \bibinfo{volume}{5} (\bibinfo{year}{2021}) \bibinfo{pages}{466}. \DOIprefix\doi{https://doi.org/10.22331/q-2021-06-04-466}.
\bibitem[{Toumi et~al.(2021)Toumi, Yeung, and de~Felice}]{toumi2021diagrammatic}
\bibinfo{author}{A.~Toumi}, \bibinfo{author}{R.~Yeung}, \bibinfo{author}{G.~de~Felice}, \bibinfo{title}{{Diagrammatic differentiation for quantum machine learning}}, \bibinfo{year}{2021}. \href{http://arxiv.org/abs/2103.07960}{{\tt arXiv:2103.07960}}.
\bibitem[{Mandra et~al.(2023)Mandra, Akbari~Asanjan, Brady, Lott, and {Bernal Neira}}]{Mandra_PySA_Fast_Simulated_2023}
\bibinfo{author}{S.~Mandra}, \bibinfo{author}{A.~Akbari~Asanjan}, \bibinfo{author}{L.~Brady}, \bibinfo{author}{A.~Lott}, \bibinfo{author}{D.~E. {Bernal Neira}}, \bibinfo{title}{{PySA: Fast Simulated Annealing in Native Python}}, \bibinfo{year}{2023}. \URLprefix \url{https://github.com/nasa/pysa}.
\bibitem[{{Maciel Xavier} et~al.(2023){Maciel Xavier}, {Ripper}, {Andrade}, {Dias Garcia}, {Maculan}, and {Bernal Neira}}]{Xavier_Qubo_jl_2023}
\bibinfo{author}{P.~{Maciel Xavier}}, \bibinfo{author}{P.~{Ripper}}, \bibinfo{author}{T.~{Andrade}}, \bibinfo{author}{J.~{Dias Garcia}}, \bibinfo{author}{N.~{Maculan}}, \bibinfo{author}{D.~E. {Bernal Neira}},
\newblock \bibinfo{title}{{QUBO.jl: A Julia Ecosystem for Quadratic Unconstrained Binary Optimization}},
\newblock \bibinfo{journal}{arXiv e-prints}  (\bibinfo{year}{2023}) \bibinfo{pages}{arXiv:2307.02577}. \DOIprefix\doi{10.48550/arXiv.2307.02577}. \href{http://arxiv.org/abs/2307.02577}{{\tt arXiv:2307.02577}}.
\bibitem[{Brown et~al.(2023)Brown, {Bernal Neira}, Venturelli, and Pavone}]{brown2023copositive}
\bibinfo{author}{R.~Brown}, \bibinfo{author}{D.~E. {Bernal Neira}}, \bibinfo{author}{D.~Venturelli}, \bibinfo{author}{M.~Pavone}, \bibinfo{title}{{A Copositive Framework for Analysis of Hybrid Ising-Classical Algorithms}}, \bibinfo{year}{2023}. \href{http://arxiv.org/abs/2207.13630}{{\tt arXiv:2207.13630}}.
\bibitem[{Brown et~al.(2024)}]{Brown2024_accelerating}
\bibinfo{author}{R.~Brown}, et~al., \bibinfo{title}{{A}ccelerating {C}ontinuous {V}ariable {C}oherent {I}sing {M}achines via {M}omentum}, \bibinfo{year}{2024}. \href{http://arxiv.org/abs/2401.12135}{{\tt arXiv:2401.12135}}.
\bibitem[{Kim et~al.(2021)Kim, Mandr\`{a}, Venturelli, and Jamieson}]{Kim2021}
\bibinfo{author}{M.~Kim}, \bibinfo{author}{S.~Mandr\`{a}}, \bibinfo{author}{D.~Venturelli}, \bibinfo{author}{K.~Jamieson},
\newblock \bibinfo{title}{{Physics-inspired heuristics for soft MIMO detection in 5G new radio and beyond}},
\newblock in: \bibinfo{booktitle}{Proceedings of the 27th Annual International Conference on Mobile Computing and Networking}, MobiCom '21, \bibinfo{publisher}{Association for Computing Machinery}, \bibinfo{address}{New York, NY, USA}, \bibinfo{year}{2021}, p. \bibinfo{pages}{42–55}. \URLprefix \url{https://doi.org/10.1145/3447993.3448619}. \DOIprefix\doi{10.1145/3447993.3448619}.
\bibitem[{Mohseni et~al.(2022)}]{mohseni2022ising}
\bibinfo{author}{N.~Mohseni}, et~al.,
\newblock \bibinfo{title}{{Ising machines as hardware solvers of combinatorial optimization problems}},
\newblock \bibinfo{journal}{Nature Reviews Physics} \bibinfo{volume}{4} (\bibinfo{year}{2022}) \bibinfo{pages}{363--379}.
\bibitem[{Singh et~al.(2021)Singh, Jamieson, Venturelli, and McMahon}]{singh2021ising}
\bibinfo{author}{A.~K. Singh}, \bibinfo{author}{K.~Jamieson}, \bibinfo{author}{D.~Venturelli}, \bibinfo{author}{P.~McMahon}, \bibinfo{title}{{Ising Machines' Dynamics and Regularization for Near-Optimal Large and Massive MIMO Detection}}, \bibinfo{year}{2021}. \href{http://arxiv.org/abs/2105.10535}{{\tt arXiv:2105.10535}}.
\bibitem[{Taassob et~al.(2023)}]{taassob23}
\bibinfo{author}{A.~Taassob}, et~al.,
\newblock \bibinfo{title}{Neural {D}eep {O}perator {N}etworks representation of {C}oherent {I}sing {M}achine {D}ynamics},
\newblock \bibinfo{journal}{Neurips 2023 Workshop Machine Learning with New Compute Paradigms}  (\bibinfo{year}{2023}).
\bibitem[{Witten(2010)}]{witten2010new}
\bibinfo{author}{E.~Witten},
\newblock \bibinfo{title}{{A new look at the path integral of quantum mechanics}},
\newblock \bibinfo{journal}{Surv. Differ. Geom.} \bibinfo{volume}{15} (\bibinfo{year}{2010}) \bibinfo{pages}{345--421}. \URLprefix \url{https://arxiv.org/abs/1009.6032}.
\bibitem[{Witten(2011)}]{witten2011analytic}
\bibinfo{author}{E.~Witten},
\newblock \bibinfo{title}{Analytic continuation of {C}hern-{S}imons theory},
\newblock \bibinfo{journal}{AMS/IP Stud. Adv. Math} \bibinfo{volume}{50} (\bibinfo{year}{2011}) \bibinfo{pages}{347}. \URLprefix \url{https://arxiv.org/abs/1001.2933}.
\bibitem[{Mooney et~al.(2022)Mooney, Bringewatt, Warrington, and Brady}]{Mooney2022}
\bibinfo{author}{T.~C. Mooney}, \bibinfo{author}{J.~Bringewatt}, \bibinfo{author}{N.~C. Warrington}, \bibinfo{author}{L.~T. Brady},
\newblock \bibinfo{title}{{Lefschetz thimble quantum Monte Carlo for spin systems}},
\newblock \bibinfo{journal}{Phys. Rev. B} \bibinfo{volume}{106} (\bibinfo{year}{2022}) \bibinfo{pages}{214416}. \URLprefix \url{https://link.aps.org/doi/10.1103/PhysRevB.106.214416}. \DOIprefix\doi{10.1103/PhysRevB.106.214416}.
\bibitem[{Bendall et~al.(2006)}]{BENDALL2006288}
\bibinfo{author}{G.~Bendall}, et~al.,
\newblock \bibinfo{title}{Greedy-type resistance of combinatorial problems},
\newblock \bibinfo{journal}{Discrete Optimization} \bibinfo{volume}{3} (\bibinfo{year}{2006}) \bibinfo{pages}{288--298}.
\bibitem[{M\"obius et~al.(1997)}]{Mobius1997}
\bibinfo{author}{A.~M\"obius}, et~al.,
\newblock \bibinfo{title}{Optimization by {T}hermal {C}ycling},
\newblock \bibinfo{journal}{PRL} \bibinfo{volume}{79} (\bibinfo{year}{1997}) \bibinfo{pages}{4297--4301}.
\bibitem[{Barzegar et~al.(2021)Barzegar, Kankani, Mandr\`a, and Katzgraber}]{Barzegar2021}
\bibinfo{author}{A.~Barzegar}, \bibinfo{author}{A.~Kankani}, \bibinfo{author}{S.~Mandr\`a}, \bibinfo{author}{H.~G. Katzgraber},
\newblock \bibinfo{title}{Optimization and benchmarking of the thermal cycling algorithm},
\newblock \bibinfo{journal}{Phys. Rev. E} \bibinfo{volume}{104} (\bibinfo{year}{2021}) \bibinfo{pages}{035302}. \URLprefix \url{https://link.aps.org/doi/10.1103/PhysRevE.104.035302}. \DOIprefix\doi{10.1103/PhysRevE.104.035302}.
\bibitem[{Zhu et~al.(2015)}]{zhu2015efficient}
\bibinfo{author}{Z.~Zhu}, et~al.,
\newblock \bibinfo{title}{{Efficient cluster algorithm for spin glasses in any space dimension}},
\newblock \bibinfo{journal}{PRL} \bibinfo{volume}{115} (\bibinfo{year}{2015}) \bibinfo{pages}{077201}.
\bibitem[{Mandra et~al.(2016)}]{mandra2016strengths}
\bibinfo{author}{S.~Mandra}, et~al.,
\newblock \bibinfo{title}{Strengths and weaknesses of weak-strong cluster problems: {A} detailed overview of state-of-the-art classical heuristics versus quantum approaches},
\newblock \bibinfo{journal}{PRA} \bibinfo{volume}{94} (\bibinfo{year}{2016}) \bibinfo{pages}{022337}.
\bibitem[{Mandra et~al.(2018)}]{mandra2018deceptive}
\bibinfo{author}{S.~Mandra}, et~al.,
\newblock \bibinfo{title}{{A deceptive step towards quantum speedup detection}},
\newblock \bibinfo{journal}{Quantum Science and Technology} \bibinfo{volume}{3} (\bibinfo{year}{2018}) \bibinfo{pages}{04LT01}.
\bibitem[{Gao et~al.(2020)Gao, Wilson, Vandal, Vinci, Nemani, and Rieffel}]{Gao2020}
\bibinfo{author}{N.~Gao}, \bibinfo{author}{M.~Wilson}, \bibinfo{author}{T.~Vandal}, \bibinfo{author}{W.~Vinci}, \bibinfo{author}{R.~Nemani}, \bibinfo{author}{E.~Rieffel},
\newblock \bibinfo{title}{{High-Dimensional Similarity Search with Quantum-Assisted Variational Autoencoder}},
\newblock in: \bibinfo{booktitle}{Proceedings of the 26th ACM SIGKDD International Conference on Knowledge Discovery \& Data Mining}, KDD '20, \bibinfo{publisher}{Association for Computing Machinery}, \bibinfo{address}{New York, NY, USA}, \bibinfo{year}{2020}, p. \bibinfo{pages}{956–964}. \URLprefix \url{https://doi.org/10.1145/3394486.3403138}. \DOIprefix\doi{10.1145/3394486.3403138}.
\bibitem[{Akbari~Asanjan et~al.(2023)Akbari~Asanjan, Memarzadeh, Lott, Rieffel, and Grabbe}]{Asanjan2023}
\bibinfo{author}{A.~Akbari~Asanjan}, \bibinfo{author}{M.~Memarzadeh}, \bibinfo{author}{P.~A. Lott}, \bibinfo{author}{E.~Rieffel}, \bibinfo{author}{S.~Grabbe},
\newblock \bibinfo{title}{{Probabilistic Wildfire Segmentation Using Supervised Deep Generative Model from Satellite Imagery}},
\newblock \bibinfo{journal}{Remote Sensing} \bibinfo{volume}{15} (\bibinfo{year}{2023}). \URLprefix \url{https://www.mdpi.com/2072-4292/15/11/2718}. \DOIprefix\doi{10.3390/rs15112718}.
\bibitem[{Wilson et~al.(2021{\natexlab{a}})Wilson, Vandal, Hogg, and Rieffel}]{Wilson2021}
\bibinfo{author}{M.~Wilson}, \bibinfo{author}{T.~Vandal}, \bibinfo{author}{T.~Hogg}, \bibinfo{author}{E.~G. Rieffel},
\newblock \bibinfo{title}{{Quantum-assisted associative adversarial network: applying quantum annealing in deep learning}},
\newblock \bibinfo{journal}{Quantum Machine Intelligence} \bibinfo{volume}{3} (\bibinfo{year}{2021}{\natexlab{a}}) \bibinfo{pages}{19}. \URLprefix \url{https://doi.org/10.1007/s42484-021-00047-9}. \DOIprefix\doi{10.1007/s42484-021-00047-9}.
\bibitem[{Wilson et~al.(2021{\natexlab{b}})Wilson, Stromswold, Wudarski, Hadfield, Tubman, and Rieffel}]{Wilson2021b}
\bibinfo{author}{M.~Wilson}, \bibinfo{author}{R.~Stromswold}, \bibinfo{author}{F.~Wudarski}, \bibinfo{author}{S.~Hadfield}, \bibinfo{author}{N.~M. Tubman}, \bibinfo{author}{E.~G. Rieffel},
\newblock \bibinfo{title}{{Optimizing quantum heuristics with meta-learning}},
\newblock \bibinfo{journal}{Quantum Machine Intelligence} \bibinfo{volume}{3} (\bibinfo{year}{2021}{\natexlab{b}}) \bibinfo{pages}{13}. \URLprefix \url{https://doi.org/10.1007/s42484-020-00022-w}. \DOIprefix\doi{10.1007/s42484-020-00022-w}.
\bibitem[{O'Connor and Vinci(2021)}]{oconnor2021rbmflow}
\bibinfo{author}{D.~O'Connor}, \bibinfo{author}{W.~Vinci}, \bibinfo{title}{{RBM-Flow and D-Flow: Invertible Flows with Discrete Energy Base Spaces}}, \bibinfo{year}{2021}. \href{http://arxiv.org/abs/2012.13196}{{\tt arXiv:2012.13196}}.
\bibitem[{Templin et~al.(2023)Templin, Memarzadeh, Vinci, Lott, Asanjan, Armenakas, and Rieffel}]{templin2023anomaly}
\bibinfo{author}{T.~Templin}, \bibinfo{author}{M.~Memarzadeh}, \bibinfo{author}{W.~Vinci}, \bibinfo{author}{P.~A. Lott}, \bibinfo{author}{A.~A. Asanjan}, \bibinfo{author}{A.~A. Armenakas}, \bibinfo{author}{E.~Rieffel}, \bibinfo{title}{{Anomaly Detection in Aeronautics Data with Quantum-compatible Discrete Deep Generative Model}}, \bibinfo{year}{2023}. \href{http://arxiv.org/abs/2303.12302}{{\tt arXiv:2303.12302}}.
\bibitem[{Kitaev(2006)}]{KITAEV20062}
\bibinfo{author}{A.~Kitaev},
\newblock \bibinfo{title}{{Anyons in an exactly solved model and beyond}},
\newblock \bibinfo{journal}{Annals of Physics} \bibinfo{volume}{321} (\bibinfo{year}{2006}) \bibinfo{pages}{2--111}. \URLprefix \url{https://www.sciencedirect.com/science/article/pii/S0003491605002381}. \DOIprefix\doi{https://doi.org/10.1016/j.aop.2005.10.005}, \bibinfo{note}{january Special Issue}.
\bibitem[{Li et~al.(2023)Li, Alam, Iadecola, Jahin, Job, Kurkcuoglu, Li, Orth, Ozguler, Perdue, and Tubman}]{Li2023}
\bibinfo{author}{A.~C.~Y. Li}, \bibinfo{author}{M.~S. Alam}, \bibinfo{author}{T.~Iadecola}, \bibinfo{author}{A.~Jahin}, \bibinfo{author}{J.~Job}, \bibinfo{author}{D.~M. Kurkcuoglu}, \bibinfo{author}{R.~Li}, \bibinfo{author}{P.~P. Orth}, \bibinfo{author}{A.~B. Ozguler}, \bibinfo{author}{G.~N. Perdue}, \bibinfo{author}{N.~M. Tubman},
\newblock \bibinfo{title}{{Benchmarking variational quantum eigensolvers for the square-octagon-lattice Kitaev model}},
\newblock \bibinfo{journal}{Phys. Rev. Res.} \bibinfo{volume}{5} (\bibinfo{year}{2023}) \bibinfo{pages}{033071}. \URLprefix \url{https://link.aps.org/doi/10.1103/PhysRevResearch.5.033071}. \DOIprefix\doi{10.1103/PhysRevResearch.5.033071}.
\bibitem[{Wecker et~al.(2015)Wecker, Hastings, and Troyer}]{PhysRevA.92.042303}
\bibinfo{author}{D.~Wecker}, \bibinfo{author}{M.~B. Hastings}, \bibinfo{author}{M.~Troyer},
\newblock \bibinfo{title}{{Progress towards practical quantum variational algorithms}},
\newblock \bibinfo{journal}{Phys. Rev. A} \bibinfo{volume}{92} (\bibinfo{year}{2015}) \bibinfo{pages}{042303}. \URLprefix \url{https://link.aps.org/doi/10.1103/PhysRevA.92.042303}. \DOIprefix\doi{10.1103/PhysRevA.92.042303}.
\bibitem[{Jahin et~al.(2022)Jahin, Li, Iadecola, Orth, Perdue, Macridin, Alam, and Tubman}]{Jahin2022}
\bibinfo{author}{A.~Jahin}, \bibinfo{author}{A.~C.~Y. Li}, \bibinfo{author}{T.~Iadecola}, \bibinfo{author}{P.~P. Orth}, \bibinfo{author}{G.~N. Perdue}, \bibinfo{author}{A.~Macridin}, \bibinfo{author}{M.~S. Alam}, \bibinfo{author}{N.~M. Tubman},
\newblock \bibinfo{title}{{Fermionic approach to variational quantum simulation of Kitaev spin models}},
\newblock \bibinfo{journal}{Phys. Rev. A} \bibinfo{volume}{106} (\bibinfo{year}{2022}) \bibinfo{pages}{022434}. \URLprefix \url{https://link.aps.org/doi/10.1103/PhysRevA.106.022434}. \DOIprefix\doi{10.1103/PhysRevA.106.022434}.
\bibitem[{Arute et~al.(2020)Arute, Arya, Babbush, Bacon, Bardin, and et~al.}]{arute2020observation}
\bibinfo{author}{F.~Arute}, \bibinfo{author}{K.~Arya}, \bibinfo{author}{R.~Babbush}, \bibinfo{author}{D.~Bacon}, \bibinfo{author}{J.~C. Bardin}, \bibinfo{author}{et~al.}, \bibinfo{title}{{Observation of separated dynamics of charge and spin in the Fermi-Hubbard model}}, \bibinfo{year}{2020}. \href{http://arxiv.org/abs/2010.07965}{{\tt arXiv:2010.07965}}.
\bibitem[{Levy et~al.(2022)Levy, Izquierdo, Wang, Marshall, Barreto, Fry-Bouriaux, O'Connor, Warburton, Wiebe, Rieffel, and Wudarski}]{levy2022solving}
\bibinfo{author}{R.~Levy}, \bibinfo{author}{Z.~G. Izquierdo}, \bibinfo{author}{Z.~Wang}, \bibinfo{author}{J.~Marshall}, \bibinfo{author}{J.~Barreto}, \bibinfo{author}{L.~Fry-Bouriaux}, \bibinfo{author}{D.~T. O'Connor}, \bibinfo{author}{P.~A. Warburton}, \bibinfo{author}{N.~Wiebe}, \bibinfo{author}{E.~Rieffel}, \bibinfo{author}{F.~A. Wudarski}, \bibinfo{title}{{Towards solving the Fermi-Hubbard model via tailored quantum annealers}}, \bibinfo{year}{2022}. \href{http://arxiv.org/abs/2207.14374}{{\tt arXiv:2207.14374}}.
\bibitem[{Kremenetski et~al.(2021)Kremenetski, Mejuto-Zaera, Cotton, and Tubman}]{Kremenetski2021a}
\bibinfo{author}{V.~Kremenetski}, \bibinfo{author}{C.~Mejuto-Zaera}, \bibinfo{author}{S.~J. Cotton}, \bibinfo{author}{N.~M. Tubman},
\newblock \bibinfo{title}{{Simulation of adiabatic quantum computing for molecular ground states}},
\newblock \bibinfo{journal}{The Journal of Chemical Physics} \bibinfo{volume}{155} (\bibinfo{year}{2021}) \bibinfo{pages}{234106}. \URLprefix \url{https://doi.org/10.1063/5.0060124}. \DOIprefix\doi{10.1063/5.0060124}.
\bibitem[{Woitzik et~al.(2020)Woitzik, Barkoutsos, Wudarski, Buchleitner, and Tavernelli}]{Woitzik_2020}
\bibinfo{author}{A.~J.~C. Woitzik}, \bibinfo{author}{P.~K. Barkoutsos}, \bibinfo{author}{F.~Wudarski}, \bibinfo{author}{A.~Buchleitner}, \bibinfo{author}{I.~Tavernelli},
\newblock \bibinfo{title}{{Entanglement production and convergence properties of the variational quantum eigensolver}},
\newblock \bibinfo{journal}{Physical Review A} \bibinfo{volume}{102} (\bibinfo{year}{2020}). \URLprefix \url{http://dx.doi.org/10.1103/PhysRevA.102.042402}. \DOIprefix\doi{10.1103/physreva.102.042402}.
\bibitem[{Khan et~al.(2023)Khan, Clark, and Tubman}]{khan2023preoptimizing}
\bibinfo{author}{A.~Khan}, \bibinfo{author}{B.~K. Clark}, \bibinfo{author}{N.~M. Tubman},
\newblock \bibinfo{title}{{Pre-optimizing variational quantum eigensolvers with tensor networks}},
\newblock \bibinfo{journal}{arXiv:2310.12965}  (\bibinfo{year}{2023}).
\bibitem[{Klymko et~al.(2021)Klymko, Mejuto-Zaera, Cotton, Wudarski, Urbanek, Hait, Head-Gordon, Whaley, Moussa, Wiebe, de~Jong, and Tubman}]{klymko2021real}
\bibinfo{author}{K.~Klymko}, \bibinfo{author}{C.~Mejuto-Zaera}, \bibinfo{author}{S.~J. Cotton}, \bibinfo{author}{F.~Wudarski}, \bibinfo{author}{M.~Urbanek}, \bibinfo{author}{D.~Hait}, \bibinfo{author}{M.~Head-Gordon}, \bibinfo{author}{K.~B. Whaley}, \bibinfo{author}{J.~Moussa}, \bibinfo{author}{N.~Wiebe}, \bibinfo{author}{W.~A. de~Jong}, \bibinfo{author}{N.~M. Tubman}, \bibinfo{title}{Real time evolution for ultracompact {H}amiltonian eigenstates on quantum hardware}, \bibinfo{year}{2021}. \href{http://arxiv.org/abs/2103.08563}{{\tt arXiv:2103.08563}}.
\bibitem[{Gustafson et~al.(2023)Gustafson, Li, Khan, Kim, Kurkcuoglu, Alam, Orth, Rahmani, and Iadecola}]{Gustafson2023c}
\bibinfo{author}{E.~J. Gustafson}, \bibinfo{author}{A.~C.~Y. Li}, \bibinfo{author}{A.~Khan}, \bibinfo{author}{J.~Kim}, \bibinfo{author}{D.~M. Kurkcuoglu}, \bibinfo{author}{M.~S. Alam}, \bibinfo{author}{P.~P. Orth}, \bibinfo{author}{A.~Rahmani}, \bibinfo{author}{T.~Iadecola},
\newblock \bibinfo{title}{{Preparing quantum many-body scar states on quantum computers}},
\newblock \bibinfo{journal}{Quantum} \bibinfo{volume}{7} (\bibinfo{year}{2023}) \bibinfo{pages}{1171}. \URLprefix \url{http://dx.doi.org/10.22331/q-2023-11-07-1171}. \DOIprefix\doi{10.22331/q-2023-11-07-1171}.
\bibitem[{Chamaki et~al.(2022)Chamaki, Hadfield, Klymko, O'Gorman, and Tubman}]{chamaki2022selfconsistent}
\bibinfo{author}{D.~B. Chamaki}, \bibinfo{author}{S.~Hadfield}, \bibinfo{author}{K.~Klymko}, \bibinfo{author}{B.~O'Gorman}, \bibinfo{author}{N.~M. Tubman}, \bibinfo{title}{{Self-consistent Quantum Iteratively Sparsified Hamiltonian method (SQuISH): A new algorithm for efficient Hamiltonian simulation and compression}}, \bibinfo{year}{2022}. \href{http://arxiv.org/abs/2211.16522}{{\tt arXiv:2211.16522}}.
\bibitem[{von Lilienfeld and Burke(2020)}]{vonLilienfeld2020}
\bibinfo{author}{O.~A. von Lilienfeld}, \bibinfo{author}{K.~Burke},
\newblock \bibinfo{title}{{Retrospective on a decade of machine learning for chemical discovery}},
\newblock \bibinfo{journal}{Nature Communications} \bibinfo{volume}{11} (\bibinfo{year}{2020}) \bibinfo{pages}{4895}. \URLprefix \url{https://doi.org/10.1038/s41467-020-18556-9}. \DOIprefix\doi{10.1038/s41467-020-18556-9}.
\bibitem[{Pfau et~al.(2020)Pfau, Spencer, Matthews, and Foulkes}]{Pfau_2020}
\bibinfo{author}{D.~Pfau}, \bibinfo{author}{J.~S. Spencer}, \bibinfo{author}{A.~G. D.~G. Matthews}, \bibinfo{author}{W.~M.~C. Foulkes},
\newblock \bibinfo{title}{{Ab initio solution of the many-electron Schrödinger equation with deep neural networks}},
\newblock \bibinfo{journal}{Physical Review Research} \bibinfo{volume}{2} (\bibinfo{year}{2020}). \URLprefix \url{http://dx.doi.org/10.1103/PhysRevResearch.2.033429}. \DOIprefix\doi{10.1103/physrevresearch.2.033429}.
\bibitem[{Wilson et~al.(2021)Wilson, Gao, Wudarski, Rieffel, and Tubman}]{wilson2021simulations}
\bibinfo{author}{M.~Wilson}, \bibinfo{author}{N.~Gao}, \bibinfo{author}{F.~Wudarski}, \bibinfo{author}{E.~Rieffel}, \bibinfo{author}{N.~M. Tubman}, \bibinfo{title}{{Simulations of state-of-the-art fermionic neural network wave functions with diffusion Monte Carlo}}, \bibinfo{year}{2021}. \href{http://arxiv.org/abs/2103.12570}{{\tt arXiv:2103.12570}}.
\bibitem[{Wilson et~al.(2023)Wilson, Moroni, Holzmann, Gao, Wudarski, Vegge, and Bhowmik}]{Wilson_2023}
\bibinfo{author}{M.~Wilson}, \bibinfo{author}{S.~Moroni}, \bibinfo{author}{M.~Holzmann}, \bibinfo{author}{N.~Gao}, \bibinfo{author}{F.~Wudarski}, \bibinfo{author}{T.~Vegge}, \bibinfo{author}{A.~Bhowmik},
\newblock \bibinfo{title}{{Neural network ansatz for periodic wave functions and the homogeneous electron gas}},
\newblock \bibinfo{journal}{Physical Review B} \bibinfo{volume}{107} (\bibinfo{year}{2023}). \URLprefix \url{http://dx.doi.org/10.1103/PhysRevB.107.235139}. \DOIprefix\doi{10.1103/physrevb.107.235139}.
\bibitem[{Bernal et~al.(2022)Bernal, Ajagekar, Harwood, Stober, Trenev, and You}]{Bernal2022}
\bibinfo{author}{D.~E. Bernal}, \bibinfo{author}{A.~Ajagekar}, \bibinfo{author}{S.~M. Harwood}, \bibinfo{author}{S.~T. Stober}, \bibinfo{author}{D.~Trenev}, \bibinfo{author}{F.~You},
\newblock \bibinfo{title}{Perspectives of quantum computing for chemical engineering},
\newblock \bibinfo{journal}{AIChE Journal} \bibinfo{volume}{68} (\bibinfo{year}{2022}) \bibinfo{pages}{e17651}. \URLprefix \url{https://aiche.onlinelibrary.wiley.com/doi/abs/10.1002/aic.17651}. \DOIprefix\doi{https://doi.org/10.1002/aic.17651}.
\bibitem[{{Bernal Neira} et~al.(2023){Bernal Neira}, Laird, Harwood, Trenev, and Venturelli}]{bernal2023impact}
\bibinfo{author}{D.~E. {Bernal Neira}}, \bibinfo{author}{C.~D. Laird}, \bibinfo{author}{S.~M. Harwood}, \bibinfo{author}{D.~Trenev}, \bibinfo{author}{D.~Venturelli},
\newblock \bibinfo{title}{{Impact of Emerging Computing Architectures and Opportunities for Process Systems Engineering Applications}},
\newblock in: \bibinfo{booktitle}{{Foundations of Computer-Aided Process Operations (FOCAPO)/Chemical Process Control (CPC) 2023}}, \bibinfo{year}{2023}.
\bibitem[{Cotton(2022)}]{Cotton2022}
\bibinfo{author}{S.~J. Cotton},
\newblock \bibinfo{title}{A truncated {{Davidson}} method for the efficient ``chemically accurate'' calculation of full configuration interaction wavefunctions without any large matrix diagonalization},
\newblock \bibinfo{journal}{The Journal of Chemical Physics} \bibinfo{volume}{157} (\bibinfo{year}{2022}) \bibinfo{pages}{224105}. \DOIprefix\doi{10.1063/5.0115796}.
\bibitem[{Suri et~al.(2023)Suri, Barreto, Hadfield, Wiebe, Wudarski, and Marshall}]{Suri2023twounitary}
\bibinfo{author}{N.~Suri}, \bibinfo{author}{J.~Barreto}, \bibinfo{author}{S.~Hadfield}, \bibinfo{author}{N.~Wiebe}, \bibinfo{author}{F.~Wudarski}, \bibinfo{author}{J.~Marshall},
\newblock \bibinfo{title}{Two-{U}nitary {D}ecomposition {A}lgorithm and {O}pen {Q}uantum {S}ystem {S}imulation},
\newblock \bibinfo{journal}{{Quantum}} \bibinfo{volume}{7} (\bibinfo{year}{2023}) \bibinfo{pages}{1002}. \URLprefix \url{https://doi.org/10.22331/q-2023-05-15-1002}. \DOIprefix\doi{10.22331/q-2023-05-15-1002}.
\bibitem[{Kurkcuoglu et~al.(2022)Kurkcuoglu, Alam, Job, Li, Macridin, Perdue, and Providence}]{kurkcuoglu2022quantum}
\bibinfo{author}{D.~M. Kurkcuoglu}, \bibinfo{author}{M.~S. Alam}, \bibinfo{author}{J.~A. Job}, \bibinfo{author}{A.~C.~Y. Li}, \bibinfo{author}{A.~Macridin}, \bibinfo{author}{G.~N. Perdue}, \bibinfo{author}{S.~Providence}, \bibinfo{title}{{Quantum simulation of $\phi^4$ theories in qudit systems}}, \bibinfo{year}{2022}. \href{http://arxiv.org/abs/2108.13357}{{\tt arXiv:2108.13357}}.
\bibitem[{Lamm et~al.(2019)Lamm, Lawrence, and Yamauchi}]{Lamm:2019bik}
\bibinfo{author}{H.~Lamm}, \bibinfo{author}{S.~Lawrence}, \bibinfo{author}{Y.~Yamauchi} (\bibinfo{collaboration}{NuQS}),
\newblock \bibinfo{title}{{General Methods for Digital Quantum Simulation of Gauge Theories}},
\newblock \bibinfo{journal}{Phys. Rev. D} \bibinfo{volume}{100} (\bibinfo{year}{2019}) \bibinfo{pages}{034518}. \DOIprefix\doi{10.1103/PhysRevD.100.034518}. \href{http://arxiv.org/abs/1903.08807}{{\tt arXiv:1903.08807}}.
\bibitem[{Alam et~al.(2022)Alam, Hadfield, Lamm, and Li}]{Alam2022b}
\bibinfo{author}{M.~S. Alam}, \bibinfo{author}{S.~Hadfield}, \bibinfo{author}{H.~Lamm}, \bibinfo{author}{A.~C. Li},
\newblock \bibinfo{title}{{Primitive quantum gates for dihedral gauge theories}},
\newblock \bibinfo{journal}{Physical Review D} \bibinfo{volume}{105} (\bibinfo{year}{2022}). \URLprefix \url{http://dx.doi.org/10.1103/PhysRevD.105.114501}. \DOIprefix\doi{10.1103/physrevd.105.114501}.
\bibitem[{Charles et~al.(2023)Charles, Gustafson, Hardt, Herren, Hogan, Lamm, Starecheski, de~Water, and Wagman}]{charles2023simulating}
\bibinfo{author}{C.~Charles}, \bibinfo{author}{E.~J. Gustafson}, \bibinfo{author}{E.~Hardt}, \bibinfo{author}{F.~Herren}, \bibinfo{author}{N.~Hogan}, \bibinfo{author}{H.~Lamm}, \bibinfo{author}{S.~Starecheski}, \bibinfo{author}{R.~S.~V. de~Water}, \bibinfo{author}{M.~L. Wagman}, \bibinfo{title}{{Simulating $\mathbb{Z}_2$ lattice gauge theory on a quantum computer}}, \bibinfo{year}{2023}. \href{http://arxiv.org/abs/2305.02361}{{\tt arXiv:2305.02361}}.
\bibitem[{Gustafson et~al.(2023)Gustafson, Lamm, and Lovelace}]{Gustafson:2023kvd}
\bibinfo{author}{E.~J. Gustafson}, \bibinfo{author}{H.~Lamm}, \bibinfo{author}{F.~Lovelace},
\newblock \bibinfo{title}{{Primitive Quantum Gates for an $SU(2)$ Discrete Subgroup: Binary Octahedral}}  (\bibinfo{year}{2023}). \href{http://arxiv.org/abs/2312.10285}{{\tt arXiv:2312.10285}}.
\bibitem[{Arute et~al.(2019)Arute, Arya, Babbush, Bacon, Bardin, and et~al.}]{Arute2019}
\bibinfo{author}{F.~Arute}, \bibinfo{author}{K.~Arya}, \bibinfo{author}{R.~Babbush}, \bibinfo{author}{D.~Bacon}, \bibinfo{author}{J.~C. Bardin}, \bibinfo{author}{et~al.},
\newblock \bibinfo{title}{{Quantum Supremacy Using a Programmable Superconducting Processor}},
\newblock \bibinfo{journal}{Nature} \bibinfo{volume}{574} (\bibinfo{year}{2019}) \bibinfo{pages}{505--510}. \URLprefix \url{https://www.nature.com/articles/s41586-019-1666-5}. \DOIprefix\doi{10.1038/s41586-019-1666-5}.
\bibitem[{Villalonga et~al.(2020)Villalonga, Lyakh, Boixo, Neven, Humble, Biswas, Rieffel, Ho, and Mandrà}]{Villalonga_2020}
\bibinfo{author}{B.~Villalonga}, \bibinfo{author}{D.~Lyakh}, \bibinfo{author}{S.~Boixo}, \bibinfo{author}{H.~Neven}, \bibinfo{author}{T.~S. Humble}, \bibinfo{author}{R.~Biswas}, \bibinfo{author}{E.~G. Rieffel}, \bibinfo{author}{A.~Ho}, \bibinfo{author}{S.~Mandrà},
\newblock \bibinfo{title}{{Establishing the quantum supremacy frontier with a 281 Pflop/s simulation}},
\newblock \bibinfo{journal}{Quantum Science and Technology} \bibinfo{volume}{5} (\bibinfo{year}{2020}) \bibinfo{pages}{034003}. \URLprefix \url{https://dx.doi.org/10.1088/2058-9565/ab7eeb}. \DOIprefix\doi{10.1088/2058-9565/ab7eeb}.
\bibitem[{Gottesman(1998)}]{gottesman1998fault}
\bibinfo{author}{D.~Gottesman},
\newblock \bibinfo{title}{{Fault-tolerant quantum computation with higher-dimensional systems}},
\newblock in: \bibinfo{booktitle}{{NASA International Conf. on Quantum Computing and Quantum Communications}}, \bibinfo{organization}{Springer}, \bibinfo{year}{1998}, pp. \bibinfo{pages}{302--313}.
\bibitem[{Aaronson and Gottesman(2004)}]{aaronson2004improved}
\bibinfo{author}{S.~Aaronson}, \bibinfo{author}{D.~Gottesman},
\newblock \bibinfo{title}{{Improved simulation of stabilizer circuits}},
\newblock \bibinfo{journal}{PRA} \bibinfo{volume}{70} (\bibinfo{year}{2004}) \bibinfo{pages}{052328}.
\bibitem[{Bravyi and Gosset(2016)}]{bravyi2016improved}
\bibinfo{author}{S.~Bravyi}, \bibinfo{author}{D.~Gosset},
\newblock \bibinfo{title}{Improved classical simulation of quantum circuits dominated by {C}lifford gates},
\newblock \bibinfo{journal}{PRL} \bibinfo{volume}{116} (\bibinfo{year}{2016}) \bibinfo{pages}{250501}.
\bibitem[{Mi et~al.(2021)Mi, Roushan, Quintana, Mandrà, Marshall, and et~al.}]{Mi2021}
\bibinfo{author}{X.~Mi}, \bibinfo{author}{P.~Roushan}, \bibinfo{author}{C.~Quintana}, \bibinfo{author}{S.~Mandrà}, \bibinfo{author}{J.~Marshall}, \bibinfo{author}{et~al.},
\newblock \bibinfo{title}{Information scrambling in quantum circuits},
\newblock \bibinfo{journal}{Science} \bibinfo{volume}{374} (\bibinfo{year}{2021}) \bibinfo{pages}{1479--1483}. \URLprefix \url{https://www.science.org/doi/abs/10.1126/science.abg5029}. \DOIprefix\doi{10.1126/science.abg5029}.
\bibitem[{Aharonov et~al.(2023)Aharonov, Gao, Landau, Liu, and Vazirani}]{aharonov2023polynomial}
\bibinfo{author}{D.~Aharonov}, \bibinfo{author}{X.~Gao}, \bibinfo{author}{Z.~Landau}, \bibinfo{author}{Y.~Liu}, \bibinfo{author}{U.~Vazirani},
\newblock \bibinfo{title}{{A polynomial-time classical algorithm for noisy random circuit sampling}},
\newblock in: \bibinfo{booktitle}{{Proc. of the 55th Annual ACM Symposium on Theory of Computing}}, \bibinfo{year}{2023}, pp. \bibinfo{pages}{945--957}.
\bibitem[{Markov and Shi(2008)}]{markov2008simulating}
\bibinfo{author}{I.~L. Markov}, \bibinfo{author}{Y.~Shi},
\newblock \bibinfo{title}{{Simulating quantum computation by contracting tensor networks}},
\newblock \bibinfo{journal}{SIAM Journal on Computing} \bibinfo{volume}{38} (\bibinfo{year}{2008}) \bibinfo{pages}{963--981}.
\bibitem[{Morvan et~al.(2023)Morvan, Villalonga, Mi, Mandrà, Bengtsson, and et~al.}]{morvan2023phase}
\bibinfo{author}{A.~Morvan}, \bibinfo{author}{B.~Villalonga}, \bibinfo{author}{X.~Mi}, \bibinfo{author}{S.~Mandrà}, \bibinfo{author}{A.~Bengtsson}, \bibinfo{author}{et~al.}, \bibinfo{title}{{Phase transition in Random Circuit Sampling}}, \bibinfo{year}{2023}. \href{http://arxiv.org/abs/2304.11119}{{\tt arXiv:2304.11119}}.
\bibitem[{Kechedzhi et~al.(2024)Kechedzhi, Isakov, Mandr{\`a}, Villalonga, Mi, Boixo, and Smelyanskiy}]{kechedzhi2024effective}
\bibinfo{author}{K.~Kechedzhi}, \bibinfo{author}{S.~Isakov}, \bibinfo{author}{S.~Mandr{\`a}}, \bibinfo{author}{B.~Villalonga}, \bibinfo{author}{X.~Mi}, \bibinfo{author}{S.~Boixo}, \bibinfo{author}{V.~Smelyanskiy},
\newblock \bibinfo{title}{{Effective quantum volume, fidelity and computational cost of noisy quantum processing experiments}},
\newblock \bibinfo{journal}{FGCS} \bibinfo{volume}{153} (\bibinfo{year}{2024}) \bibinfo{pages}{431--441}.
\bibitem[{Bertini et~al.(2019)Bertini, Kos, and Prosen}]{PhysRevX.9.021033}
\bibinfo{author}{B.~Bertini}, \bibinfo{author}{P.~Kos}, \bibinfo{author}{T.~c.~v. Prosen},
\newblock \bibinfo{title}{{Entanglement Spreading in a Minimal Model of Maximal Many-Body Quantum Chaos}},
\newblock \bibinfo{journal}{Phys. Rev. X} \bibinfo{volume}{9} (\bibinfo{year}{2019}) \bibinfo{pages}{021033}. \URLprefix \url{https://link.aps.org/doi/10.1103/PhysRevX.9.021033}. \DOIprefix\doi{10.1103/PhysRevX.9.021033}.
\bibitem[{Gray and Kourtis(2021)}]{gray2021hyper}
\bibinfo{author}{J.~Gray}, \bibinfo{author}{S.~Kourtis},
\newblock \bibinfo{title}{{Hyper-optimized tensor network contraction}},
\newblock \bibinfo{journal}{Quantum} \bibinfo{volume}{5} (\bibinfo{year}{2021}) \bibinfo{pages}{410}.
\bibitem[{Pan et~al.(2022)Pan, Chen, and Zhang}]{pan2022solving}
\bibinfo{author}{F.~Pan}, \bibinfo{author}{K.~Chen}, \bibinfo{author}{P.~Zhang},
\newblock \bibinfo{title}{{Solving the Sampling Problem of the Sycamore Quantum Circuits}},
\newblock \bibinfo{journal}{PRL} \bibinfo{volume}{129} (\bibinfo{year}{2022}) \bibinfo{pages}{090502}.
\bibitem[{Arnborg et~al.(1987)Arnborg, Corneil, and Proskurowski}]{arnborg1987complexity}
\bibinfo{author}{S.~Arnborg}, \bibinfo{author}{D.~G. Corneil}, \bibinfo{author}{A.~Proskurowski},
\newblock \bibinfo{title}{{Complexity of finding embeddings in ak-tree}},
\newblock \bibinfo{journal}{SIAM Journal on Algebraic Discrete Methods} \bibinfo{volume}{8} (\bibinfo{year}{1987}) \bibinfo{pages}{277--284}.
\bibitem[{Verstraete et~al.(2008)Verstraete, Murg, and Cirac}]{verstraete2008matrix}
\bibinfo{author}{F.~Verstraete}, \bibinfo{author}{V.~Murg}, \bibinfo{author}{J.~I. Cirac},
\newblock \bibinfo{title}{{Matrix product states, projected entangled pair states, and variational renormalization group methods for quantum spin systems}},
\newblock \bibinfo{journal}{Advances in physics} \bibinfo{volume}{57} (\bibinfo{year}{2008}) \bibinfo{pages}{143--224}. \DOIprefix\doi{https://doi.org/10.1080/14789940801912366}.
\bibitem[{Vidal(2008)}]{vidal2008class}
\bibinfo{author}{G.~Vidal},
\newblock \bibinfo{title}{{Class of quantum many-body states that can be efficiently simulated}},
\newblock \bibinfo{journal}{PRL} \bibinfo{volume}{101} (\bibinfo{year}{2008}) \bibinfo{pages}{110501}. \DOIprefix\doi{10.1103/PhysRevLett.101.110501}.
\bibitem[{Marshall and Kafri(2023)}]{Marshall2023b}
\bibinfo{author}{J.~Marshall}, \bibinfo{author}{D.~Kafri}, \bibinfo{title}{Incoherent {{Approximation}} of {{Leakage}} in {{Quantum Error Correction}}}, \bibinfo{year}{2023}. \URLprefix \url{http://arxiv.org/abs/2312.10277}. \DOIprefix\doi{10.48550/arXiv.2312.10277}. \href{http://arxiv.org/abs/2312.10277}{{\tt arXiv:2312.10277}}.
\bibitem[{Renema et~al.(2018)Renema, Menssen, Clements, Triginer, Kolthammer, and Walmsley}]{Renema2018}
\bibinfo{author}{J.~J. Renema}, \bibinfo{author}{A.~Menssen}, \bibinfo{author}{W.~R. Clements}, \bibinfo{author}{G.~Triginer}, \bibinfo{author}{W.~S. Kolthammer}, \bibinfo{author}{I.~A. Walmsley},
\newblock \bibinfo{title}{{Efficient Classical Algorithm for Boson Sampling with Partially Distinguishable Photons}},
\newblock \bibinfo{journal}{Phys. Rev. Lett.} \bibinfo{volume}{120} (\bibinfo{year}{2018}) \bibinfo{pages}{220502}. \URLprefix \url{https://link.aps.org/doi/10.1103/PhysRevLett.120.220502}. \DOIprefix\doi{10.1103/PhysRevLett.120.220502}.
\bibitem[{Sparrow(2017)}]{Sparrow2017}
\bibinfo{author}{C.~Sparrow},
\newblock \bibinfo{title}{{Quantum Interference in Universal Linear Optical Devices for Quantum Computation and Simulation}}  (\bibinfo{year}{2017}). \URLprefix \url{http://spiral.imperial.ac.uk/handle/10044/1/67638}. \DOIprefix\doi{10.25560/67638}.
\bibitem[{Marshall(2022)}]{Marshall2022b}
\bibinfo{author}{J.~Marshall},
\newblock \bibinfo{title}{{Distillation of Indistinguishable Photons}},
\newblock \bibinfo{journal}{Phys. Rev. Lett.} \bibinfo{volume}{129} (\bibinfo{year}{2022}) \bibinfo{pages}{213601}. \URLprefix \url{https://link.aps.org/doi/10.1103/PhysRevLett.129.213601}. \DOIprefix\doi{10.1103/PhysRevLett.129.213601}.
\bibitem[{Saied et~al.(2024)Saied, Marshall, Anand, and Rieffel}]{saied2024general}
\bibinfo{author}{J.~Saied}, \bibinfo{author}{J.~Marshall}, \bibinfo{author}{N.~Anand}, \bibinfo{author}{E.~G. Rieffel},
\newblock \bibinfo{title}{{General protocols for the efficient distillation of indistinguishable photons}},
\newblock \bibinfo{journal}{arXiv preprint arXiv:2404.14217}  (\bibinfo{year}{2024}). \DOIprefix\doi{10.48550/arXiv.2404.14217}.
\bibitem[{Somhorst et~al.(2024)Somhorst, Sau{\"e}r, van~den Hoven, and Renema}]{somhorst2024photon}
\bibinfo{author}{F.~Somhorst}, \bibinfo{author}{B.~K. Sau{\"e}r}, \bibinfo{author}{S.~van~den Hoven}, \bibinfo{author}{J.~J. Renema},
\newblock \bibinfo{title}{{Photon distillation schemes with reduced resource costs based on multiphoton Fourier interference}},
\newblock \bibinfo{journal}{arXiv preprint arXiv:2404.14262}  (\bibinfo{year}{2024}). \DOIprefix\doi{10.48550/arXiv.2404.14262}.
\bibitem[{{Rahimi-Keshari} et~al.(2016){Rahimi-Keshari}, Ralph, and Caves}]{Rahimi-Keshari2016}
\bibinfo{author}{S.~{Rahimi-Keshari}}, \bibinfo{author}{T.~C. Ralph}, \bibinfo{author}{C.~M. Caves},
\newblock \bibinfo{title}{Sufficient {{Conditions}} for {{Efficient Classical Simulation}} of {{Quantum Optics}}},
\newblock \bibinfo{journal}{Physical Review X} \bibinfo{volume}{6} (\bibinfo{year}{2016}) \bibinfo{pages}{021039}. \URLprefix \url{https://link.aps.org/doi/10.1103/PhysRevX.6.021039}. \DOIprefix\doi{10.1103/PhysRevX.6.021039}.
\bibitem[{Marshall and Anand(2023)}]{Marshall2023a}
\bibinfo{author}{J.~Marshall}, \bibinfo{author}{N.~Anand},
\newblock \bibinfo{title}{{Simulation of Quantum Optics by Coherent State Decomposition}},
\newblock \bibinfo{journal}{Optica Quantum} \bibinfo{volume}{1} (\bibinfo{year}{2023}) \bibinfo{pages}{78--93}. \URLprefix \url{https://opg.optica.org/opticaq/abstract.cfm?uri=opticaq-1-2-78}. \DOIprefix\doi{10.1364/OPTICAQ.504311}.
\bibitem[{Mele(2024)}]{mele2023introduction}
\bibinfo{author}{A.~A. Mele},
\newblock \bibinfo{title}{Introduction to {H}aar {M}easure {T}ools in {Q}uantum {I}nformation: {A} {B}eginner's {T}utorial},
\newblock \bibinfo{journal}{{Quantum}} \bibinfo{volume}{8} (\bibinfo{year}{2024}) \bibinfo{pages}{1340}. \URLprefix \url{https://doi.org/10.22331/q-2024-05-08-1340}. \DOIprefix\doi{10.22331/q-2024-05-08-1340}.
\bibitem[{Dankert et~al.(2009)Dankert, Cleve, Emerson, and Livine}]{dankert2009exact}
\bibinfo{author}{C.~Dankert}, \bibinfo{author}{R.~Cleve}, \bibinfo{author}{J.~Emerson}, \bibinfo{author}{E.~Livine},
\newblock \bibinfo{title}{{Exact and approximate unitary 2-designs and their application to fidelity estimation}},
\newblock \bibinfo{journal}{Phys. Rev. A} \bibinfo{volume}{80} (\bibinfo{year}{2009}) \bibinfo{pages}{012304}. \URLprefix \url{https://link.aps.org/doi/10.1103/PhysRevA.80.012304}. \DOIprefix\doi{10.1103/PhysRevA.80.012304}.
\bibitem[{Huang et~al.(2020)Huang, Kueng, and Preskill}]{huang_predicting_2020}
\bibinfo{author}{H.-Y. Huang}, \bibinfo{author}{R.~Kueng}, \bibinfo{author}{J.~Preskill},
\newblock \bibinfo{title}{{Predicting many properties of a quantum system from very few measurements}},
\newblock \bibinfo{journal}{Nature Physics} \bibinfo{volume}{16} (\bibinfo{year}{2020}) \bibinfo{pages}{1050–1057}. \URLprefix \url{http://dx.doi.org/10.1038/s41567-020-0932-7}. \DOIprefix\doi{10.1038/s41567-020-0932-7}.
\bibitem[{Saied et~al.(2024)Saied, Marshall, Anand, Grabbe, and Rieffel}]{Saied2024_spie}
\bibinfo{author}{J.~Saied}, \bibinfo{author}{J.~Marshall}, \bibinfo{author}{N.~Anand}, \bibinfo{author}{S.~Grabbe}, \bibinfo{author}{E.~G. Rieffel},
\newblock \bibinfo{title}{{Advancing quantum networking: some tools and protocols for ideal and noisy photonic systems}},
\newblock in: \bibinfo{editor}{P.~R. Hemmer}, \bibinfo{editor}{A.~L. Migdall} (Eds.), \bibinfo{booktitle}{{Quantum Computing, Communication, and Simulation IV}}, \bibinfo{publisher}{SPIE}, \bibinfo{year}{2024}. \URLprefix \url{http://dx.doi.org/10.1117/12.3023843}. \DOIprefix\doi{10.1117/12.3023843}.
\bibitem[{Sawaya et~al.(2023)Sawaya, Marti-Dafcik, Ho, Tabor, {Bernal Neira}, Magann, Premaratne, Dubey, Matsuura, Bishop et~al.}]{sawaya2023hamlib}
\bibinfo{author}{N.~P. Sawaya}, \bibinfo{author}{D.~Marti-Dafcik}, \bibinfo{author}{Y.~Ho}, \bibinfo{author}{D.~P. Tabor}, \bibinfo{author}{D.~E. {Bernal Neira}}, \bibinfo{author}{A.~B. Magann}, \bibinfo{author}{S.~Premaratne}, \bibinfo{author}{P.~Dubey}, \bibinfo{author}{A.~Matsuura}, \bibinfo{author}{N.~Bishop}, et~al.,
\newblock \bibinfo{title}{{HamLib: A library of Hamiltonians for benchmarking quantum algorithms and hardware}},
\newblock in: \bibinfo{booktitle}{{2023 IEEE International Conference on Quantum Computing and Engineering (QCE)}}, volume~\bibinfo{volume}{2}, \bibinfo{organization}{IEEE}, \bibinfo{year}{2023}, pp. \bibinfo{pages}{389--390}. \DOIprefix\doi{10.1109/QCE57702.2023.10296}.
\bibitem[{Lubinski et~al.(2023)Lubinski, Coffrin, McGeoch, Sathe, Apanavicius, and {Bernal Neira}}]{lubinski2023optimization}
\bibinfo{author}{T.~Lubinski}, \bibinfo{author}{C.~Coffrin}, \bibinfo{author}{C.~McGeoch}, \bibinfo{author}{P.~Sathe}, \bibinfo{author}{J.~Apanavicius}, \bibinfo{author}{D.~E. {Bernal Neira}}, \bibinfo{title}{{Optimization Applications as Quantum Performance Benchmarks}}, \bibinfo{year}{2023}. \href{http://arxiv.org/abs/2302.02278}{{\tt arXiv:2302.02278}}.
\bibitem[{{Bernal Neira} et~al.(2024)}]{bernal2024benchmarking_short}
\bibinfo{author}{D.~E. {Bernal Neira}}, et~al.,
\newblock \bibinfo{title}{{Benchmarking the {O}peration of {Q}uantum {H}euristics and {I}sing {M}achines: {S}coring {P}arameter {S}etting {S}trategies on {O}ptimization {A}pplications}},
\newblock \bibinfo{journal}{arXiv:2402.10255}  (\bibinfo{year}{2024}).
\bibitem[{Bernal~Neira et~al.(2023)Bernal~Neira, Brown, Sathe, and Venturelli}]{E_Bernal_Neira_Stochastic_Benchmark_toolkit_2023}
\bibinfo{author}{D.~E. Bernal~Neira}, \bibinfo{author}{R.~Brown}, \bibinfo{author}{P.~Sathe}, \bibinfo{author}{D.~Venturelli}, \bibinfo{title}{{Stochastic Benchmark: toolkit for performance evaluation and parameter tuning of stochastic parameterized stochastic optimization solvers}}, \bibinfo{year}{2023}. \URLprefix \url{https://github.com/usra-riacs/stochastic-benchmark}.
\bibitem[{Wudarski et~al.(2020)Wudarski, Marshall, Petukhov, and Rieffel}]{Wudarski2020}
\bibinfo{author}{F.~Wudarski}, \bibinfo{author}{J.~Marshall}, \bibinfo{author}{A.~Petukhov}, \bibinfo{author}{E.~Rieffel},
\newblock \bibinfo{title}{Augmented fidelities for single-qubit gates},
\newblock \bibinfo{journal}{Phys. Rev. A} \bibinfo{volume}{102} (\bibinfo{year}{2020}) \bibinfo{pages}{052612}. \URLprefix \url{https://link.aps.org/doi/10.1103/PhysRevA.102.052612}. \DOIprefix\doi{10.1103/PhysRevA.102.052612}.
\bibitem[{Sud et~al.(2022)Sud, Marshall, Wang, Rieffel, and Wudarski}]{Sud2022}
\bibinfo{author}{J.~Sud}, \bibinfo{author}{J.~Marshall}, \bibinfo{author}{Z.~Wang}, \bibinfo{author}{E.~Rieffel}, \bibinfo{author}{F.~A. Wudarski},
\newblock \bibinfo{title}{Dual-map framework for noise characterization of quantum computers},
\newblock \bibinfo{journal}{Phys. Rev. A} \bibinfo{volume}{106} (\bibinfo{year}{2022}) \bibinfo{pages}{012606}. \URLprefix \url{https://link.aps.org/doi/10.1103/PhysRevA.106.012606}. \DOIprefix\doi{10.1103/PhysRevA.106.012606}.
\bibitem[{Marshall et~al.(2020)Marshall, Wudarski, Hadfield, and Hogg}]{Marshall2020b}
\bibinfo{author}{J.~Marshall}, \bibinfo{author}{F.~Wudarski}, \bibinfo{author}{S.~Hadfield}, \bibinfo{author}{T.~Hogg},
\newblock \bibinfo{title}{Characterizing local noise in {QAOA} circuits},
\newblock \bibinfo{journal}{IOP SciNotes} \bibinfo{volume}{1} (\bibinfo{year}{2020}) \bibinfo{pages}{025208}. \URLprefix \url{https://dx.doi.org/10.1088/2633-1357/abb0d7}. \DOIprefix\doi{10.1088/2633-1357/abb0d7}.
\bibitem[{Wudarski et~al.(2023{\natexlab{a}})Wudarski, Zhang, and Dykman}]{Wudarski2023a}
\bibinfo{author}{F.~Wudarski}, \bibinfo{author}{Y.~Zhang}, \bibinfo{author}{M.~I. Dykman},
\newblock \bibinfo{title}{{Nonergodic Measurements of Qubit Frequency Noise}},
\newblock \bibinfo{journal}{Phys. Rev. Lett.} \bibinfo{volume}{131} (\bibinfo{year}{2023}{\natexlab{a}}) \bibinfo{pages}{230201}. \URLprefix \url{https://link.aps.org/doi/10.1103/PhysRevLett.131.230201}. \DOIprefix\doi{10.1103/PhysRevLett.131.230201}.
\bibitem[{Wudarski et~al.(2023{\natexlab{b}})Wudarski, Zhang, Korotkov, Petukhov, and Dykman}]{Wudarski2023b}
\bibinfo{author}{F.~Wudarski}, \bibinfo{author}{Y.~Zhang}, \bibinfo{author}{A.~N. Korotkov}, \bibinfo{author}{A.~Petukhov}, \bibinfo{author}{M.~Dykman},
\newblock \bibinfo{title}{{Characterizing Low-Frequency Qubit Noise}},
\newblock \bibinfo{journal}{Phys. Rev. Appl.} \bibinfo{volume}{19} (\bibinfo{year}{2023}{\natexlab{b}}) \bibinfo{pages}{064066}. \URLprefix \url{https://link.aps.org/doi/10.1103/PhysRevApplied.19.064066}. \DOIprefix\doi{10.1103/PhysRevApplied.19.064066}.
\bibitem[{McCourt et~al.(2023)McCourt, Neill, Lee, Quintana, Chen, Kelly, Marshall, Smelyanskiy, Dykman, Korotkov, Chuang, and Petukhov}]{McCourt2023}
\bibinfo{author}{T.~McCourt}, \bibinfo{author}{C.~Neill}, \bibinfo{author}{K.~Lee}, \bibinfo{author}{C.~Quintana}, \bibinfo{author}{Y.~Chen}, \bibinfo{author}{J.~Kelly}, \bibinfo{author}{J.~Marshall}, \bibinfo{author}{V.~N. Smelyanskiy}, \bibinfo{author}{M.~I. Dykman}, \bibinfo{author}{A.~Korotkov}, \bibinfo{author}{I.~L. Chuang}, \bibinfo{author}{A.~G. Petukhov},
\newblock \bibinfo{title}{Learning noise via dynamical decoupling of entangled qubits},
\newblock \bibinfo{journal}{Phys. Rev. A} \bibinfo{volume}{107} (\bibinfo{year}{2023}) \bibinfo{pages}{052610}. \URLprefix \url{https://link.aps.org/doi/10.1103/PhysRevA.107.052610}. \DOIprefix\doi{10.1103/PhysRevA.107.052610}.
\bibitem[{Evert et~al.(2024)Evert, Izquierdo, Sud, Hu, Grabbe, Rieffel, Reagor, and Wang}]{evert2024syncopated}
\bibinfo{author}{B.~Evert}, \bibinfo{author}{Z.~G. Izquierdo}, \bibinfo{author}{J.~Sud}, \bibinfo{author}{H.-Y. Hu}, \bibinfo{author}{S.~Grabbe}, \bibinfo{author}{E.~G. Rieffel}, \bibinfo{author}{M.~J. Reagor}, \bibinfo{author}{Z.~Wang}, \bibinfo{title}{{Syncopated Dynamical Decoupling for Suppressing Crosstalk in Quantum Circuits}}, \bibinfo{year}{2024}. \href{http://arxiv.org/abs/2403.07836}{{\tt arXiv:2403.07836}}.
\bibitem[{Acharya et~al.(2023)Acharya, Aleiner, Allen, Andersen, Ansmann, and et~al.}]{Acharya2023}
\bibinfo{author}{R.~Acharya}, \bibinfo{author}{I.~Aleiner}, \bibinfo{author}{R.~Allen}, \bibinfo{author}{T.~I. Andersen}, \bibinfo{author}{M.~Ansmann}, \bibinfo{author}{et~al.},
\newblock \bibinfo{title}{{Suppressing quantum errors by scaling a surface code logical qubit}},
\newblock \bibinfo{journal}{Nature} \bibinfo{volume}{614} (\bibinfo{year}{2023}) \bibinfo{pages}{676--681}. \URLprefix \url{https://doi.org/10.1038/s41586-022-05434-1}. \DOIprefix\doi{10.1038/s41586-022-05434-1}.
\bibitem[{Alam et~al.(2024)}]{Alam2024Dynamical}
\bibinfo{author}{S.~Alam}, et~al.,
\newblock \bibinfo{title}{{Dynamical Logical Qubits in the Bacon-Shor Code}},
\newblock \bibinfo{journal}{arXiv}  (\bibinfo{year}{2024}).
\bibitem[{Hu et~al.(2022)Hu, LaRose, You, Rieffel, and Wang}]{hu2022logical}
\bibinfo{author}{H.-Y. Hu}, \bibinfo{author}{R.~LaRose}, \bibinfo{author}{Y.-Z. You}, \bibinfo{author}{E.~Rieffel}, \bibinfo{author}{Z.~Wang}, \bibinfo{title}{{Logical shadow tomography: Efficient estimation of error-mitigated observables}}, \bibinfo{year}{2022}. \href{http://arxiv.org/abs/2203.07263}{{\tt arXiv:2203.07263}}.
\bibitem[{Akhtar et~al.(2024)Akhtar, Anand, Marshall, and You}]{akhtar2024dualunitary}
\bibinfo{author}{A.~A. Akhtar}, \bibinfo{author}{N.~Anand}, \bibinfo{author}{J.~Marshall}, \bibinfo{author}{Y.-Z. You}, \bibinfo{title}{{Dual-Unitary Classical Shadow Tomography}}, \bibinfo{year}{2024}. \href{http://arxiv.org/abs/2404.01068}{{\tt arXiv:2404.01068}}.
\bibitem[{Suri et~al.(2024)Suri, Saied, and Venturelli}]{Suri2024}
\bibinfo{author}{N.~Suri}, \bibinfo{author}{J.~Saied}, \bibinfo{author}{D.~Venturelli}, \bibinfo{title}{{Uniformly Decaying Subspaces for Error Mitigated Quantum Computation}}, \bibinfo{year}{2024}. \href{http://arxiv.org/abs/2403.00163}{{\tt arXiv:2403.00163}}.
\bibitem[{Fei et~al.(2023{\natexlab{a}})Fei, Brady, Larson, Leyffer, and Shen}]{Fei2023binarycontrolpulse}
\bibinfo{author}{X.~Fei}, \bibinfo{author}{L.~T. Brady}, \bibinfo{author}{J.~Larson}, \bibinfo{author}{S.~Leyffer}, \bibinfo{author}{S.~Shen},
\newblock \bibinfo{title}{Binary {C}ontrol {P}ulse {O}ptimization for {Q}uantum {S}ystems},
\newblock \bibinfo{journal}{{Quantum}} \bibinfo{volume}{7} (\bibinfo{year}{2023}{\natexlab{a}}) \bibinfo{pages}{892}. \URLprefix \url{https://doi.org/10.22331/q-2023-01-04-892}. \DOIprefix\doi{10.22331/q-2023-01-04-892}.
\bibitem[{Fei et~al.(2023{\natexlab{b}})Fei, Brady, Larson, Leyffer, and Shen}]{fei2023switching}
\bibinfo{author}{X.~Fei}, \bibinfo{author}{L.~T. Brady}, \bibinfo{author}{J.~Larson}, \bibinfo{author}{S.~Leyffer}, \bibinfo{author}{S.~Shen}, \bibinfo{title}{{Switching Time Optimization for Binary Quantum Optimal Control}}, \bibinfo{year}{2023}{\natexlab{b}}. \href{http://arxiv.org/abs/2308.03132}{{\tt arXiv:2308.03132}}.
\bibitem[{Fei et~al.(2024)Fei, Brady, Larson, Leyffer, and Shen}]{fei2024binary}
\bibinfo{author}{X.~Fei}, \bibinfo{author}{L.~T. Brady}, \bibinfo{author}{J.~Larson}, \bibinfo{author}{S.~Leyffer}, \bibinfo{author}{S.~Shen}, \bibinfo{title}{{Binary Quantum Control Optimization with Uncertain Hamiltonians}}, \bibinfo{year}{2024}. \href{http://arxiv.org/abs/2401.10120}{{\tt arXiv:2401.10120}}.
\bibitem[{{\"O}zg{\"u}ler and Venturelli(2022)}]{ozguler2022numerical}
\bibinfo{author}{A.~B. {\"O}zg{\"u}ler}, \bibinfo{author}{D.~Venturelli},
\newblock \bibinfo{title}{{Numerical gate synthesis for quantum heuristics on bosonic quantum processors}},
\newblock \bibinfo{journal}{Frontiers in Physics}  (\bibinfo{year}{2022}) \bibinfo{pages}{724}.
\bibitem[{Alam et~al.(2022)Alam, Belomestnykh, Bornman, Cancelo, Chao, Checchin, Dinh, Grassellino, Gustafson, Harnik, McRae, Huang, Kapoor, Kim, Kowalkowski, Kramer, Krasnikova, Kumar, Kurkcuoglu, Lamm, Lyon, Milathianaki, Murthy, Mutus, Nekrashevich, Oh, Özgüler, Perdue, Reagor, Romanenko, Sauls, Stefanazzi, Tubman, Venturelli, Wang, You, van Zanten, Zhou, Zhu, and Zorzetti}]{alam2022quantum}
\bibinfo{author}{M.~S. Alam}, \bibinfo{author}{S.~Belomestnykh}, \bibinfo{author}{N.~Bornman}, \bibinfo{author}{G.~Cancelo}, \bibinfo{author}{Y.-C. Chao}, \bibinfo{author}{M.~Checchin}, \bibinfo{author}{V.~S. Dinh}, \bibinfo{author}{A.~Grassellino}, \bibinfo{author}{E.~J. Gustafson}, \bibinfo{author}{R.~Harnik}, \bibinfo{author}{C.~R.~H. McRae}, \bibinfo{author}{Z.~Huang}, \bibinfo{author}{K.~Kapoor}, \bibinfo{author}{T.~Kim}, \bibinfo{author}{J.~B. Kowalkowski}, \bibinfo{author}{M.~J. Kramer}, \bibinfo{author}{Y.~Krasnikova}, \bibinfo{author}{P.~Kumar}, \bibinfo{author}{D.~M. Kurkcuoglu}, \bibinfo{author}{H.~Lamm}, \bibinfo{author}{A.~L. Lyon}, \bibinfo{author}{D.~Milathianaki}, \bibinfo{author}{A.~Murthy}, \bibinfo{author}{J.~Mutus}, \bibinfo{author}{I.~Nekrashevich}, \bibinfo{author}{J.~Oh}, \bibinfo{author}{A.~B. Özgüler}, \bibinfo{author}{G.~N. Perdue}, \bibinfo{author}{M.~Reagor}, \bibinfo{author}{A.~Romanenko}, \bibinfo{author}{J.~A. Sauls}, \bibinfo{author}{L.~Stefanazzi}, \bibinfo{author}{N.~M.
  Tubman}, \bibinfo{author}{D.~Venturelli}, \bibinfo{author}{C.~Wang}, \bibinfo{author}{X.~You}, \bibinfo{author}{D.~M.~T. van Zanten}, \bibinfo{author}{L.~Zhou}, \bibinfo{author}{S.~Zhu}, \bibinfo{author}{S.~Zorzetti}, \bibinfo{title}{{Quantum computing hardware for HEP algorithms and sensing}}, \bibinfo{year}{2022}. \href{http://arxiv.org/abs/2204.08605}{{\tt arXiv:2204.08605}}.
\bibitem[{Xu et~al.(2022)Xu, {\"O}zg{\"u}ler, Di~Guglielmo, Tran, Perdue, Carloni, and Fahim}]{xu2022neural}
\bibinfo{author}{D.~Xu}, \bibinfo{author}{A.~B. {\"O}zg{\"u}ler}, \bibinfo{author}{G.~Di~Guglielmo}, \bibinfo{author}{N.~Tran}, \bibinfo{author}{G.~N. Perdue}, \bibinfo{author}{L.~Carloni}, \bibinfo{author}{F.~Fahim},
\newblock \bibinfo{title}{{Neural network accelerator for quantum control}},
\newblock in: \bibinfo{booktitle}{{2022 IEEE/ACM Third International Workshop on Quantum Computing Software (QCS)}}, \bibinfo{organization}{IEEE}, \bibinfo{year}{2022}, pp. \bibinfo{pages}{43--49}. \DOIprefix\doi{10.1109/QCS56647.2022.00010}.
\bibitem[{Kim et~al.(2023)Kim, Eddins, Anand, Wei, Van Den~Berg, Rosenblatt, Nayfeh, Wu, Zaletel, Temme et~al.}]{kim2023evidence}
\bibinfo{author}{Y.~Kim}, \bibinfo{author}{A.~Eddins}, \bibinfo{author}{S.~Anand}, \bibinfo{author}{K.~X. Wei}, \bibinfo{author}{E.~Van Den~Berg}, \bibinfo{author}{S.~Rosenblatt}, \bibinfo{author}{H.~Nayfeh}, \bibinfo{author}{Y.~Wu}, \bibinfo{author}{M.~Zaletel}, \bibinfo{author}{K.~Temme}, et~al.,
\newblock \bibinfo{title}{{Evidence for the utility of quantum computing before fault tolerance}},
\newblock \bibinfo{journal}{Nature} \bibinfo{volume}{618} (\bibinfo{year}{2023}) \bibinfo{pages}{500--505}. \DOIprefix\doi{https://doi.org/10.1038/s41586-023-06096-3}.
\bibitem[{Kim et~al.(2021)Kim, Mandr{\`a}, Venturelli, and Jamieson}]{kim2021physics}
\bibinfo{author}{M.~Kim}, \bibinfo{author}{S.~Mandr{\`a}}, \bibinfo{author}{D.~Venturelli}, \bibinfo{author}{K.~Jamieson},
\newblock \bibinfo{title}{{Physics-inspired heuristics for soft MIMO detection in 5G new radio and beyond}},
\newblock in: \bibinfo{booktitle}{{Proceedings of the 27th Annual International Conference on Mobile Computing and Networking}}, \bibinfo{year}{2021}, pp. \bibinfo{pages}{42--55}. \DOIprefix\doi{https://doi.org/10.1145/3447993.3448619}.
\bibitem[{Aramon et~al.(2019)Aramon, Rosenberg, Valiante, Miyazawa, Tamura, and Katzgraber}]{aramon2019physics}
\bibinfo{author}{M.~Aramon}, \bibinfo{author}{G.~Rosenberg}, \bibinfo{author}{E.~Valiante}, \bibinfo{author}{T.~Miyazawa}, \bibinfo{author}{H.~Tamura}, \bibinfo{author}{H.~G. Katzgraber},
\newblock \bibinfo{title}{{Physics-inspired optimization for quadratic unconstrained problems using a digital annealer}},
\newblock \bibinfo{journal}{Frontiers in Physics} \bibinfo{volume}{7} (\bibinfo{year}{2019}) \bibinfo{pages}{48}. \DOIprefix\doi{https://doi.org/10.3389/fphy.2019.00048}.
\bibitem[{Mertens et~al.(2006)Mertens, M{\'e}zard, and Zecchina}]{mertens2006threshold}
\bibinfo{author}{S.~Mertens}, \bibinfo{author}{M.~M{\'e}zard}, \bibinfo{author}{R.~Zecchina},
\newblock \bibinfo{title}{{Threshold values of random K-SAT from the cavity method}},
\newblock \bibinfo{journal}{Random Structures \& Algorithms} \bibinfo{volume}{28} (\bibinfo{year}{2006}) \bibinfo{pages}{340--373}. \DOIprefix\doi{https://doi.org/10.1002/rsa.20090}.
\bibitem[{M{\'e}zard et~al.(2002)M{\'e}zard, Parisi, and Zecchina}]{mezard2002analytic}
\bibinfo{author}{M.~M{\'e}zard}, \bibinfo{author}{G.~Parisi}, \bibinfo{author}{R.~Zecchina},
\newblock \bibinfo{title}{{Analytic and algorithmic solution of random satisfiability problems}},
\newblock \bibinfo{journal}{Science} \bibinfo{volume}{297} (\bibinfo{year}{2002}) \bibinfo{pages}{812--815}. \DOIprefix\doi{10.1126/science.10732}.
\bibitem[{M{\'e}zard and Zecchina(2002)}]{mezard2002random}
\bibinfo{author}{M.~M{\'e}zard}, \bibinfo{author}{R.~Zecchina},
\newblock \bibinfo{title}{{Random k-satisfiability problem: From an analytic solution to an efficient algorithm}},
\newblock \bibinfo{journal}{Physical Review E} \bibinfo{volume}{66} (\bibinfo{year}{2002}) \bibinfo{pages}{056126}. \DOIprefix\doi{10.1103/PhysRevE.66.056126}.
\bibitem[{Kirkpatrick and Selman(1994)}]{kirkpatrick1994critical}
\bibinfo{author}{S.~Kirkpatrick}, \bibinfo{author}{B.~Selman},
\newblock \bibinfo{title}{{Critical behavior in the satisfiability of random boolean expressions}},
\newblock \bibinfo{journal}{Science} \bibinfo{volume}{264} (\bibinfo{year}{1994}) \bibinfo{pages}{1297--1301}. \DOIprefix\doi{10.1126/science.264.5163.12}.
\bibitem[{Qu and Potkonjak(1999)}]{qu1999hiding}
\bibinfo{author}{G.~Qu}, \bibinfo{author}{M.~Potkonjak},
\newblock \bibinfo{title}{{Hiding signatures in graph coloring solutions}},
\newblock in: \bibinfo{booktitle}{{International Workshop on Information Hiding}}, \bibinfo{organization}{Springer}, \bibinfo{year}{1999}, pp. \bibinfo{pages}{348--367}.
\bibitem[{Barthel et~al.(2002)Barthel, Hartmann, Leone, Ricci-Tersenghi, Weigt, and Zecchina}]{barthel2002hiding}
\bibinfo{author}{W.~Barthel}, \bibinfo{author}{A.~K. Hartmann}, \bibinfo{author}{M.~Leone}, \bibinfo{author}{F.~Ricci-Tersenghi}, \bibinfo{author}{M.~Weigt}, \bibinfo{author}{R.~Zecchina},
\newblock \bibinfo{title}{{Hiding solutions in random satisfiability problems: A statistical mechanics approach}},
\newblock \bibinfo{journal}{Physical review letters} \bibinfo{volume}{88} (\bibinfo{year}{2002}) \bibinfo{pages}{188701}. \DOIprefix\doi{10.1103/PhysRevLett.88.188701}.
\bibitem[{Krzakala and Zdeborov{\'a}(2009)}]{krzakala2009hiding}
\bibinfo{author}{F.~Krzakala}, \bibinfo{author}{L.~Zdeborov{\'a}},
\newblock \bibinfo{title}{{Hiding quiet solutions in random constraint satisfaction problems}},
\newblock \bibinfo{journal}{Physical review letters} \bibinfo{volume}{102} (\bibinfo{year}{2009}) \bibinfo{pages}{238701}. \DOIprefix\doi{10.1103/PhysRevLett.102.238701}.
\bibitem[{Zdeborov{\'a} and Krzakala(2011)}]{zdeborova2011quiet}
\bibinfo{author}{L.~Zdeborov{\'a}}, \bibinfo{author}{F.~Krzakala},
\newblock \bibinfo{title}{{Quiet planting in the locked constraint satisfaction problems}},
\newblock \bibinfo{journal}{SIAM Journal on Discrete Mathematics} \bibinfo{volume}{25} (\bibinfo{year}{2011}) \bibinfo{pages}{750--770}. \DOIprefix\doi{https://doi.org/10.1137/090750755}.
\bibitem[{Krzakala et~al.(2012)Krzakala, M{\'e}zard, and Zdeborov{\'a}}]{krzakala2012reweighted}
\bibinfo{author}{F.~Krzakala}, \bibinfo{author}{M.~M{\'e}zard}, \bibinfo{author}{L.~Zdeborov{\'a}},
\newblock \bibinfo{title}{{Reweighted belief propagation and quiet planting for random k-sat}},
\newblock \bibinfo{journal}{Journal on Satisfiability, Boolean Modeling and Computation} \bibinfo{volume}{8} (\bibinfo{year}{2012}) \bibinfo{pages}{149--171}.
\bibitem[{Sicuro and Zdeborov{\'a}(2021)}]{sicuro2021planted}
\bibinfo{author}{G.~Sicuro}, \bibinfo{author}{L.~Zdeborov{\'a}},
\newblock \bibinfo{title}{{The planted k-factor problem}},
\newblock \bibinfo{journal}{Journal of Physics A: Mathematical and Theoretical} \bibinfo{volume}{54} (\bibinfo{year}{2021}) \bibinfo{pages}{175002}. \DOIprefix\doi{10.1088/1751-8121/abee9d}.
\bibitem[{Hen et~al.(2015)Hen, Job, Albash, R\o{}nnow, Troyer, and Lidar}]{Hen2015}
\bibinfo{author}{I.~Hen}, \bibinfo{author}{J.~Job}, \bibinfo{author}{T.~Albash}, \bibinfo{author}{T.~F. R\o{}nnow}, \bibinfo{author}{M.~Troyer}, \bibinfo{author}{D.~A. Lidar},
\newblock \bibinfo{title}{Probing for quantum speedup in spin-glass problems with planted solutions},
\newblock \bibinfo{journal}{Phys. Rev. A} \bibinfo{volume}{92} (\bibinfo{year}{2015}) \bibinfo{pages}{042325}. \URLprefix \url{https://link.aps.org/doi/10.1103/PhysRevA.92.042325}. \DOIprefix\doi{10.1103/PhysRevA.92.042325}.
\bibitem[{Marshall et~al.(2016)Marshall, Martin-Mayor, and Hen}]{Marshall2016}
\bibinfo{author}{J.~Marshall}, \bibinfo{author}{V.~Martin-Mayor}, \bibinfo{author}{I.~Hen},
\newblock \bibinfo{title}{Practical engineering of hard spin-glass instances},
\newblock \bibinfo{journal}{Phys. Rev. A} \bibinfo{volume}{94} (\bibinfo{year}{2016}) \bibinfo{pages}{012320}. \URLprefix \url{https://link.aps.org/doi/10.1103/PhysRevA.94.012320}. \DOIprefix\doi{10.1103/PhysRevA.94.012320}.
\bibitem[{Mandrà et~al.(2023)Mandrà, Mossi, and Rieffel}]{mandrà2023generating}
\bibinfo{author}{S.~Mandrà}, \bibinfo{author}{G.~Mossi}, \bibinfo{author}{E.~G. Rieffel}, \bibinfo{title}{{Generating Hard Ising Instances With Planted Solutions Using Post-Quantum Cryptographic Protocols}}, \bibinfo{year}{2023}. \href{http://arxiv.org/abs/2308.09704}{{\tt arXiv:2308.09704}}.
\bibitem[{Hen et~al.(2014)Hen, Rieffel, Do, and Venturelli}]{hen2014phase}
\bibinfo{author}{I.~Hen}, \bibinfo{author}{E.~G. Rieffel}, \bibinfo{author}{M.~Do}, \bibinfo{author}{D.~Venturelli},
\newblock \bibinfo{title}{Phase {T}ransitions in {P}lanning {P}roblems: {D}esign and {A}nalysis of {P}arameterized {F}amilies of {H}ard {P}lanning {P}roblems},
\newblock in: \bibinfo{booktitle}{{28th AIAA Conference on Artificial Intelligence}}, \bibinfo{number}{ARC-E-DAA-TN13195}, \bibinfo{year}{2014}.
\bibitem[{{McEliece}(1978)}]{McEliece}
\bibinfo{author}{R.~J. {McEliece}},
\newblock \bibinfo{title}{{A Public-Key Cryptosystem Based On Algebraic Coding Theory}},
\newblock \bibinfo{journal}{Deep Space Network Progress Report} \bibinfo{volume}{44} (\bibinfo{year}{1978}) \bibinfo{pages}{114--116}. \URLprefix \url{https://api.semanticscholar.org/CorpusID:56502909}.
\bibitem[{Mezard and Montanari(2009)}]{MezardMontanariBook}
\bibinfo{author}{M.~Mezard}, \bibinfo{author}{A.~Montanari}, \bibinfo{title}{{Information, Physics, and Computation}}, \bibinfo{publisher}{Oxford University Press, Inc.}, \bibinfo{address}{USA}, \bibinfo{year}{2009}. \DOIprefix\doi{https://doi.org/10.1093/acprof:oso/9780198570837.001.0001}.
\bibitem[{Perera et~al.(2021)Perera, Akpabio, Hamze, Mandra, Rose, Aramon, and Katzgraber}]{perera2021chook}
\bibinfo{author}{D.~Perera}, \bibinfo{author}{I.~Akpabio}, \bibinfo{author}{F.~Hamze}, \bibinfo{author}{S.~Mandra}, \bibinfo{author}{N.~Rose}, \bibinfo{author}{M.~Aramon}, \bibinfo{author}{H.~G. Katzgraber}, \bibinfo{title}{{Chook -- A comprehensive suite for generating binary optimization problems with planted solutions}}, \bibinfo{year}{2021}. \href{http://arxiv.org/abs/2005.14344}{{\tt arXiv:2005.14344}}.
\bibitem[{Hamze et~al.(2020)Hamze, Raymond, Pattison, Biswas, and Katzgraber}]{hamze2020wishart}
\bibinfo{author}{F.~Hamze}, \bibinfo{author}{J.~Raymond}, \bibinfo{author}{C.~A. Pattison}, \bibinfo{author}{K.~Biswas}, \bibinfo{author}{H.~G. Katzgraber},
\newblock \bibinfo{title}{{Wishart planted ensemble: A tunably rugged pairwise Ising model with a first-order phase transition}},
\newblock \bibinfo{journal}{Physical Review E} \bibinfo{volume}{101} (\bibinfo{year}{2020}) \bibinfo{pages}{052102}. \DOIprefix\doi{10.1103/PhysRevE.101.052102}.
\bibitem[{Hen(2019)}]{hen2019equation}
\bibinfo{author}{I.~Hen},
\newblock \bibinfo{title}{{Equation planting: a tool for benchmarking Ising machines}},
\newblock \bibinfo{journal}{Physical Review Applied} \bibinfo{volume}{12} (\bibinfo{year}{2019}) \bibinfo{pages}{011003}. \DOIprefix\doi{10.1103/PhysRevApplied.12.011003}.
\bibitem[{Claes et~al.(2021)Claes, Rieffel, and Wang}]{Claes2021}
\bibinfo{author}{J.~Claes}, \bibinfo{author}{E.~Rieffel}, \bibinfo{author}{Z.~Wang},
\newblock \bibinfo{title}{{Character Randomized Benchmarking for Non-Multiplicity-Free Groups With Applications to Subspace, Leakage, and Matchgate Randomized Benchmarking}},
\newblock \bibinfo{journal}{PRX Quantum} \bibinfo{volume}{2} (\bibinfo{year}{2021}) \bibinfo{pages}{010351}. \URLprefix \url{https://link.aps.org/doi/10.1103/PRXQuantum.2.010351}. \DOIprefix\doi{10.1103/PRXQuantum.2.010351}.
\bibitem[{Zhu(2017)}]{zhu_multiqubit_2017}
\bibinfo{author}{H.~Zhu},
\newblock \bibinfo{title}{Multiqubit {Clifford} groups are unitary 3-designs},
\newblock \bibinfo{journal}{Physical Review A} \bibinfo{volume}{96} (\bibinfo{year}{2017}) \bibinfo{pages}{062336}. \DOIprefix\doi{10.1103/PhysRevA.96.062336}.
\bibitem[{Webb(2016)}]{webb_clifford_2016}
\bibinfo{author}{Z.~Webb}, \bibinfo{title}{The {Clifford} group forms a unitary 3-design}, \bibinfo{year}{2016}. \href{http://arxiv.org/abs/1510.02769}{{\tt arXiv:1510.02769}}.
\bibitem[{Kueng et~al.(2015)}]{kueng_qubit_2015}
\bibinfo{author}{R.~Kueng}, et~al., \bibinfo{title}{{Qubit stabilizer states are complex projective 3-designs}}, \bibinfo{year}{2015}. \href{http://arxiv.org/abs/1510.02767}{{\tt arXiv:1510.02767}}.
\bibitem[{Anand and et. al.(2024)}]{Anand2024_qudit_designs}
\bibinfo{author}{N.~Anand}, \bibinfo{author}{et. al.}, \bibinfo{title}{{Qudit designs with applications to coherent states}}, \bibinfo{year}{2024}. \bibinfo{note}{In preparation}.
\bibitem[{Marshall et~al.(2017)Marshall, Campos~Venuti, and Zanardi}]{Marshall2017}
\bibinfo{author}{J.~Marshall}, \bibinfo{author}{L.~Campos~Venuti}, \bibinfo{author}{P.~Zanardi},
\newblock \bibinfo{title}{{Noise Suppression via Generalized-{{Markovian}} Processes}},
\newblock \bibinfo{journal}{Physical Review A} \bibinfo{volume}{96} (\bibinfo{year}{2017}) \bibinfo{pages}{052113}. \URLprefix \url{https://link.aps.org/doi/10.1103/PhysRevA.96.052113}. \DOIprefix\doi{10.1103/PhysRevA.96.052113}.
\bibitem[{Vlachos et~al.(2022)Vlachos, Zhang, Maurya, Marshall, Albash, and Levenson-Falk}]{Vlachos2022}
\bibinfo{author}{E.~Vlachos}, \bibinfo{author}{H.~Zhang}, \bibinfo{author}{V.~Maurya}, \bibinfo{author}{J.~Marshall}, \bibinfo{author}{T.~Albash}, \bibinfo{author}{E.~M. Levenson-Falk},
\newblock \bibinfo{title}{Master equation emulation and coherence preservation with classical control of a superconducting qubit},
\newblock \bibinfo{journal}{Phys. Rev. A} \bibinfo{volume}{106} (\bibinfo{year}{2022}) \bibinfo{pages}{062620}. \URLprefix \url{https://link.aps.org/doi/10.1103/PhysRevA.106.062620}. \DOIprefix\doi{10.1103/PhysRevA.106.062620}.
\bibitem[{Wiseman et~al.(2023)Wiseman, Cavalcanti, and Rieffel}]{Wiseman2023thoughtfullocal}
\bibinfo{author}{H.~M. Wiseman}, \bibinfo{author}{E.~G. Cavalcanti}, \bibinfo{author}{E.~G. Rieffel},
\newblock \bibinfo{title}{A "thoughtful" {L}ocal {F}riendliness no-go theorem: a prospective experiment with new assumptions to suit},
\newblock \bibinfo{journal}{{Quantum}} \bibinfo{volume}{7} (\bibinfo{year}{2023}) \bibinfo{pages}{1112}. \URLprefix \url{https://doi.org/10.22331/q-2023-09-14-1112}. \DOIprefix\doi{10.22331/q-2023-09-14-1112}.
\bibitem[{Swingle(2018)}]{swingle_unscrambling_2018}
\bibinfo{author}{B.~Swingle},
\newblock \bibinfo{title}{{Unscrambling the Physics of Out-of-Time-Order Correlators}},
\newblock \bibinfo{journal}{Nature Physics} \bibinfo{volume}{14} (\bibinfo{year}{2018}) \bibinfo{pages}{988--990}. \DOIprefix\doi{10.1038/s41567-018-0295-5}.
\bibitem[{Xu and Swingle(2024)}]{xu2023scrambling}
\bibinfo{author}{S.~Xu}, \bibinfo{author}{B.~Swingle},
\newblock \bibinfo{title}{{Scrambling Dynamics and Out-of-Time-Ordered Correlators in Quantum Many-Body Systems}},
\newblock \bibinfo{journal}{PRX Quantum} \bibinfo{volume}{5} (\bibinfo{year}{2024}) \bibinfo{pages}{010201}. \URLprefix \url{https://link.aps.org/doi/10.1103/PRXQuantum.5.010201}. \DOIprefix\doi{10.1103/PRXQuantum.5.010201}.
\bibitem[{Kitaev(2015)}]{kitaev_simple_2015}
\bibinfo{author}{A.~Kitaev}, \bibinfo{title}{A simple model of quantum holography (part 1)}, \bibinfo{year}{2015}. \URLprefix \url{http://online.kitp.ucsb.edu/online/entangled15/kitaev/}.
\bibitem[{Nahum et~al.(2018)Nahum, Vijay, and Haah}]{nahum2018operator}
\bibinfo{author}{A.~Nahum}, \bibinfo{author}{S.~Vijay}, \bibinfo{author}{J.~Haah},
\newblock \bibinfo{title}{{Operator Spreading in Random Unitary Circuits}},
\newblock \bibinfo{journal}{Phys. Rev. X} \bibinfo{volume}{8} (\bibinfo{year}{2018}) \bibinfo{pages}{021014}. \URLprefix \url{https://link.aps.org/doi/10.1103/PhysRevX.8.021014}. \DOIprefix\doi{10.1103/PhysRevX.8.021014}.
\bibitem[{Zanardi(2001)}]{zanardi_entanglement_2001}
\bibinfo{author}{P.~Zanardi},
\newblock \bibinfo{title}{Entanglement of quantum evolutions},
\newblock \bibinfo{journal}{Phys. Rev. A} \bibinfo{volume}{63} (\bibinfo{year}{2001}) \bibinfo{pages}{040304}. \URLprefix \url{https://link.aps.org/doi/10.1103/PhysRevA.63.040304}. \DOIprefix\doi{10.1103/PhysRevA.63.040304}.
\bibitem[{Hayden and Preskill(2007)}]{Hayden_2007}
\bibinfo{author}{P.~Hayden}, \bibinfo{author}{J.~Preskill},
\newblock \bibinfo{title}{{Black holes as mirrors: quantum information in random subsystems}},
\newblock \bibinfo{journal}{Journal of High Energy Physics} \bibinfo{volume}{2007} (\bibinfo{year}{2007}) \bibinfo{pages}{120–120}. \URLprefix \url{http://dx.doi.org/10.1088/1126-6708/2007/09/120}. \DOIprefix\doi{10.1088/1126-6708/2007/09/120}.
\bibitem[{Dubail(2017)}]{Dubail_2017}
\bibinfo{author}{J.~Dubail},
\newblock \bibinfo{title}{{Entanglement scaling of operators: a conformal field theory approach, with a glimpse of simulability of long-time dynamics in 1+1d}},
\newblock \bibinfo{journal}{Journal of Physics A: Mathematical and Theoretical} \bibinfo{volume}{50} (\bibinfo{year}{2017}) \bibinfo{pages}{234001}. \URLprefix \url{http://dx.doi.org/10.1088/1751-8121/aa6f38}. \DOIprefix\doi{10.1088/1751-8121/aa6f38}.
\bibitem[{Bravyi et~al.(2006)Bravyi, Hastings, and Verstraete}]{bravyi-lieb-robinson-2006}
\bibinfo{author}{S.~Bravyi}, \bibinfo{author}{M.~B. Hastings}, \bibinfo{author}{F.~Verstraete},
\newblock \bibinfo{title}{{Lieb-Robinson Bounds and the Generation of Correlations and Topological Quantum Order}},
\newblock \bibinfo{journal}{Phys. Rev. Lett.} \bibinfo{volume}{97} (\bibinfo{year}{2006}) \bibinfo{pages}{050401}. \URLprefix \url{https://link.aps.org/doi/10.1103/PhysRevLett.97.050401}. \DOIprefix\doi{10.1103/PhysRevLett.97.050401}.
\bibitem[{Mandrà et~al.(2021)Mandrà, Marshall, Rieffel, and Biswas}]{Mandra2021}
\bibinfo{author}{S.~Mandrà}, \bibinfo{author}{J.~Marshall}, \bibinfo{author}{E.~G. Rieffel}, \bibinfo{author}{R.~Biswas},
\newblock \bibinfo{title}{{HybridQ: A Hybrid Simulator for Quantum Circuits}},
\newblock in: \bibinfo{booktitle}{{2021 IEEE/ACM Second International Workshop on Quantum Computing Software (QCS)}}, \bibinfo{year}{2021}, pp. \bibinfo{pages}{99--109}. \DOIprefix\doi{10.1109/QCS54837.2021.00015}.
\bibitem[{Styliaris et~al.(2021)Styliaris, Anand, and Zanardi}]{styliaris_information_2021}
\bibinfo{author}{G.~Styliaris}, \bibinfo{author}{N.~Anand}, \bibinfo{author}{P.~Zanardi},
\newblock \bibinfo{title}{{Information Scrambling over Bipartitions: Equilibration, Entropy Production, and Typicality}},
\newblock \bibinfo{journal}{Phys. Rev. Lett.} \bibinfo{volume}{126} (\bibinfo{year}{2021}) \bibinfo{pages}{030601}. \URLprefix \url{https://link.aps.org/doi/10.1103/PhysRevLett.126.030601}. \DOIprefix\doi{10.1103/PhysRevLett.126.030601}.
\bibitem[{Zanardi and Anand(2021)}]{PhysRevA.103.062214}
\bibinfo{author}{P.~Zanardi}, \bibinfo{author}{N.~Anand},
\newblock \bibinfo{title}{{Information scrambling and chaos in open quantum systems}},
\newblock \bibinfo{journal}{Phys. Rev. A} \bibinfo{volume}{103} (\bibinfo{year}{2021}) \bibinfo{pages}{062214}. \URLprefix \url{https://link.aps.org/doi/10.1103/PhysRevA.103.062214}. \DOIprefix\doi{10.1103/PhysRevA.103.062214}.
\bibitem[{Anand and Zanardi(2022)}]{Anand2022brotocsquantum}
\bibinfo{author}{N.~Anand}, \bibinfo{author}{P.~Zanardi},
\newblock \bibinfo{title}{{BROTOC}s and {Q}uantum {I}nformation {S}crambling at {F}inite {T}emperature},
\newblock \bibinfo{journal}{{Quantum}} \bibinfo{volume}{6} (\bibinfo{year}{2022}) \bibinfo{pages}{746}. \URLprefix \url{https://doi.org/10.22331/q-2022-06-27-746}. \DOIprefix\doi{10.22331/q-2022-06-27-746}.
\bibitem[{Andreadakis et~al.(2023)Andreadakis, Anand, and Zanardi}]{PhysRevA.107.042217}
\bibinfo{author}{F.~Andreadakis}, \bibinfo{author}{N.~Anand}, \bibinfo{author}{P.~Zanardi},
\newblock \bibinfo{title}{{Scrambling of algebras in open quantum systems}},
\newblock \bibinfo{journal}{Phys. Rev. A} \bibinfo{volume}{107} (\bibinfo{year}{2023}) \bibinfo{pages}{042217}. \URLprefix \url{https://link.aps.org/doi/10.1103/PhysRevA.107.042217}. \DOIprefix\doi{10.1103/PhysRevA.107.042217}.
\bibitem[{Barch et~al.(2023)Barch, Anand, Marshall, Rieffel, and Zanardi}]{Barch2023}
\bibinfo{author}{B.~Barch}, \bibinfo{author}{N.~Anand}, \bibinfo{author}{J.~Marshall}, \bibinfo{author}{E.~Rieffel}, \bibinfo{author}{P.~Zanardi},
\newblock \bibinfo{title}{Scrambling and operator entanglement in local non-hermitian quantum systems},
\newblock \bibinfo{journal}{Phys. Rev. B} \bibinfo{volume}{108} (\bibinfo{year}{2023}) \bibinfo{pages}{134305}. \URLprefix \url{https://link.aps.org/doi/10.1103/PhysRevB.108.134305}. \DOIprefix\doi{10.1103/PhysRevB.108.134305}.
\bibitem[{Marvian et~al.(2019)Marvian, Lidar, and Hen}]{Marvian_2019}
\bibinfo{author}{M.~Marvian}, \bibinfo{author}{D.~A. Lidar}, \bibinfo{author}{I.~Hen},
\newblock \bibinfo{title}{{On the computational complexity of curing non-stoquastic Hamiltonians}},
\newblock \bibinfo{journal}{Nature Communications} \bibinfo{volume}{10} (\bibinfo{year}{2019}). \URLprefix \url{http://dx.doi.org/10.1038/s41467-019-09501-6}. \DOIprefix\doi{10.1038/s41467-019-09501-6}.
\bibitem[{Bringewatt and Brady(2022)}]{Bringewatt2022}
\bibinfo{author}{J.~Bringewatt}, \bibinfo{author}{L.~T. Brady},
\newblock \bibinfo{title}{Simultaneous stoquasticity},
\newblock \bibinfo{journal}{Phys. Rev. A} \bibinfo{volume}{105} (\bibinfo{year}{2022}) \bibinfo{pages}{062601}. \URLprefix \url{https://link.aps.org/doi/10.1103/PhysRevA.105.062601}. \DOIprefix\doi{10.1103/PhysRevA.105.062601}.
\bibitem[{Marshall et~al.(2019)Marshall, Venturelli, Hen, and Rieffel}]{Marshall2019}
\bibinfo{author}{J.~Marshall}, \bibinfo{author}{D.~Venturelli}, \bibinfo{author}{I.~Hen}, \bibinfo{author}{E.~G. Rieffel},
\newblock \bibinfo{title}{Power of {{Pausing}}: {{Advancing Understanding}} of {{Thermalization}} in {{Experimental Quantum Annealers}}},
\newblock \bibinfo{journal}{Physical Review Applied} \bibinfo{volume}{11} (\bibinfo{year}{2019}) \bibinfo{pages}{044083}. \URLprefix \url{https://link.aps.org/doi/10.1103/PhysRevApplied.11.044083}. \DOIprefix\doi{10.1103/PhysRevApplied.11.044083}.
\bibitem[{Albash and Marshall(2021)}]{Albash2021}
\bibinfo{author}{T.~Albash}, \bibinfo{author}{J.~Marshall},
\newblock \bibinfo{title}{{Comparing Relaxation Mechanisms in Quantum and Classical Transverse-Field Annealing}},
\newblock \bibinfo{journal}{Phys. Rev. Appl.} \bibinfo{volume}{15} (\bibinfo{year}{2021}) \bibinfo{pages}{014029}. \URLprefix \url{https://link.aps.org/doi/10.1103/PhysRevApplied.15.014029}. \DOIprefix\doi{10.1103/PhysRevApplied.15.014029}.
\bibitem[{Kapit and Oganesyan(2021)}]{Kapit_2021}
\bibinfo{author}{E.~Kapit}, \bibinfo{author}{V.~Oganesyan},
\newblock \bibinfo{title}{{Noise-tolerant quantum speedups in quantum annealing without fine tuning}},
\newblock \bibinfo{journal}{Quantum Science and Technology} \bibinfo{volume}{6} (\bibinfo{year}{2021}) \bibinfo{pages}{025013}. \URLprefix \url{https://dx.doi.org/10.1088/2058-9565/abd59a}. \DOIprefix\doi{10.1088/2058-9565/abd59a}.
\bibitem[{Mossi et~al.(2023)Mossi, Oganesyan, and Kapit}]{2306.10632}
\bibinfo{author}{G.~Mossi}, \bibinfo{author}{V.~Oganesyan}, \bibinfo{author}{E.~Kapit}, \bibinfo{title}{{Embedding quantum optimization problems using AC driven quantum ferromagnets}}, \bibinfo{year}{2023}. \href{http://arxiv.org/abs/arXiv:2306.10632}{{\tt arXiv:arXiv:2306.10632}}.
\bibitem[{Grattan et~al.(2023)Grattan, Barton, Feeney, Mossi, Patnaik, Sagal, Carr, Oganesyan, and Kapit}]{2311.17814}
\bibinfo{author}{G.~Grattan}, \bibinfo{author}{B.~A. Barton}, \bibinfo{author}{S.~Feeney}, \bibinfo{author}{G.~Mossi}, \bibinfo{author}{P.~Patnaik}, \bibinfo{author}{J.~C. Sagal}, \bibinfo{author}{L.~D. Carr}, \bibinfo{author}{V.~Oganesyan}, \bibinfo{author}{E.~Kapit}, \bibinfo{title}{{Exponential acceleration of macroscopic quantum tunneling in a Floquet Ising model}}, \bibinfo{year}{2023}. \href{http://arxiv.org/abs/arXiv:2311.17814}{{\tt arXiv:arXiv:2311.17814}}.
\bibitem[{Suri et~al.(2023)Suri, Wang, Hunt, and Xiao}]{PhysRevB.108.155409}
\bibinfo{author}{N.~Suri}, \bibinfo{author}{C.~Wang}, \bibinfo{author}{B.~M. Hunt}, \bibinfo{author}{D.~Xiao},
\newblock \bibinfo{title}{{Superlattice engineering of topology in massive Dirac fermions}},
\newblock \bibinfo{journal}{Phys. Rev. B} \bibinfo{volume}{108} (\bibinfo{year}{2023}) \bibinfo{pages}{155409}. \URLprefix \url{https://link.aps.org/doi/10.1103/PhysRevB.108.155409}. \DOIprefix\doi{10.1103/PhysRevB.108.155409}.
\bibitem[{Do et~al.(2020)Do, Wang, O'Gorman, Venturelli, Rieffel, and Frank}]{Do2020}
\bibinfo{author}{M.~Do}, \bibinfo{author}{Z.~Wang}, \bibinfo{author}{B.~O'Gorman}, \bibinfo{author}{D.~Venturelli}, \bibinfo{author}{E.~Rieffel}, \bibinfo{author}{J.~Frank}, \bibinfo{title}{{Planning for Compilation of a Quantum Algorithm for Graph Coloring}}, \bibinfo{year}{2020}. \href{http://arxiv.org/abs/2002.10917}{{\tt arXiv:2002.10917}}.
\bibitem[{Venturelli et~al.(2017)Venturelli, Do, Rieffel, and Frank}]{venturelli2017_ijcai2017p620}
\bibinfo{author}{D.~Venturelli}, \bibinfo{author}{M.~Do}, \bibinfo{author}{E.~Rieffel}, \bibinfo{author}{J.~Frank},
\newblock \bibinfo{title}{{Temporal Planning for Compilation of Quantum Approximate Optimization Circuits}},
\newblock in: \bibinfo{booktitle}{Proceedings of the Twenty-Sixth International Joint Conference on Artificial Intelligence, {IJCAI-17}}, \bibinfo{year}{2017}, pp. \bibinfo{pages}{4440--4446}. \URLprefix \url{https://doi.org/10.24963/ijcai.2017/620}. \DOIprefix\doi{10.24963/ijcai.2017/620}.
\bibitem[{Alam et~al.(2023)Alam, Berthusen, and Orth}]{Alam2023}
\bibinfo{author}{M.~S. Alam}, \bibinfo{author}{N.~F. Berthusen}, \bibinfo{author}{P.~P. Orth},
\newblock \bibinfo{title}{{Quantum logic gate synthesis as a Markov decision process}},
\newblock \bibinfo{journal}{npj Quantum Information} \bibinfo{volume}{9} (\bibinfo{year}{2023}). \URLprefix \url{https://doi.org/10.1038/s41534-023-00766-w}. \DOIprefix\doi{10.1038/s41534-023-00766-w}.
\bibitem[{Gold et~al.(2021)Gold, Paquette, Stockklauser, Reagor, Alam, Bestwick, Didier, Nersisyan, Oruc, Razavi, Scharmann, Sete, Sur, Venturelli, Winkleblack, Wudarski, Harburn, and Rigetti}]{gold2021entanglement}
\bibinfo{author}{A.~Gold}, \bibinfo{author}{J.~Paquette}, \bibinfo{author}{A.~Stockklauser}, \bibinfo{author}{M.~J. Reagor}, \bibinfo{author}{M.~S. Alam}, \bibinfo{author}{A.~Bestwick}, \bibinfo{author}{N.~Didier}, \bibinfo{author}{A.~Nersisyan}, \bibinfo{author}{F.~Oruc}, \bibinfo{author}{A.~Razavi}, \bibinfo{author}{B.~Scharmann}, \bibinfo{author}{E.~A. Sete}, \bibinfo{author}{B.~Sur}, \bibinfo{author}{D.~Venturelli}, \bibinfo{author}{C.~J. Winkleblack}, \bibinfo{author}{F.~Wudarski}, \bibinfo{author}{M.~Harburn}, \bibinfo{author}{C.~Rigetti}, \bibinfo{title}{{Entanglement Across Separate Silicon Dies in a Modular Superconducting Qubit Device}}, \bibinfo{year}{2021}. \href{http://arxiv.org/abs/2102.13293}{{\tt arXiv:2102.13293}}.

\end{thebibliography}

\label{sec:References}

\end{document}